\documentclass[preprint,authoryear,11pt]{elsarticle}
\usepackage{amsmath,float}
\usepackage{ccaption,rotating}
\usepackage{array}
\usepackage{subfig}
\usepackage{graphicx}
\def\thesubfigure{\arabic{subfigure}}
   \makeatletter
     \def\@thesubfigure{\thesubfigure:\space}
     \let\p@subfigure\relax

\def\fgas{$f_{gas}$}
\def\fbaryon{$f_{baryon}$}

\def\nclu{38 }
\def\critaccel{$1.17e^{-8} cm/s^{2}$}
\def\mondfbaryonmean{$\langle f_{baryon}\rangle=0.164 $}
\def\newtonfbaryonmean{$\langle f_{baryon}\rangle=0.1305 $}
\newcommand{\chandra}{{\it Chandra }}
\newcommand{\Xo}{\mbox{$S_{x0}$}}
\newcommand{\no}{\mbox{$n_{e0}$}}
\def\meanao{$1.17\pm 0.02 \times 10^{-7}$}
\def\meana0uncertainty{$1.8e^{-9}$}
\def\apj{Astrophysical Journal}
\def\mnras{Mon. Not. Royal Astronomical Soc.}
\def\apjs{Astrophysical Journal Supplement}
\def\araa{Annual Rev. Astron. Astroph.}
\def\aap{Astronomy and Astrophysics}
\def\apjl{Astrophysical Journal Lett.}

\journal{New Astronomy}

\begin{document}

\begin{frontmatter}

\author[label1]{Carl Blaksley}
\author[label2]{Massimiliano Bonamente}

\address[label1]{Department of Physics, University of Alabama, Huntsville, AL, USA and\\
Arnold Sommerfeld Center For Theoretical Physics, Faculty of Physics, Ludwig-Maximilians Universit\"{a}t,
M\"{u}nchen, Germany}
\address[label2]{Department of Physics, University of Alabama, Huntsville, AL, USA and\\
NASA National Space and Science Technology Center, Huntsville, AL, USA}

\title{Dark matter and  Modified Newtonian Dynamics in a sample of high-redshift galaxy
clusters observed with Chandra}

\begin{abstract}
We compare the measurement of the gravitational mass of \nclu\ high-redshift 
galaxy clusters observed by $\chandra$ 
using Modified Newtonian Dynamics (MOND) and standard Newtonian gravity. 
Our analysis confirms earlier findings that MOND cannot explain the difference between
the baryonic mass and the total mass inferred from the assumption of hydrostatic equilibrium. 
We also find that the baryon fraction at $r_{2500}$ using MOND is consistent with the 
Wilkinson Microwave Anisotropy Probe (WMAP) value of $\Omega_{B}/\Omega_{M}$.
\end{abstract}

\end{frontmatter}

\section{Introduction}
Since the introduction by Milgrom in 1983 \citep{milgrom1983}, MOND had success in explaining galaxy rotation curves using only the mass-to-light ratio as a free parameter, and was able to predict the applicability of the Tully-Fisher relation to low surface brightness galaxies before dynamic information on them was available \citep{Scarpareview}. These successes are not surprising, given the fact that MOND was created as a phenomenological theory in order to eliminate the need for dark matter in galaxies. 

While there have been numerous applications of MOND to individual galaxies, it is also important to study MOND on those scales where dark matter is believed to dominate, in particular that of galaxy clusters. There have been far fewer studies of MOND on this scale, with previous work on the subject including \citet{sanders1999}, \citet{aguirre}, \citet{Silk2005} and \citet{angus2007}. These results indicate that MOND does not eliminate the need for dark matter in galaxy clusters.
In this paper we use MOND to calculate gravitational and baryonic masses for a sample of \nclu\ galaxy clusters. 
The motivation of this work is the need to confirm the previous results by using a larger sample of massive clusters at high redshift (z=0.14-0.89).

We begin with a brief overview of the data and the models used, found in Section \ref{datasection}. Section \ref{mtotalcalc} describes the calculation of the gas mass using X-ray observation, the derivation of the MOND acceleration, and the calculation of total masses from hydrostatic equilibrium.  We then present our results in Section \ref{results} and our conclusions in Section \ref{conclusion}.  The cosmological parameters $h=0.7$, $\Omega_M$=0.3 and $\Omega_{\Lambda}$=0.7 are used throughout this work.

\section{Chandra X-ray Data And Data Modeling}
\label{datasection}
We analyze \chandra X-ray data from \nclu\ clusters in the redshift 
range $z$=0.14---0.89 (Table \ref{resultssummarytable}). 
The \chandra\ observations and the data modeling with the isothermal $\beta$ model
are presented in \citet{bonamente2006,laroque2006,bonamente2008}.
Here those aspects of the data
modeling and analysis that are relevant to the present investigation.

The electron density model is based on the isothermal spherical $\beta$-model
\citep{cavaliere1976, cavaliere1978}, which has the form
\begin{eqnarray}\label{eq:n_e}
n_e(r) = \no \left ( 1 + \frac{r^2}{r_c^2} \right )^{-3\beta/2},
\end{eqnarray}
where $n_{e0}$ is the central electron number density, 
$r_c$ is a core radius, and $\beta$ is a power-law index.
The radial profile of the X-ray surface brightness is obtained
via  integration along the line of sight, and results in the following
analytical expression:
\begin{equation}
S_x  =  \Xo \left ( 1 + \frac{\theta^2}{\theta_c^2} \right
)^{(1-6\beta)/2} \label{eq:Sx}.
\end{equation}

Best-fit model parameters and confidence intervals for all model
parameters are obtained using a Markov chain Monte Carlo (MCMC) method
described in detail by \citet{bonamente2004} and \citet{laroque2006}. 
 For each cluster, the Markov chain constrains the
parameters $S_{x0}$, $\beta$, $\theta_c$, $T_e$, and
the chemical abundance (see \citealt{laroque2006} for best-fit values).  We use the
cosmological parameters $h=0.7$, $\Omega_M$=0.3 and $\Omega_{\Lambda}$=0.7 to calculate each
cluster's angular diameter distance $D_A$
\citep[e.g.,][]{carroll1992}. 

All of our calculations are done out to a maxium radius of $r_{2500}$, the radius at which the cluster mass density is 2500 times the critical density,
which is also approximately the radius out to which our \chandra\ data
are sensitive without any extrapolation of the models.
In \citet{laroque2006} we have compared masses obtained using the simple
isothermal $\beta$ model with those obtained 
using a more complex non-isothermal $\beta$ model, and shown that the two methods yield
the same ratio of baryonic to total mass. 
The non-isothermal model included a double $\beta$ model distribution for the density
(Equation 7 in \citealt{laroque2006}),
and an arbitrary temperature profile that was constrained by assuming that
the plasma is in hydrostatic equilibrium with an NFW potential (Equation 8 in \citealt{laroque2006}). 
We therefore expect no significant bias in the study of MOND
masses using this simple model, which has the advantage of analytical expressions for the
observables and the masses.

\section{Calculation of gas, Newtonian and MOND Masses}
\label{mtotalcalc}
 For the isothermal $\beta$ model, the gas mass enclosed within a given cluster radius is given by \citep[e.g.][]{laroque2006}
\begin{equation}\label{eq:Mgas}
 M_{gas}(r)=A\int^{r/D_{A}}_{0}\Big(1+\dfrac{\theta^{2}}{\theta_{c}^{2}}\Big)^{-3\beta/2}\theta^{2}d\theta,
\end{equation}
 where $A=4\pi\mu_{e}n_{e_{\circ}}m_{p}D_{A}^{3}$, $\mu_{e}$ is the mean electron weight 
(calculated from the \chandra\ data, with typical value of $\mu_{e}=1.17$), 
and $D_{A}$ is the cluster distance. 
The central electron density $n_{e_{\circ}}$ is calculated from the 
X-ray surface brightness \citep{laroque2006, Birkinshaw1991} as
\begin{equation}\label{eq:ne0}
 n_{e_{\circ}}=\Bigg[\dfrac{S_{x_{\circ}}4\pi(1+z)^{4}(\mu_{H}/\mu_{e})\Gamma(3\beta)}{\Lambda_{eH}D_{A}\pi^{1/2}(3\beta-1/2)\theta_{c}}\Bigg]^{1/2},
\end{equation}
in which $\mu_{H}$ is the mean hydrogen weight (typical value $\mu_{H}=1.4$), and $\Lambda_{eH}$ is
the emissivity of the plasma.
The gas mass in Equation \eqref{eq:Mgas} is measured directly from observables and does not depend on the law of gravity in any way.

For the total mass, $M_{total}$, we solve the hydrostatic equilibrium equation using 
Newtonian gravitation and the ideal gas law, to obtain the Newtonian mass:
\begin{equation}\label{eq:mtotalunsimplified}
 M_{total}(r)=-\dfrac{kr^{2}}{G n_e(r) \mu_{tot}m_{p}}\Bigg[T_{e}(r)\dfrac{dn_{e}(r)}{dr}+n_{e}(r)\dfrac{dT_{e}(r)}{dr}\Bigg],
\end{equation}
in which $\mu_{tot}$ is the mean molecular weight (typical value $\mu_{tot}=0.61$).
Owing to our use of an isothermal model, the second term goes to zero and using Equation \eqref{eq:n_e} it becomes
\begin{equation}\label{eq:mtotalsimplified}
 M_{total}(r)=\dfrac{3\beta k T_{e}}{G\mu m_{p}}\dfrac{r^{3}}{r_{c}^{2}+r^{2}}.
\end{equation}

Starting with the most basic form of MOND \citep{Silk2005, Scarpareview},
we derive the gravitational mass based on the MOND law of gravity. 
The MOND acceleration ($a$) is related to the Newtonian acceleration ($a_{N}$) by the relation
\begin{equation} \label{eq:a}
 \frac{G M(r)}{r^2} = a_{N}=a \cdot \mu(\dfrac{a}{a_{\circ}}).
\end{equation}
where $a_{\circ}$ is a  critical acceleration that is meant to be a new physical constant and represents the acceleration below which Newtonian gravity is no longer valid \citep{milgrom1983, Scarpareview}; the function
$\mu$ is any interpolating function that provides a transition between the two regimes.
 We take the critical acceleration to be $a_0=$\critaccel\ 
as used in \cite{angus2007}; this value is derived from galaxy rotation curve fits \cite{Scarpareview}. 
In Section \ref{results} we also discuss the results obtained using a variable
value of the critical acceleration.

 The interpolating function $\mu(x)$ is any function which yields the appropriate asymptotic behavior,  
\begin{equation*}
\mu(x)=\left\{ \begin{array}{ll}
1 & \textrm{ for $x\gg1$}\\
x & \textrm{ for $x\ll1$}
 \end{array} \right.
\end{equation*}
The most used interpolating function \citep{Scarpareview, Silk2005}, and the one which we use here, is 
\begin{equation} \label{eq:mu}
 \mu(x)=\dfrac{x}{\sqrt{1+x^{2}}}.
\end{equation}

Using Equation \eqref{eq:mu} in Equation \eqref{eq:a} and solving for $a$, gives us the relation
\begin{equation}\label{eq:amond}
 a^{2}=a_{N}^{2}\Bigg(\dfrac{1}{2}+\dfrac{1}{2}\sqrt{1+\dfrac{4a_{\circ}^{2}}{a_{N}^{2}}}\Bigg).
\end{equation}
Equation \eqref{eq:amond} is the MOND acceleration as a function of the Newtonian acceleration and the critical acceleration parameter \citep[see also][]{Silk2005}. 
We use the hydrostatic equilibrium equation according to MOND,
\begin{equation} \label{eq:hydrostatic}
 -\dfrac{1}{\rho_g(r)}\dfrac{dP_g}{dr}=a(r),
\end{equation}
($\rho_g(r)$ is the gas density and $P_g$ its pressure) 
in order to determine the MOND equation for total mass analogous to Equation \eqref{eq:mtotalsimplified} for Newtonian gravity. 
Using Equations \eqref{eq:amond} and \eqref{eq:hydrostatic} 
with  $a_N=G M(r)/r^2$ leads to the elimination of the Newtonian acceleration $a_{N}$:
\begin{equation}
 \dfrac{GM(r)}{r^{2}}=\dfrac{\Bigg(\dfrac{-1}{\rho(r)}\dfrac{dP}{dr}\Bigg)^{2}}{\sqrt{\Bigg(a_{\circ}^{2}+\Bigg(\dfrac{-1}{\rho(r)}\dfrac{dP}{dr}\Bigg)^{2}\Bigg)}}.
\end{equation}
For the $\beta$ model profile of density we obtain:
\begin{equation}\label{eq:M_MONDunsimplified}
M_{MOND}(r)=\dfrac{\dfrac{r^2}{G}\Bigg(\dfrac{3\beta KT}{\mu m_{p}}\dfrac{r}{r_{c}^{2}+r^{2}}\Bigg)}{\sqrt{a_{\circ}^{2}+\Bigg(\dfrac{3\beta KT}{\mu m_{p}}\dfrac{r}{r_{c}^{2}+r^{2}}\Bigg)^2}}.
\end{equation}
This equation can be further simplified when one realizes that the numerator 
is the total mass according to Newtonian dynamics, Equation \eqref{eq:mtotalsimplified}. 
This simplification gives us a final equation for the total 
mass based on the hydrostatic equilibrium condition and using MOND gravity.

\begin{equation}\label{eq:M_MOND}
{M_{MOND}(r)=\dfrac{M_{Newton}(r)}{\sqrt{1+\Bigg(\dfrac{a_{\circ}\mu m_{p}}{3\beta KT}\dfrac{(r_{c}^{2}+r^{2})}{r}\Bigg)^2}}}
\end{equation}
This equation displays the appropriate asymptotic behavior, namely that for a 
critical acceleration of zero we recover the Newtonian $M_{total}$. 

\section{Results}

\label{results}

\subsection{Comparison of Newtonian and MOND gravitational masses }
%


Having found an equation for the MOND total mass ($M_{MOND}$)
 we calculate $M_{gas}$, $M_{total}$ and $M_{MOND}$ for each of the $\nclu\ $ 
galaxy clusters in our sample out to a radius of $r_{2500}$.  
In each case $M_{gas}$ was calculated using Equation \eqref{eq:Mgas}, 
$M_{total}$ by means of Equation \eqref{eq:mtotalsimplified}, 
and $M_{MOND}$ using Equation \eqref{eq:M_MOND}. 
For each cluster, the MOND gravitational mass is lower than the the
Newtonian gravitational mass by just $\sim 20-30$\%, and both are significantly larger  than the gas mass
$M_{gas}$ (see Table \ref{resultssummarytable}).
The radial profiles of the masses are shown for all clusters in Appendix A.

In order to calculate the baryonic mass we approximate the stellar contribution by 
$M_{*} = 0.15 M_{gas}$, as is done in \cite{Silk2005}. 
We therefore estimate the total baryonic mass as $M_{baryon}=M_{gas}+M_{*}$. 
We find that the average $f_{gas}(r_{2500})$ for MOND is 14.2$\pm$0.2\%, and that
the average  $f_{baryon}(r_{2500})$ is 16.4$\pm$0.2\% (Table \ref{massfractiontable}). 
These results agree with the results obtained by \citet{Silk2005}, 
which report a value of $f_{baryon}$=20.2$\pm$2.0\%
at a larger radius of $r_{1000}$.
We find that,
even with the inclusion of $M_{*}$, the baryon mass is still significantly lower than 
the total MOND mass for every cluster, confirming the fact that MOND is incapable of 
eliminating the need for dark matter in clusters. 
The mean \fgas\ and \fbaryon\ in Table \ref{massfractiontable}
are calculated as the weighted mean of all measurement, in which the weight is
the standard error of each measurement. We also calculate the
un-weighted means and their root-mean-square errors as \fgas=$0.109\pm0.003$~(Newton) 
and \fgas=$0.139\pm0.004$~(MOND), and \fbaryon=$0.125\pm0.004$~(Newton)
and \fbaryon=$0.160\pm0.005$~(MOND).

\begin{table}
\centering
\caption{Cluster Parameters And Masses For The Isothermal $\beta$-Model \label{resultssummarytable}}
\resizebox{15cm}{!}{
\begin{tabular}{lcccccr}\hline
\hline
Cluster &z & $r_{2500}$ & $D_{A}$& $M_{gas}$ & $M_{total}$ & $M_{Mond}$ \\
 &$$&$(arcsec)$&$(Gpc)$&$(10^{13} M_{\odot})$&$(10^{14} M_{\odot})$&$(10^{14} M_{\odot})$\\
\hline
Abell1413\dotfill &$0.143$& $201.6_{-4.6}^{+5.2}$& $0.52$ & $2.63_{-0.1}^{+0.11}$ & $2.15_{-0.14}^{+0.17}$  & $1.51_{-0.12}^{+0.14}$\\ 
Abell1689\dotfill &$0.18$& $217.1_{-5.5}^{+5.1}$& $0.63$ & $5.1_{-0.17}^{+0.17}$ & $4.95_{-0.36}^{+0.36}$  & $3.97_{-0.33}^{+0.32}$\\ 
Abell1835\dotfill &$0.25$& $171.1_{-4.4}^{+5}$& $0.81$ & $5.81_{-0.2}^{+0.23}$ & $5.55_{-0.42}^{+0.5}$  & $4.59_{-0.38}^{+0.46}$\\ 
Abell1914\dotfill &$0.17$& $226.7_{-4}^{+4.6}$& $0.6$ & $4.87_{-0.1}^{+0.12}$ & $4.82_{-0.25}^{+0.3}$  & $3.85_{-0.22}^{+0.27}$\\ 
Abell1995\dotfill &$0.32$& $133.4_{-4.7}^{+4.5}$& $0.96$ & $3.51_{-0.14}^{+0.14}$ & $4.74_{-0.48}^{+0.5}$  & $3.91_{-0.44}^{+0.46}$\\ 
Abell2111\dotfill &$0.23$& $140.5_{-8.9}^{+9.8}$& $0.76$ & $2.19_{-0.23}^{+0.27}$ & $2.49_{-0.44}^{+0.56}$  & $1.84_{-0.37}^{+0.48}$\\ 
Abell2163\dotfill &$0.2$& $206.7_{-2.9}^{+3}$& $0.68$ & $8.08_{-0.2}^{+0.21}$ & $5.49_{-0.23}^{+0.24}$  & $4.47_{-0.21}^{+0.22}$\\ 
Abell2204\dotfill &$0.15$& $256.1_{-11}^{+13}$& $0.54$ & $4.74_{-0.24}^{+0.27}$ & $4.96_{-0.64}^{+0.8}$  & $3.95_{-0.57}^{+0.72}$\\ 
Abell2218\dotfill &$0.18$& $190_{-5.4}^{+5.7}$& $0.63$ & $3.02_{-0.12}^{+0.13}$ & $3.32_{-0.27}^{+0.31}$  & $2.53_{-0.24}^{+0.27}$\\ 
Abell2259\dotfill &$0.16$& $172.1_{-7.9}^{+8.3}$& $0.57$ & $1.82_{-0.14}^{+0.14}$ & $1.79_{-0.24}^{+0.27}$  & $1.23_{-0.19}^{+0.22}$\\ 
Abell2261\dotfill &$0.22$& $148.3_{-6.3}^{+6.7}$& $0.73$ & $3.03_{-0.19}^{+0.2}$ & $2.56_{-0.31}^{+0.36}$  & $1.9_{-0.26}^{+0.31}$\\ 
Abell267\dotfill &$0.23$& $131.3_{-7.5}^{+8.5}$& $0.76$ & $2.24_{-0.18}^{+0.2}$ & $2.03_{-0.33}^{+0.42}$  & $1.46_{-0.27}^{+0.35}$\\ 
Abell370\dotfill &$0.38$& $97.95_{-4.1}^{+4.1}$& $1.07$ & $2.78_{-0.2}^{+0.21}$ & $2.78_{-0.33}^{+0.37}$  & $2.19_{-0.29}^{+0.33}$\\ 
Abell586\dotfill &$0.17$& $181.6_{-7}^{+8}$& $0.6$ & $2.27_{-0.11}^{+0.13}$ & $2.48_{-0.28}^{+0.34}$  & $1.8_{-0.23}^{+0.29}$\\ 
Abell611\dotfill &$0.29$& $110.3_{-3.5}^{+3.6}$& $0.9$ & $2.37_{-0.11}^{+0.11}$ & $2.13_{-0.2}^{+0.22}$  & $1.58_{-0.17}^{+0.19}$\\ 
Abell665\dotfill &$0.18$& $160.5_{-3.3}^{+3.6}$& $0.63$ & $2.63_{-0.095}^{+0.1}$ & $2_{-0.12}^{+0.14}$  & $1.41_{-0.099}^{+0.11}$\\ 
Abell68\dotfill &$0.26$& $153_{-9.3}^{+10}$& $0.83$ & $3.65_{-0.31}^{+0.34}$ & $4.32_{-0.74}^{+0.91}$  & $3.48_{-0.66}^{+0.82}$\\ 
Abell697\dotfill &$0.28$& $133.1_{-4.9}^{+5}$& $0.88$ & $4.4_{-0.26}^{+0.28}$ & $3.46_{-0.37}^{+0.4}$  & $2.73_{-0.32}^{+0.36}$\\ 
Abell773\dotfill &$0.22$& $148.9_{-5.3}^{+5.6}$& $0.73$ & $2.74_{-0.15}^{+0.17}$ & $2.59_{-0.27}^{+0.3}$  & $1.93_{-0.23}^{+0.26}$\\ 
CLJ0016+1609\dotfill &$0.54$& $79.85_{-3}^{+3}$& $1.31$ & $4.38_{-0.29}^{+0.29}$ & $3.33_{-0.37}^{+0.39}$  & $2.8_{-0.34}^{+0.36}$\\ 
CLJ1226+3332\dotfill &$0.89$& $66.02_{-6.5}^{+7.4}$& $1.6$ & $3.89_{-0.46}^{+0.51}$ & $5.21_{-1.4}^{+2}$  & $4.8_{-1.3}^{+1.9}$\\ 
MACSJ0647.7+7015\dotfill &$0.58$& $91.94_{-5.7}^{+6.2}$& $1.36$ & $4.91_{-0.42}^{+0.47}$ & $5.97_{-1.1}^{+1.3}$  & $5.31_{-1}^{+1.2}$\\ 
MACSJ0744.8+3927\dotfill &$0.69$& $58.88_{-3.2}^{+3.5}$& $1.47$ & $3.08_{-0.25}^{+0.27}$ & $2.26_{-0.35}^{+0.42}$  & $1.89_{-0.32}^{+0.39}$\\ 
MACSJ1149.5+2223\dotfill &$0.54$& $70.64_{-3.6}^{+3.9}$& $1.31$ & $3.09_{-0.3}^{+0.34}$ & $2.31_{-0.34}^{+0.4}$  & $1.86_{-0.3}^{+0.37}$\\ 
MACSJ1311.0-0310\dotfill &$0.49$& $73.92_{-6.6}^{+7.6}$& $1.25$ & $2.12_{-0.21}^{+0.25}$ & $2.17_{-0.53}^{+0.74}$  & $1.71_{-0.47}^{+0.66}$\\ 
MACSJ1423.8+2404\dotfill &$0.55$& $65.93_{-2}^{+2.1}$& $1.32$ & $2.26_{-0.096}^{+0.1}$ & $1.94_{-0.18}^{+0.2}$  & $1.54_{-0.16}^{+0.17}$\\ 
MACSJ2129.4-0741\dotfill &$0.57$& $72.42_{-4.1}^{+4.6}$& $1.35$ & $3.27_{-0.26}^{+0.3}$ & $2.82_{-0.45}^{+0.57}$  & $2.35_{-0.41}^{+0.53}$\\ 
MACSJ2214.9-1359\dotfill &$0.482$& $91.31_{-4.9}^{+5.1}$& $1.23$ & $3.94_{-0.31}^{+0.32}$ & $3.85_{-0.59}^{+0.68}$  & $3.24_{-0.54}^{+0.63}$\\ 
MACSJ2228.5+2036\dotfill &$0.41$& $83.61_{-3.8}^{+4.1}$& $1.12$ & $2.78_{-0.21}^{+0.23}$ & $2.05_{-0.27}^{+0.32}$  & $1.57_{-0.23}^{+0.28}$\\ 
MS0451.6-0305\dotfill &$0.55$& $82.18_{-3.5}^{+3.7}$& $1.32$ & $4.8_{-0.3}^{+0.32}$ & $3.76_{-0.46}^{+0.53}$  & $3.2_{-0.43}^{+0.49}$\\ 
MS1054.5-0321\dotfill &$0.83$& $33.74_{-8.5}^{+7.1}$& $1.57$ & $0.905_{-0.49}^{+0.55}$ & $0.612_{-0.36}^{+0.47}$  & $0.454_{-0.29}^{+0.41}$\\ 
MS1137.5+6625\dotfill &$0.78$& $41.63_{-2.9}^{+3.3}$& $1.54$ & $1.26_{-0.14}^{+0.15}$ & $1.02_{-0.2}^{+0.26}$  & $0.802_{-0.18}^{+0.23}$\\ 
MS1358.4+6245\dotfill &$0.33$& $113.4_{-5.2}^{+5.8}$& $0.98$ & $2.53_{-0.17}^{+0.19}$ & $3.13_{-0.41}^{+0.51}$  & $2.47_{-0.36}^{+0.45}$\\ 
MS2053.7-0449\dotfill &$0.58$& $54.2_{-4.6}^{+5.1}$& $1.36$ & $0.933_{-0.12}^{+0.13}$ & $1.22_{-0.29}^{+0.38}$  & $0.921_{-0.24}^{+0.33}$\\ 
RXJ1347.5-1145\dotfill &$0.45$& $122.3_{-3.6}^{+3.8}$& $1.19$ & $8.86_{-0.35}^{+0.37}$ & $8.08_{-0.69}^{+0.77}$  & $7.18_{-0.65}^{+0.73}$\\ 
RXJ1716.4+6708\dotfill &$0.81$& $45.01_{-3.9}^{+4.4}$& $1.56$ & $1.23_{-0.16}^{+0.19}$ & $1.39_{-0.33}^{+0.45}$  & $1.14_{-0.3}^{+0.41}$\\ 
RXJ2129.7+0005\dotfill &$0.24$& $128.5_{-4.9}^{+5.3}$& $0.78$ & $2.57_{-0.14}^{+0.16}$ & $2.08_{-0.23}^{+0.27}$  & $1.5_{-0.19}^{+0.22}$\\ 
ZW3146\dotfill &$0.29$& $131.5_{-2.5}^{+2.6}$& $0.9$ & $4.44_{-0.11}^{+0.11}$ & $3.62_{-0.2}^{+0.22}$  & $2.87_{-0.18}^{+0.19}$\\ 
\hline
\end{tabular}
}\end{table}
\begin{table}
\centering
\caption{Newtonian And MOND Mass Fractions\label{massfractiontable}}
\resizebox{15cm}{!}{
\begin{tabular}{lcc|cr}\hline
\hline
Cluster & $f_{\{gas\}}$&&$f_{\{baryon\}}$\\ 
& Newton& MOND &Newton & MOND\\
\hline
Abell1413\dotfill & $0.123_{-0.0045}^{+0.0041}$& $0.175_{-0.0085}^{+0.008}$& $0.141_{-0.0052}^{+0.0048}$& $0.201_{-0.0098}^{+0.0092}$ \\ 
Abell1689\dotfill & $0.103_{-0.0041}^{+0.0045}$& $0.129_{-0.0061}^{+0.0068}$& $0.118_{-0.0047}^{+0.0051}$& $0.148_{-0.007}^{+0.0078}$ \\ 
Abell1835\dotfill & $0.105_{-0.005}^{+0.0047}$& $0.127_{-0.0071}^{+0.0068}$& $0.12_{-0.0057}^{+0.0054}$& $0.146_{-0.0082}^{+0.0078}$ \\ 
Abell1914\dotfill & $0.101_{-0.0037}^{+0.0034}$& $0.127_{-0.0055}^{+0.0051}$& $0.116_{-0.0042}^{+0.0039}$& $0.146_{-0.0063}^{+0.0058}$ \\ 
Abell1995\dotfill & $0.0741_{-0.0046}^{+0.0051}$& $0.0897_{-0.0064}^{+0.0073}$& $0.0852_{-0.0053}^{+0.0059}$& $0.103_{-0.0074}^{+0.0084}$ \\ 
Abell2111\dotfill & $0.0883_{-0.0077}^{+0.0078}$& $0.119_{-0.014}^{+0.014}$& $0.101_{-0.0088}^{+0.009}$& $0.137_{-0.016}^{+0.017}$ \\ 
Abell2163\dotfill & $0.147_{-0.0025}^{+0.0026}$& $0.18_{-0.004}^{+0.004}$& $0.169_{-0.0029}^{+0.003}$& $0.208_{-0.0046}^{+0.0047}$ \\ 
Abell2204\dotfill & $0.0957_{-0.0099}^{+0.0094}$& $0.12_{-0.014}^{+0.014}$& $0.11_{-0.011}^{+0.011}$& $0.138_{-0.016}^{+0.016}$ \\ 
Abell2218\dotfill & $0.0909_{-0.0044}^{+0.0042}$& $0.119_{-0.0071}^{+0.0071}$& $0.105_{-0.005}^{+0.0049}$& $0.137_{-0.0082}^{+0.0081}$ \\ 
Abell2259\dotfill & $0.102_{-0.0069}^{+0.0069}$& $0.148_{-0.013}^{+0.014}$& $0.117_{-0.008}^{+0.0079}$& $0.17_{-0.015}^{+0.016}$ \\ 
Abell2261\dotfill & $0.118_{-0.0082}^{+0.0084}$& $0.159_{-0.014}^{+0.015}$& $0.136_{-0.0095}^{+0.0096}$& $0.183_{-0.016}^{+0.017}$ \\ 
Abell267\dotfill & $0.111_{-0.011}^{+0.011}$& $0.154_{-0.019}^{+0.02}$& $0.127_{-0.013}^{+0.013}$& $0.177_{-0.022}^{+0.023}$ \\ 
Abell370\dotfill & $0.1_{-0.0052}^{+0.0054}$& $0.127_{-0.0084}^{+0.009}$& $0.115_{-0.0059}^{+0.0062}$& $0.146_{-0.0096}^{+0.01}$ \\ 
Abell586\dotfill & $0.0916_{-0.0071}^{+0.0068}$& $0.126_{-0.012}^{+0.012}$& $0.105_{-0.0081}^{+0.0078}$& $0.145_{-0.014}^{+0.014}$ \\ 
Abell611\dotfill & $0.111_{-0.0057}^{+0.0058}$& $0.15_{-0.0097}^{+0.01}$& $0.128_{-0.0066}^{+0.0067}$& $0.173_{-0.011}^{+0.012}$ \\ 
Abell665\dotfill & $0.131_{-0.0038}^{+0.0036}$& $0.187_{-0.0074}^{+0.0072}$& $0.151_{-0.0044}^{+0.0042}$& $0.215_{-0.0085}^{+0.0083}$ \\ 
Abell68\dotfill & $0.0843_{-0.0083}^{+0.0089}$& $0.105_{-0.012}^{+0.014}$& $0.097_{-0.0095}^{+0.01}$& $0.12_{-0.014}^{+0.016}$ \\ 
Abell697\dotfill & $0.127_{-0.0063}^{+0.0066}$& $0.161_{-0.01}^{+0.011}$& $0.146_{-0.0073}^{+0.0076}$& $0.186_{-0.012}^{+0.012}$ \\ 
Abell773\dotfill & $0.106_{-0.0055}^{+0.0058}$& $0.142_{-0.0095}^{+0.01}$& $0.121_{-0.0063}^{+0.0067}$& $0.163_{-0.011}^{+0.012}$ \\ 
CLJ0016+1609\dotfill & $0.131_{-0.0063}^{+0.0067}$& $0.156_{-0.0091}^{+0.0099}$& $0.151_{-0.0073}^{+0.0078}$& $0.18_{-0.01}^{+0.011}$ \\ 
CLJ1226+3332\dotfill & $0.0746_{-0.014}^{+0.015}$& $0.081_{-0.016}^{+0.018}$& $0.0858_{-0.016}^{+0.018}$& $0.0931_{-0.018}^{+0.021}$ \\ 
MACSJ0647.7+7015\dotfill & $0.0821_{-0.0084}^{+0.0091}$& $0.0925_{-0.011}^{+0.012}$& $0.0945_{-0.0097}^{+0.011}$& $0.106_{-0.012}^{+0.014}$ \\ 
MACSJ0744.8+3927\dotfill & $0.136_{-0.012}^{+0.012}$& $0.163_{-0.016}^{+0.018}$& $0.157_{-0.013}^{+0.014}$& $0.187_{-0.019}^{+0.02}$ \\ 
MACSJ1149.5+2223\dotfill & $0.134_{-0.0078}^{+0.0077}$& $0.166_{-0.013}^{+0.013}$& $0.154_{-0.009}^{+0.0088}$& $0.191_{-0.014}^{+0.015}$ \\ 
MACSJ1311.0-0310\dotfill & $0.0979_{-0.017}^{+0.019}$& $0.124_{-0.025}^{+0.03}$& $0.113_{-0.019}^{+0.022}$& $0.142_{-0.028}^{+0.035}$ \\ 
MACSJ1423.8+2404\dotfill & $0.116_{-0.0061}^{+0.0064}$& $0.147_{-0.0093}^{+0.0098}$& $0.134_{-0.007}^{+0.0073}$& $0.168_{-0.011}^{+0.011}$ \\ 
MACSJ2129.4-0741\dotfill & $0.116_{-0.011}^{+0.011}$& $0.14_{-0.016}^{+0.016}$& $0.134_{-0.013}^{+0.013}$& $0.161_{-0.018}^{+0.019}$ \\ 
MACSJ2214.9-1359\dotfill & $0.102_{-0.0084}^{+0.0088}$& $0.122_{-0.012}^{+0.013}$& $0.118_{-0.0096}^{+0.01}$& $0.14_{-0.013}^{+0.015}$ \\ 
MACSJ2228.5+2036\dotfill & $0.135_{-0.0087}^{+0.0087}$& $0.177_{-0.014}^{+0.015}$& $0.155_{-0.01}^{+0.01}$& $0.203_{-0.017}^{+0.018}$ \\ 
MS0451.6-0305\dotfill & $0.127_{-0.0085}^{+0.0088}$& $0.15_{-0.012}^{+0.012}$& $0.147_{-0.0098}^{+0.01}$& $0.172_{-0.013}^{+0.014}$ \\ 
MS1054.5-0321\dotfill & $0.145_{-0.012}^{+0.015}$& $0.191_{-0.027}^{+0.051}$& $0.166_{-0.014}^{+0.018}$& $0.219_{-0.031}^{+0.059}$ \\ 
MS1137.5+6625\dotfill & $0.123_{-0.014}^{+0.015}$& $0.157_{-0.022}^{+0.024}$& $0.141_{-0.016}^{+0.017}$& $0.18_{-0.025}^{+0.028}$ \\ 
MS1358.4+6245\dotfill & $0.0809_{-0.006}^{+0.0061}$& $0.102_{-0.0093}^{+0.0097}$& $0.093_{-0.0069}^{+0.007}$& $0.118_{-0.011}^{+0.011}$ \\ 
MS2053.7-0449\dotfill & $0.0761_{-0.011}^{+0.012}$& $0.101_{-0.017}^{+0.02}$& $0.0875_{-0.012}^{+0.013}$& $0.116_{-0.02}^{+0.023}$ \\ 
RXJ1347.5-1145\dotfill & $0.11_{-0.0055}^{+0.0056}$& $0.123_{-0.0069}^{+0.0071}$& $0.126_{-0.0063}^{+0.0065}$& $0.142_{-0.008}^{+0.0082}$ \\ 
RXJ1716.4+6708\dotfill & $0.0883_{-0.011}^{+0.013}$& $0.108_{-0.016}^{+0.019}$& $0.102_{-0.013}^{+0.014}$& $0.124_{-0.019}^{+0.022}$ \\ 
RXJ2129.7+0005\dotfill & $0.124_{-0.0077}^{+0.0082}$& $0.171_{-0.014}^{+0.015}$& $0.142_{-0.0088}^{+0.0094}$& $0.197_{-0.016}^{+0.017}$ \\ 
ZW3146\dotfill & $0.123_{-0.0043}^{+0.0044}$& $0.154_{-0.0064}^{+0.0066}$& $0.141_{-0.0049}^{+0.005}$& $0.178_{-0.0073}^{+0.0076}$ \\ 
 \hline
Weighted Mean
 & $0.114\pm0.001$ & $0.142\pm0.002$ & $0.131\pm0.001$ & $0.164\pm0.002$ \\
\hline
\end{tabular}
}\end{table}

\begin{figure}[!h]
\includegraphics[width=2.5in]{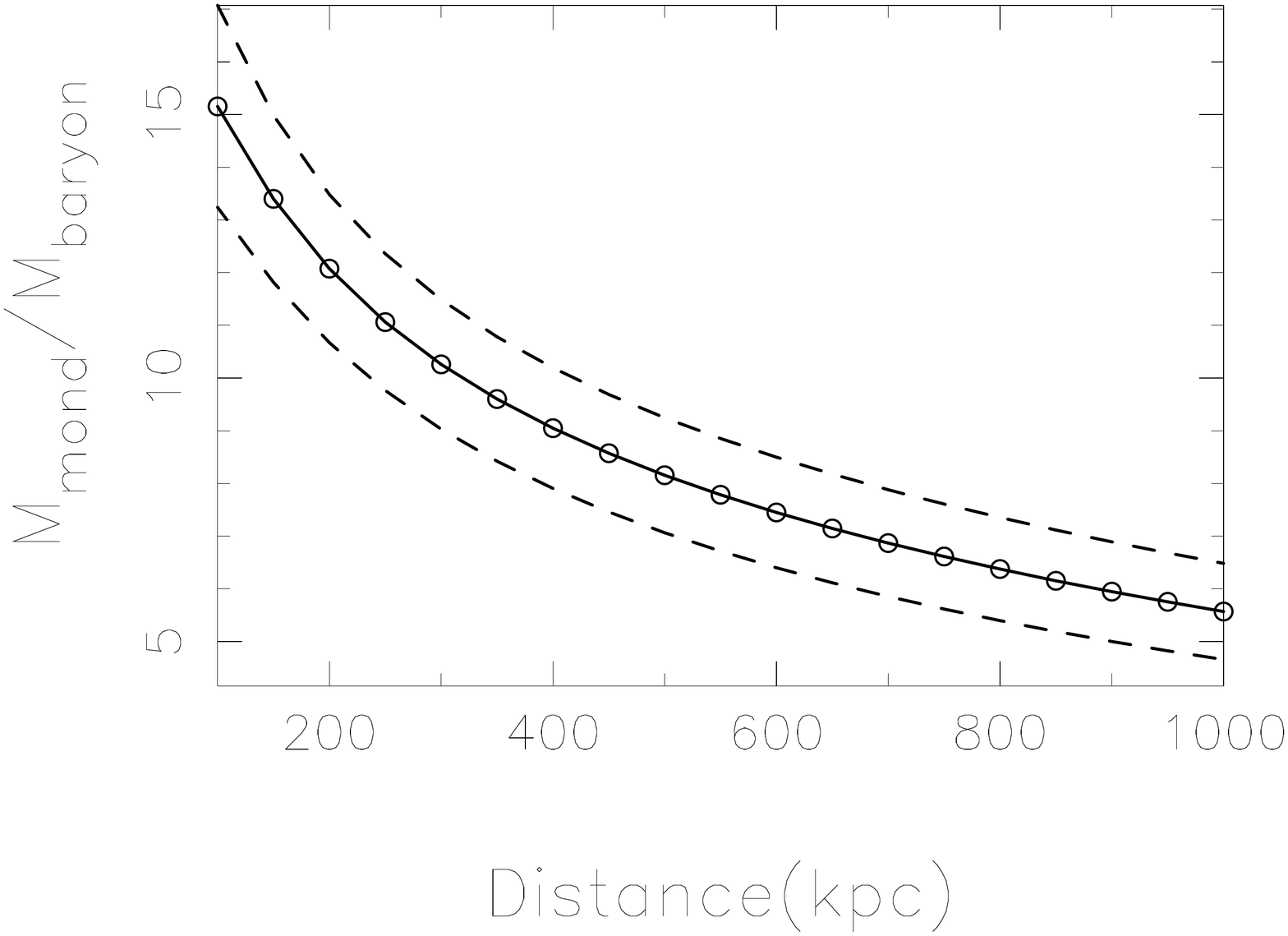}
\includegraphics[width=2.5in]{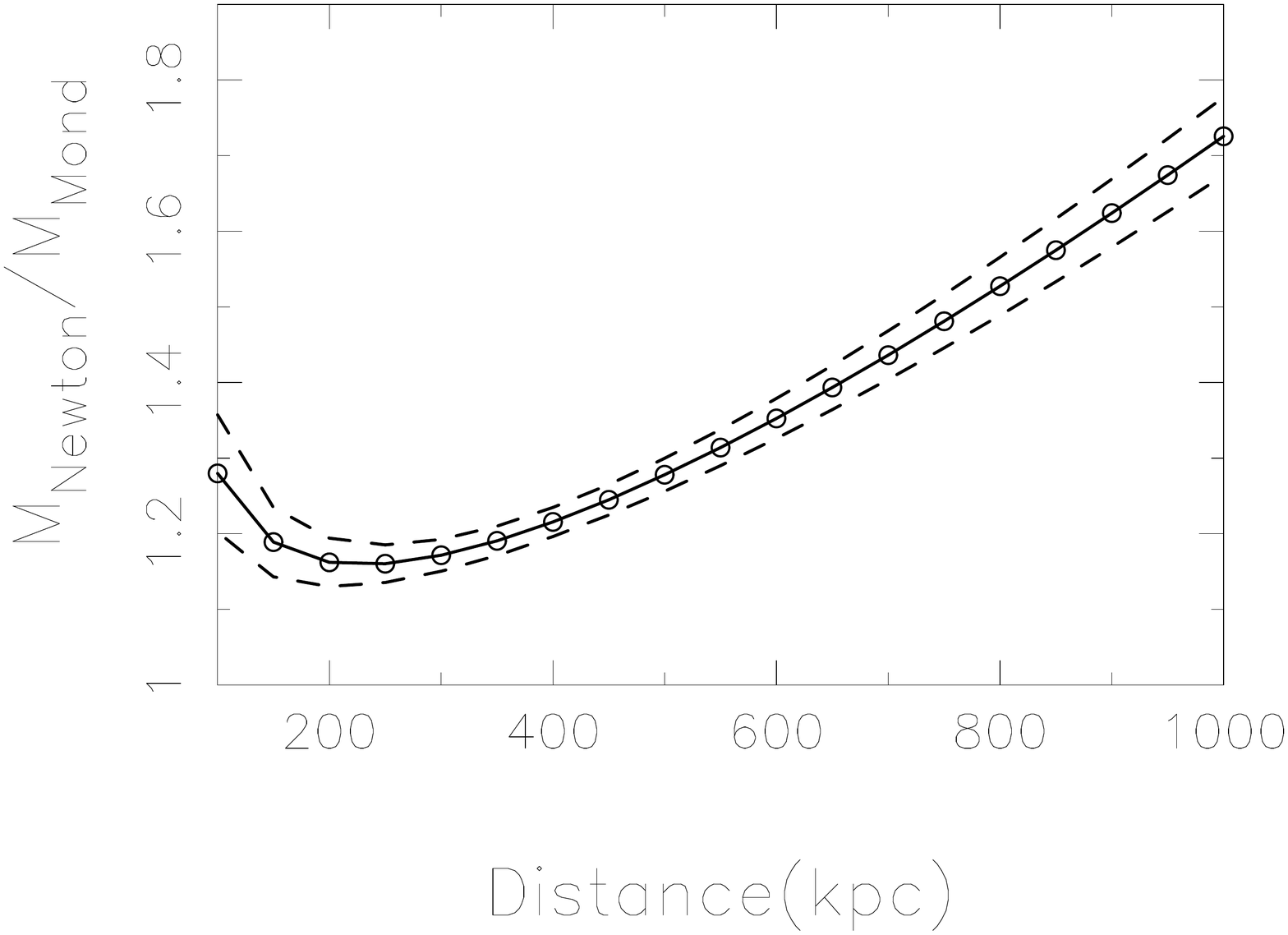}
\caption{Average distribution of $f_{baryon}$ and $M_{total}/M_{MOND}$ for the
38 clusters; the dashed lines represent the root-mean-square uncertainty of the 38 measurements
at each radius \label{fig:average-masses}}.
\end{figure}

We show the average distributions of $f_{baryon}$ and $M_{total}/M_{MOND}$
as function of physical radius in Figure \ref{fig:average-masses}. The distribution
of $f_{baryon}$ shows that the dynamical mass is always significantly higher
than the baryonic mass. The distribution of $M_{total}/M_{MOND}$ shows that
the difference between the Newtonian and MOND masses are largest near the cluster
core and at the largest radii, where the gravitational accelerations are lowest
and therefore the MOND correction largest. Typical values of $r_{2500}$ are
between 0.5 and 1 Mpc, therefore we do not extrapolate these plots beyond 1~Mpc.
The \chandra\ data presented in this paper were fit with a 100~kpc core cut
\citep{laroque2006}, and therefore we do not extrapolate our models below this radius either.

The  cluster baryon fraction calculated 
via the standard Newtonian dynamics in this paper (\newtonfbaryonmean) and 
by other authors \citep[e.g.,][ for a review]{Mccarthy2007} 
are systematically lower than the WMAP value
of $\Omega_{B}/\Omega_{M}$=$0.176\pm0.02$ \citep{Bennett2003}, indicating
the presence of undetected baryons in clusters.
Within MOND, the results of this paper (\mondfbaryonmean) and those of \citet{Silk2005}
are in statistical agreement with the WMAP value of $\Omega_{B}/\Omega_{M}$.
We therefore speculate that MOND can in principle reconcile the 
cluster baryon fraction with the
cosmic value
of $\Omega_{B}/\Omega_{M}$, without the need for additional
undetected baryons in clusters. 


\subsection{Results for a variable value of the MOND critical acceleration}
We also consider the possibility of $a_0$ being a free parameter of the MOND theory,
and use our data to determine it. 
To this end, we begin with Equation \eqref{eq:M_MOND} for the MOND total mass, and
calculate the value of the critical acceleration at each radius
 such that $M_{MOND}(r)=M_{baryon}(r)$, 
\begin{equation}\label{eq:M_MOND=M_gas}
{M_{baryon}(r)=M_{MOND}(r)=
\dfrac{M_{Newton}(r)}{\sqrt{1+\Bigg(\dfrac{a_{\circ}\mu m_{p}}{3\beta KT}\dfrac{r_{c}^{2}+r^{2}}{r}\Bigg)^2}}}. 
\end{equation}

We then solve equation \eqref{eq:M_MOND=M_gas} for $a_{\circ}$ and find that
\begin{equation}\label{eq:a0}  
a_{\circ}(r)=\dfrac{3\beta KT}{\mu m_{p}}\dfrac{r}{r_{c}^{2}+r^{2}}\sqrt{\dfrac{M_{Newton}^{2}}{M_{baryon}^{2}}-1},
\end{equation}    
where $M_{Newton}$ is given by equation \eqref{eq:mtotalsimplified}. 
The radial distribution of $a_{\circ}$ is shown for all clusters in Appendix B.

In Figure \ref{a0r2500} we show the value of the 
critical acceleration calculated at $r_{2500}$ for all \nclu\ clusters,
with a mean value of $a_{\circ}=1.49\pm0.09 \times 10^{-7}$~cm/s$^2$.
The main result is that the value of the "free" $a_{\circ}$ at $r_{2500}$
differs from that measured
from galaxy rotation curves \citep[e.g.,][]{Scarpareview} by one order of magnitude.
Moerover, we find statistically significant differences in the measurement of $a_{\circ}$
at different radii for a given cluster (see Appendix B), and for different clusters
at the same radius of $r_{2500}$ (Figure \ref{a0r2500}).

%

\begin{figure}[ht!]
\centering
\includegraphics[width=3in,angle=0]{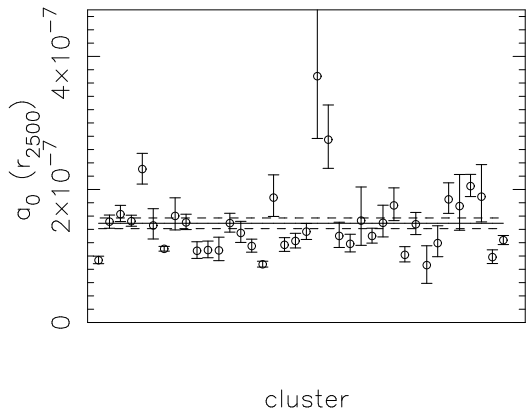}
\caption{Calculation of $a_{\circ}(r_{2500})$. 
The horizontal line is the mean of $a_{\circ}$ and its root-mean-square error, calculated
as the un-weighted mean of the 38 measurements.  The weighted mean
of the 38 measurements is $a_{\circ}$=\meanao~cm/s$^2$. \label{a0r2500}}
\end{figure}

\begin{figure}[ht!]
\centering
\includegraphics[width=3in,angle=0]{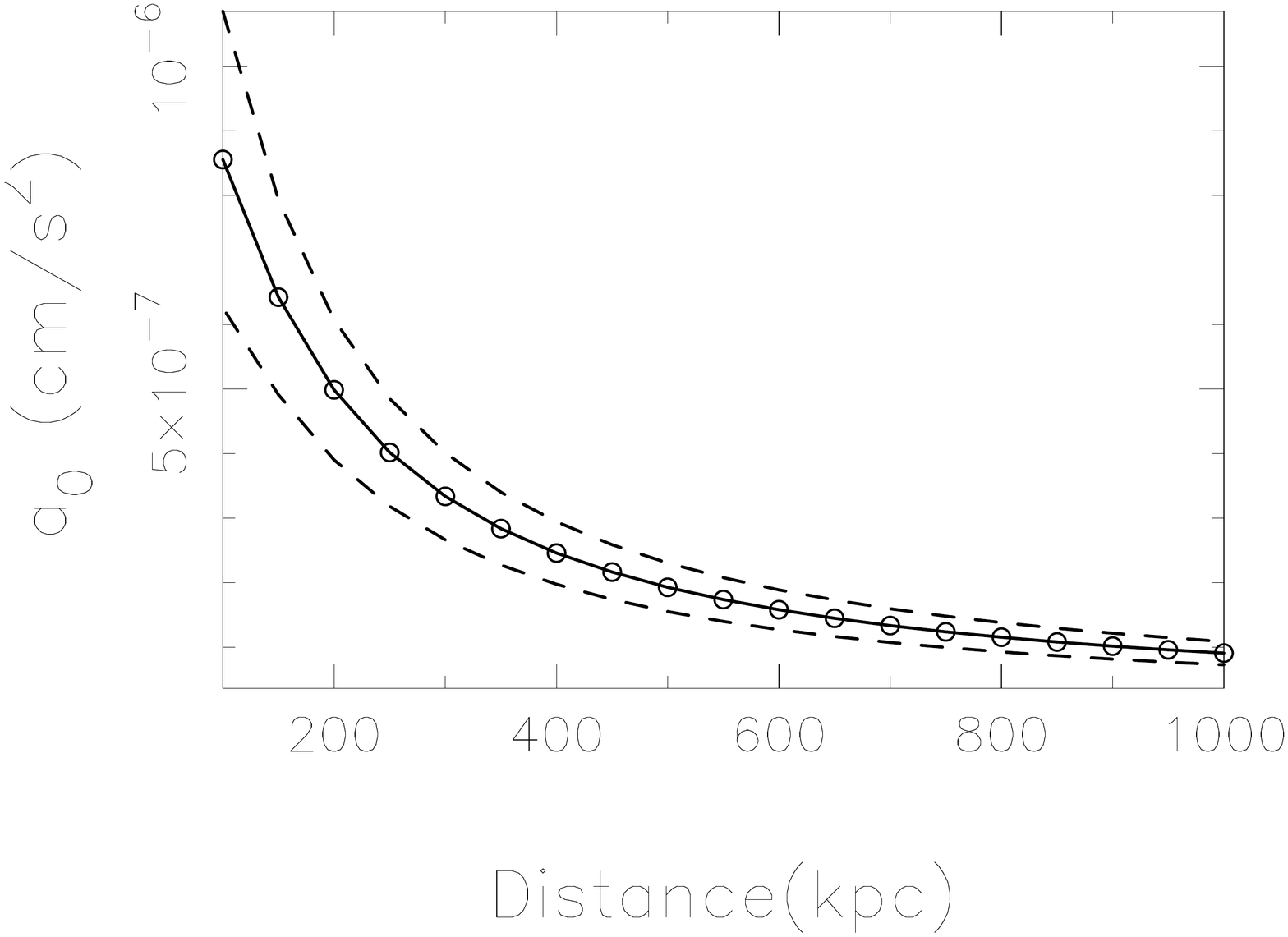}
\caption{Average distribution of $a_{\circ}$ for the
38 clusters; the dashed lines represent the root-mean-square uncertainty of the 38 measurements
at each radius \label{fig:average-a0}}
\end{figure}

We also show the average values of the best-fit critical acceleration $a_{\circ}$ for the entire
sample, as function of radius, in Figure \ref{fig:average-a0}. At all radii, the critical acceleration
required to explain the hydrostatic MOND mass is at least one order of magnitude larger than the
canonical value obtained from galaxy rotation curves \citep[e.g.,][]{Scarpareview}.

\section{Conclusions}
\label{conclusion}
In this paper we have measured gravitational and gas masses at $r_{2500}$ 
for a sample of 38 high-redshift galaxy clusters
using standard Newtonian gravity and Modified Newtonian Dynamics. This is the largest sample of galaxy clusters
to which the MOND theory of gravity is applied to date.

We initially used a fixed value of the MOND critical acceleration $a_{\circ}$= \critaccel\, 
 and measured an average \mondfbaryonmean; this result
confirms that MOND does not eliminate the need for dark matter in galaxy clusters. Our work therefore
confirms the  previous  reports by \citet{angus2007} and \citet{Silk2005}.
Further evidence that the MOND theory of gravity is unable to explain  X-ray
observations of  galaxy clusters
is found by a fit of our X-ray data to a MOND model with varying $a_{\circ}$.
The value of $a_{\circ}$ required
to achieve a baryon fraction of unity varies both with cluster radius, and between clusters; moreover,
the mean value of this free $a_{\circ}$  is one order of magnitude larger than that required
by galaxy dynamics. We therefore conclude
that X-ray observations of the hot cluster plasma do require strong presence of
dark matter even when MOND is used.

This analysis also finds that the  $f_{baryon}$ at $r_{2500}$ 
calculated using MOND is in statistical agreement with 
the WMAP value of $\Omega_b/\Omega_M$, similar to the results of \citet{Silk2005}. 
\clearpage

\section*{Appendix A}
In this appendix we provide the radial profiles of the gas, Newtonian and MOND masses for all clusters,
calculated between 100~kpc and $r_{2500}$.
\label{appendix1}
\begin{figure}[!h]
 \caption{\footnotesize Distribution of $M_{gas}$,$M_{total}$, and $M_{MOND}$ as a function of radius(arcsec) for all 38 clusters.
The symbols are used as follows: $\bigcirc$ for $ M_{total}$, $\times$ for $M_{MOND}$, and $\bigtriangleup$ for $M_{gas}$ \label{massplotappendix}}
\centering
\subfloat[Abell1413]{\label{Abell1413masses}\includegraphics[width=3.0cm,angle=270]{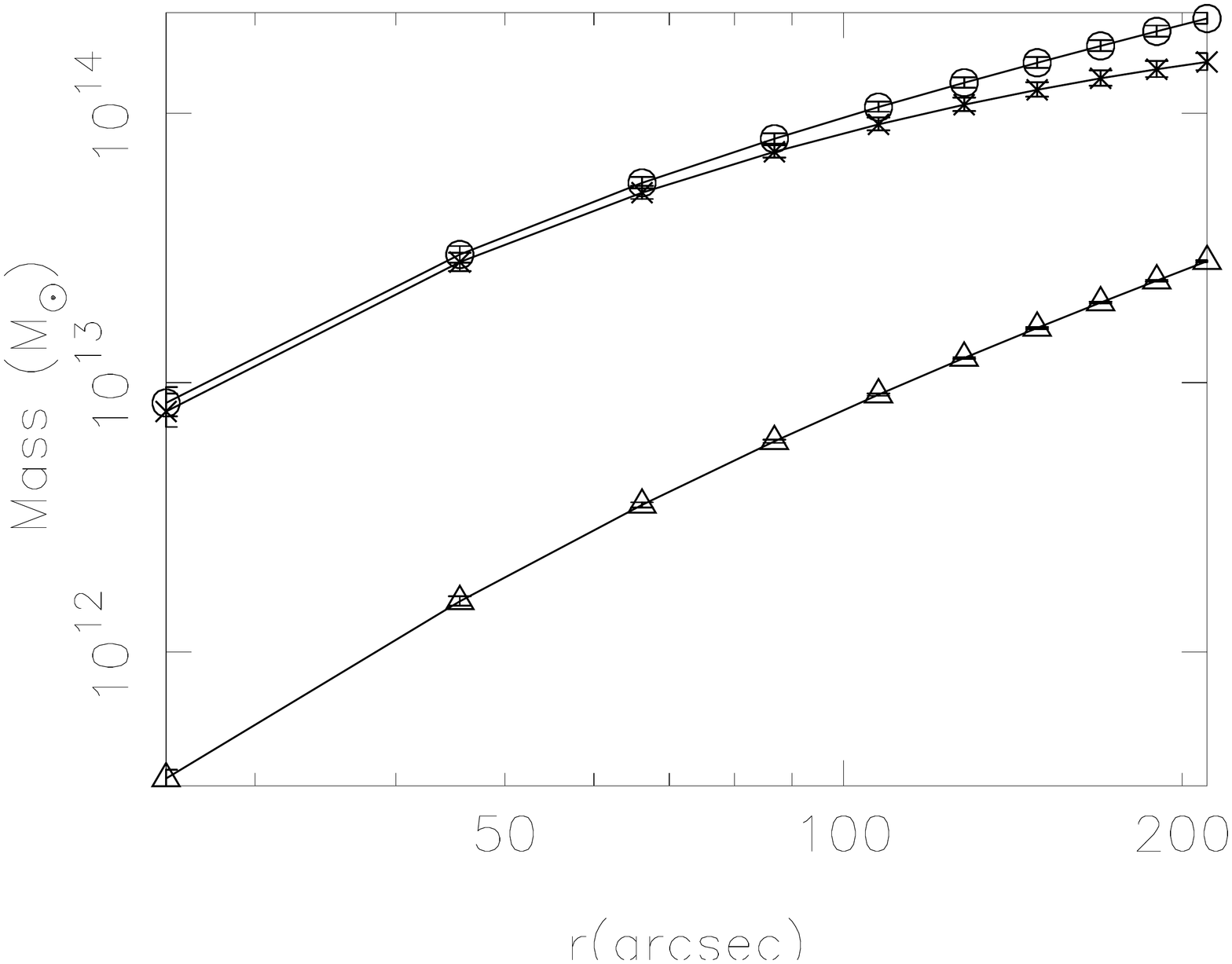}}
\subfloat[Abell1689]{\label{Abell1689masses}\includegraphics[width=3.0cm,angle=270]{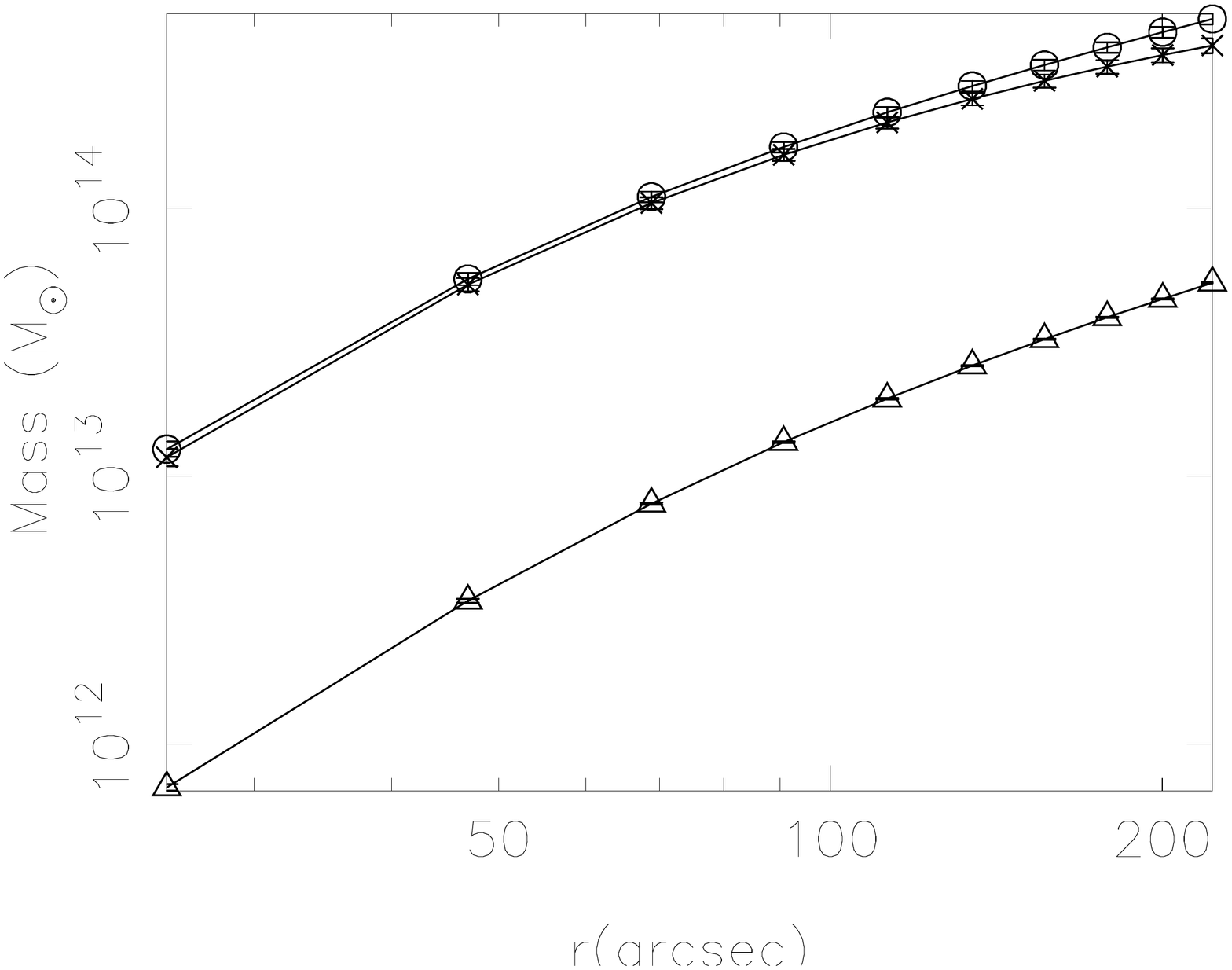}}
\subfloat[Abell1835]{\label{Abell1835masses}\includegraphics[width=3.0cm,angle=270]{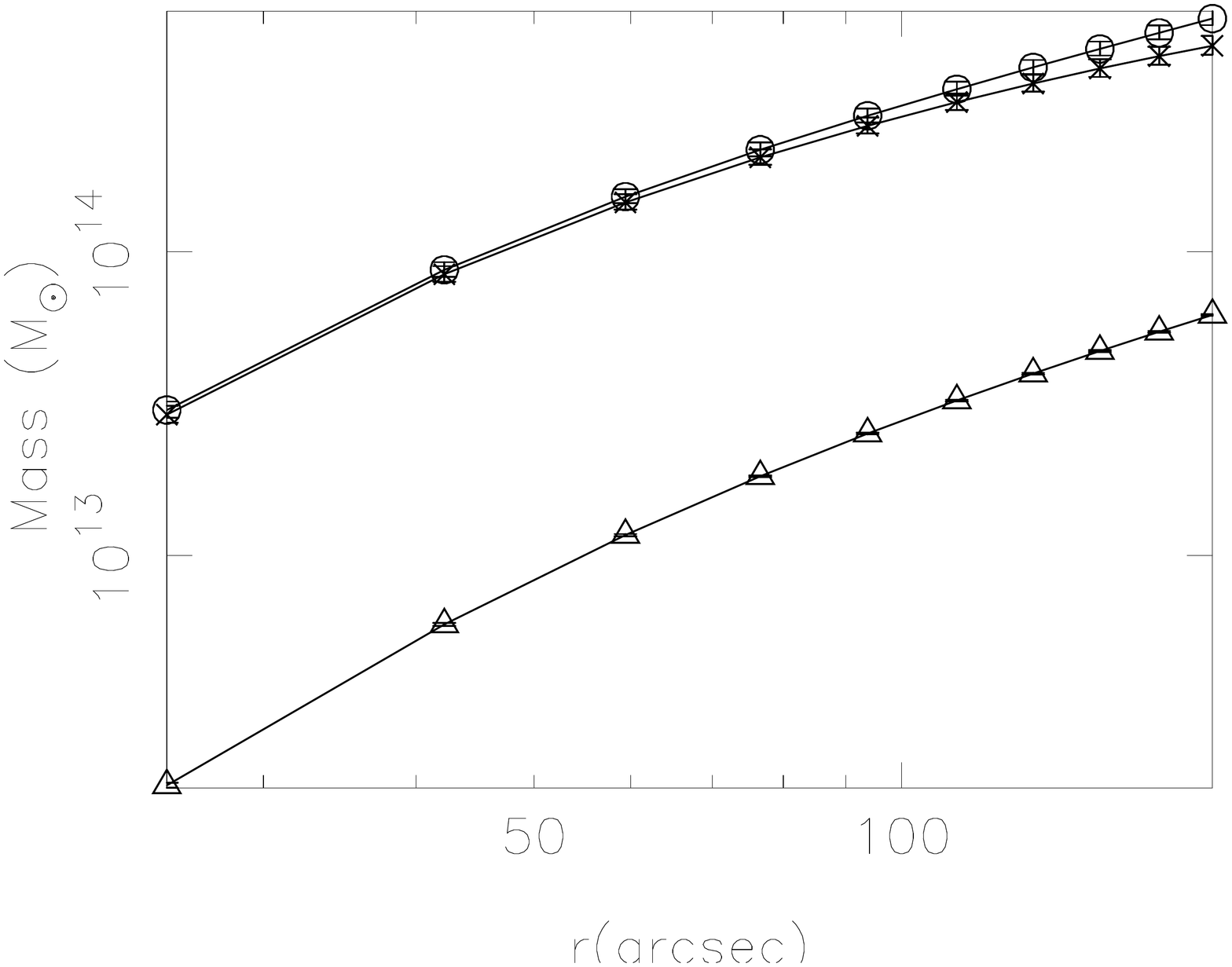}}

\subfloat[Abell1914]{\label{Abell1914masses}\includegraphics[width=3.0cm,angle=270]{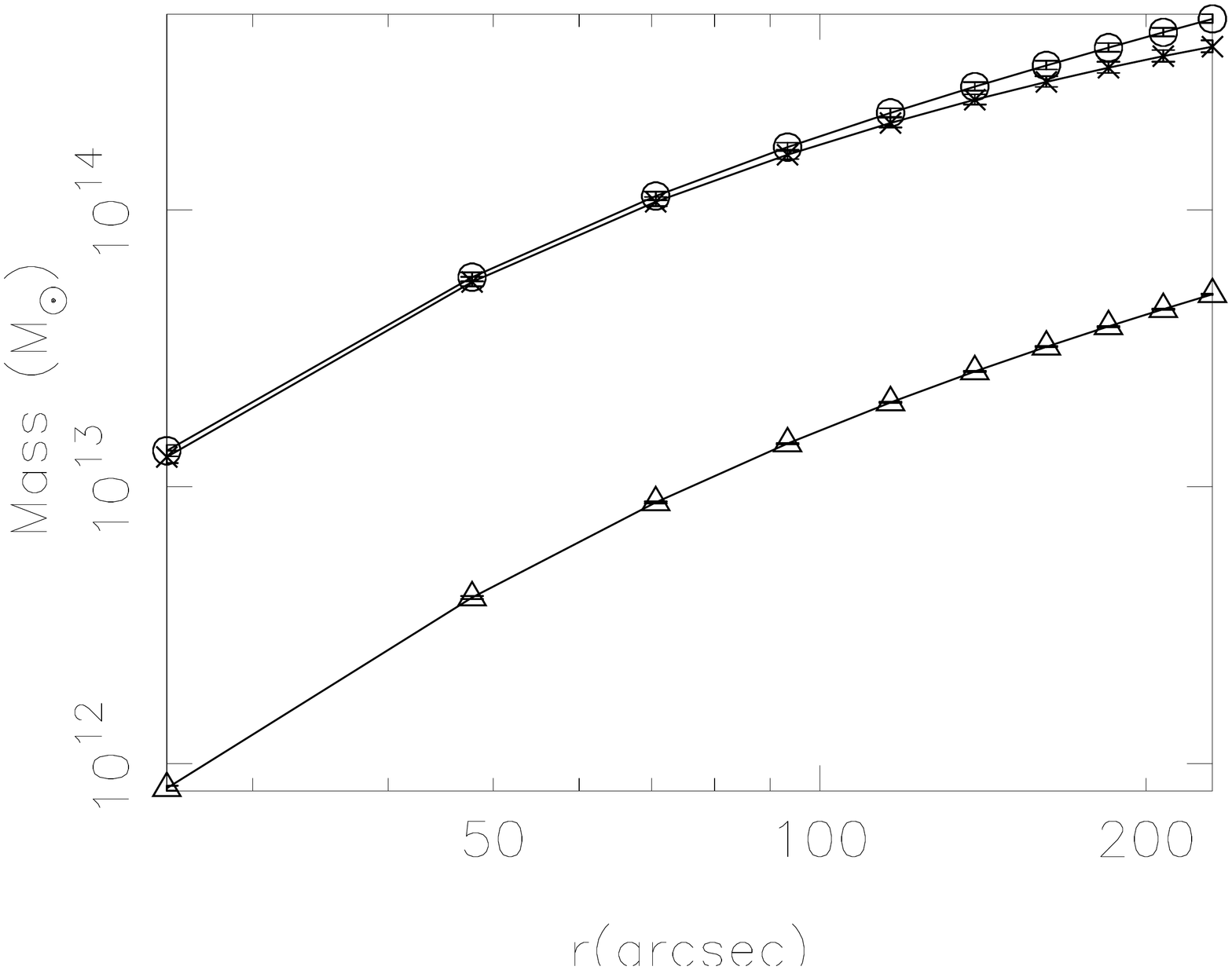}}
\subfloat[Abell1995]{\label{Abell1995masses}\includegraphics[width=3.0cm,angle=270]{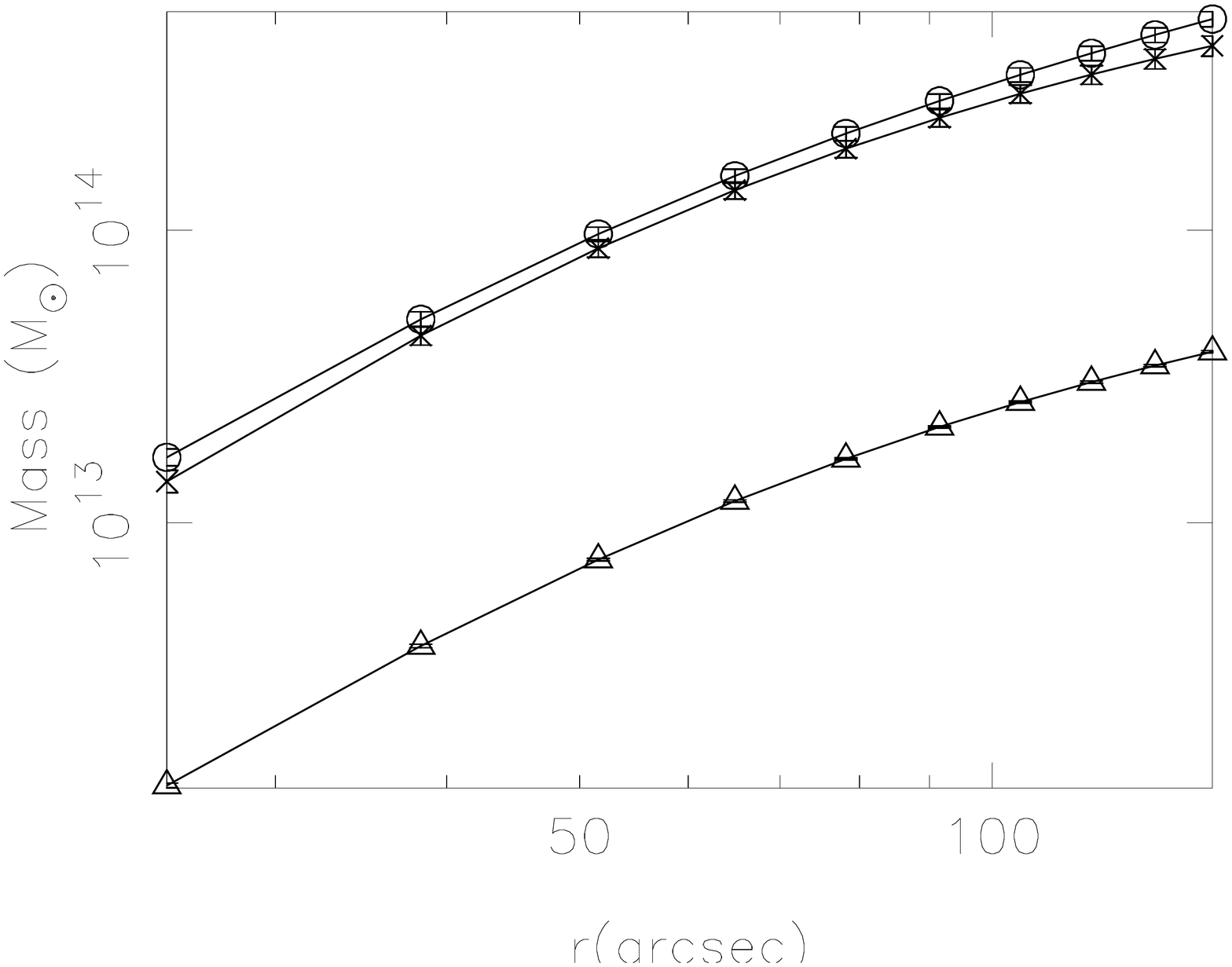}}
\subfloat[Abell2111]{\label{Abell2111masses}\includegraphics[width=3.0cm,angle=270]{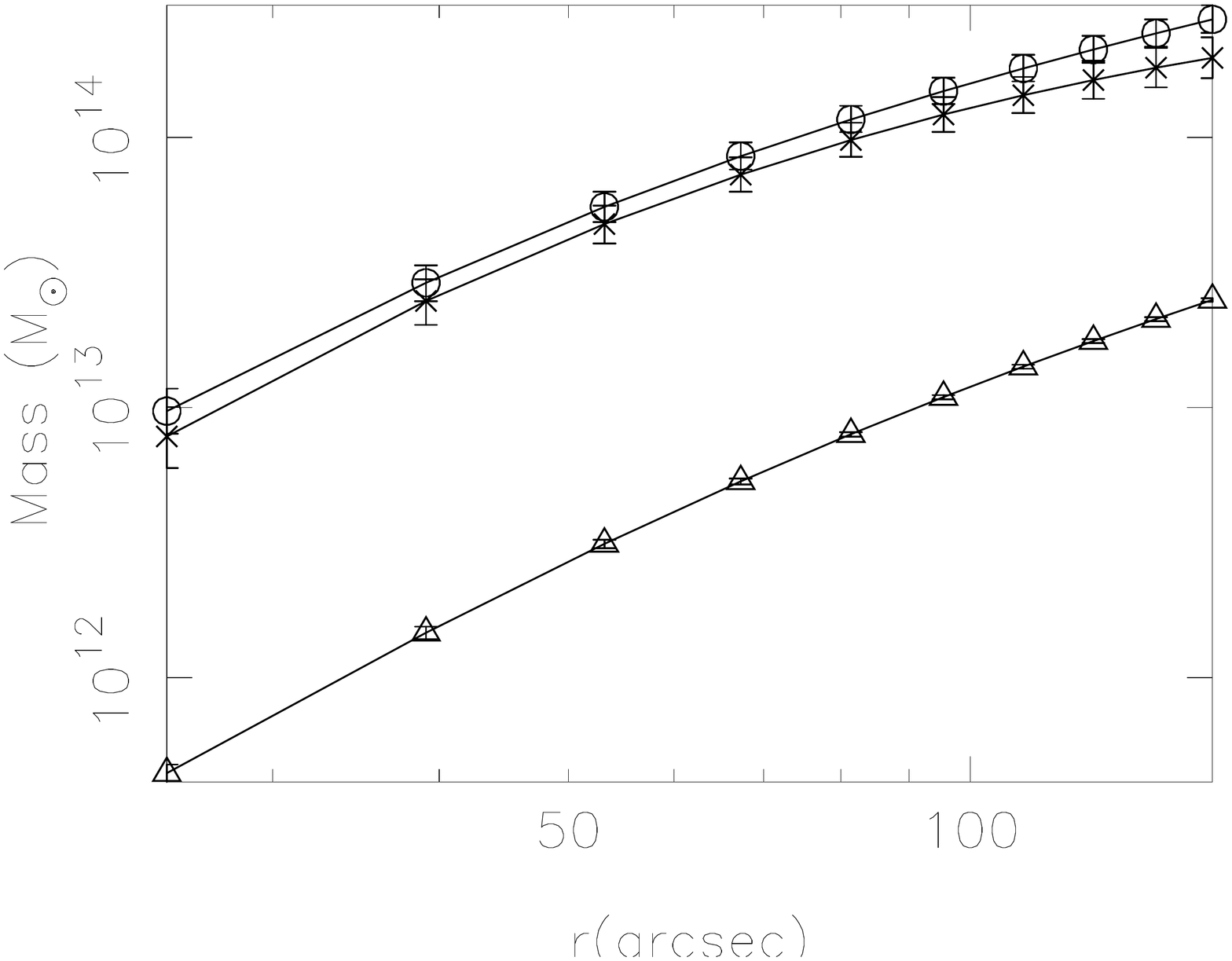}}

\subfloat[Abell2163]{\label{Abell2163masses}\includegraphics[width=3.0cm,angle=270]{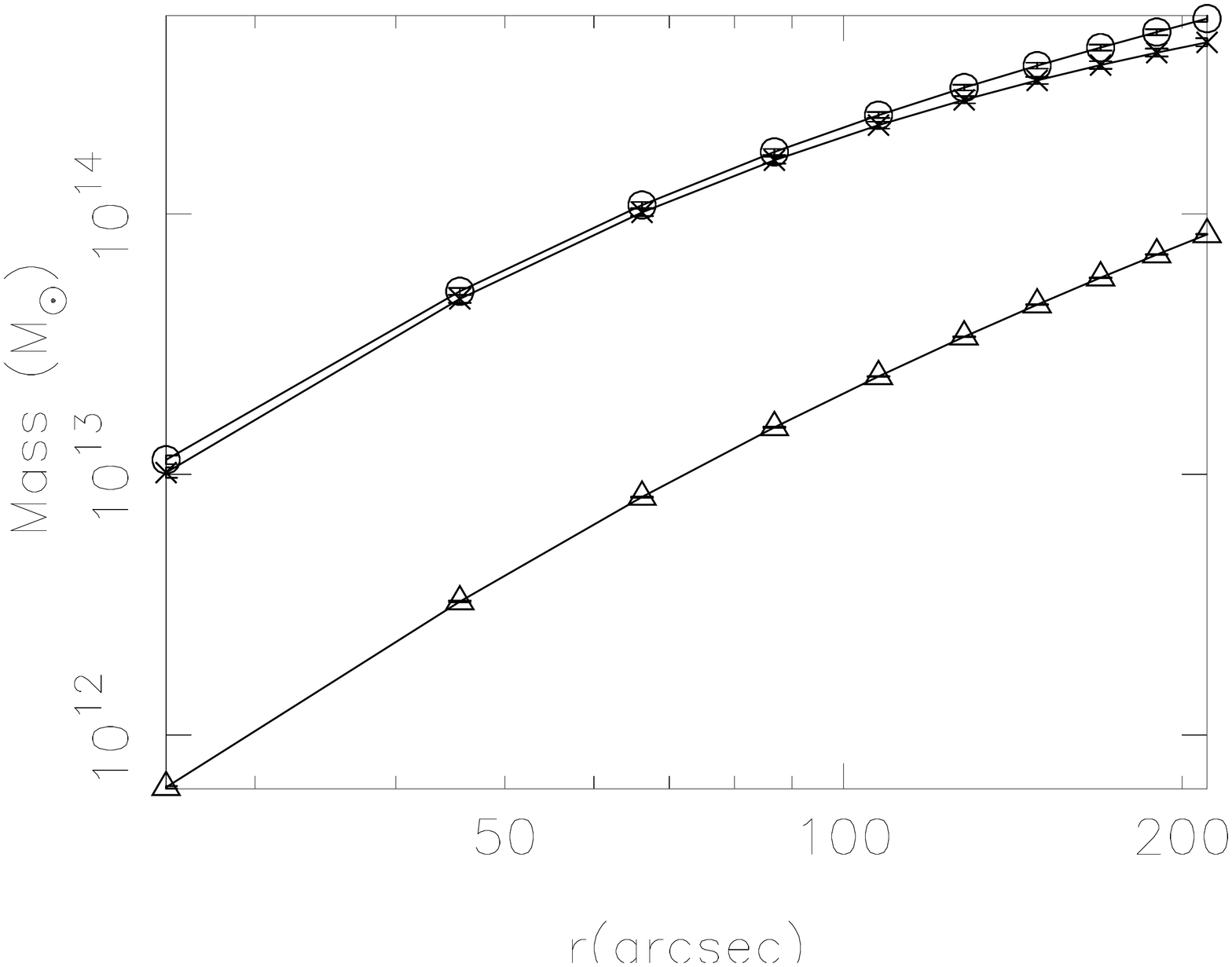}}
\subfloat[Abell2204]{\label{Abell2204masses}\includegraphics[width=3.0cm,angle=270]{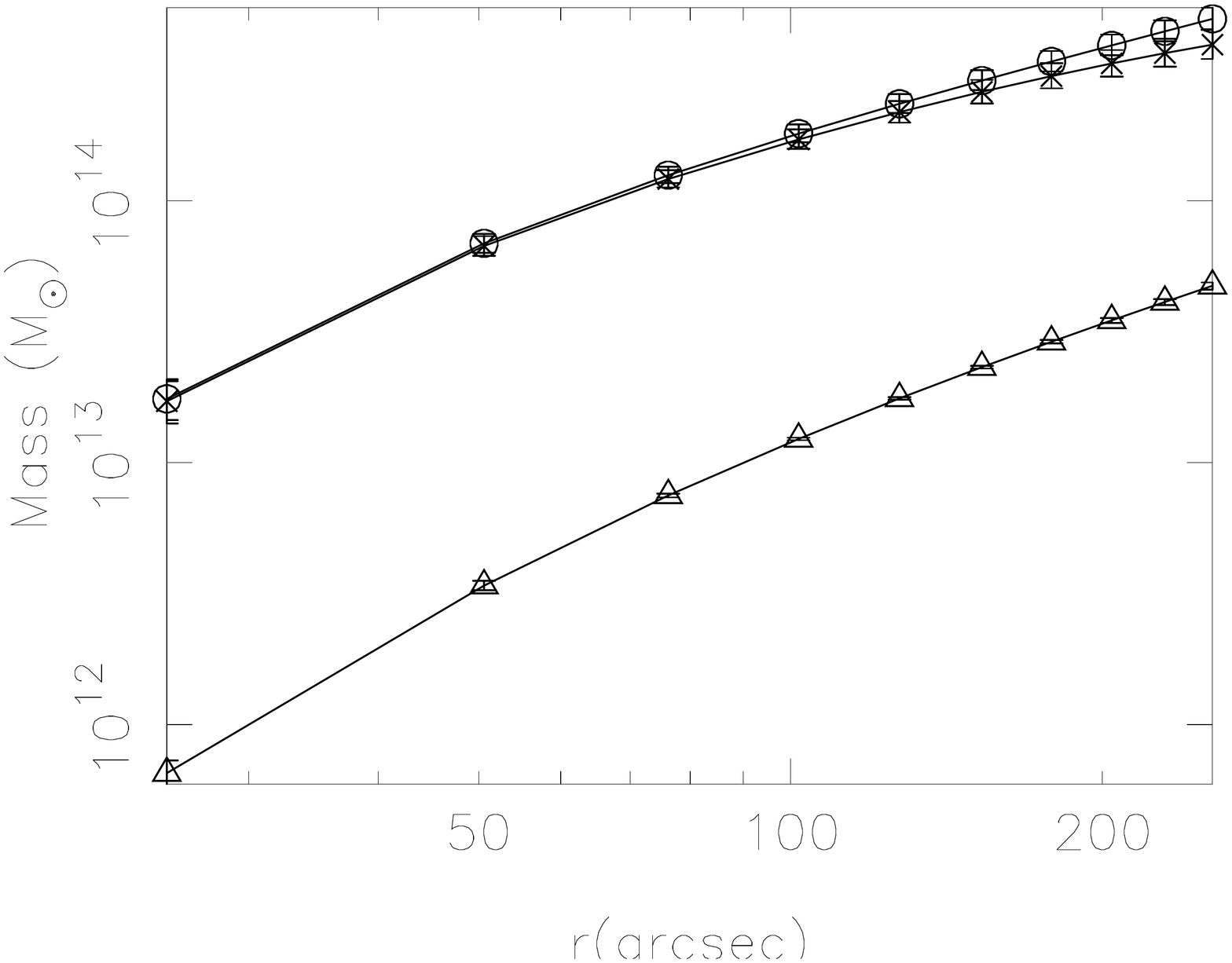}}
\subfloat[Abell2218]{\label{Abell2218masses}\includegraphics[width=3.0cm,angle=270]{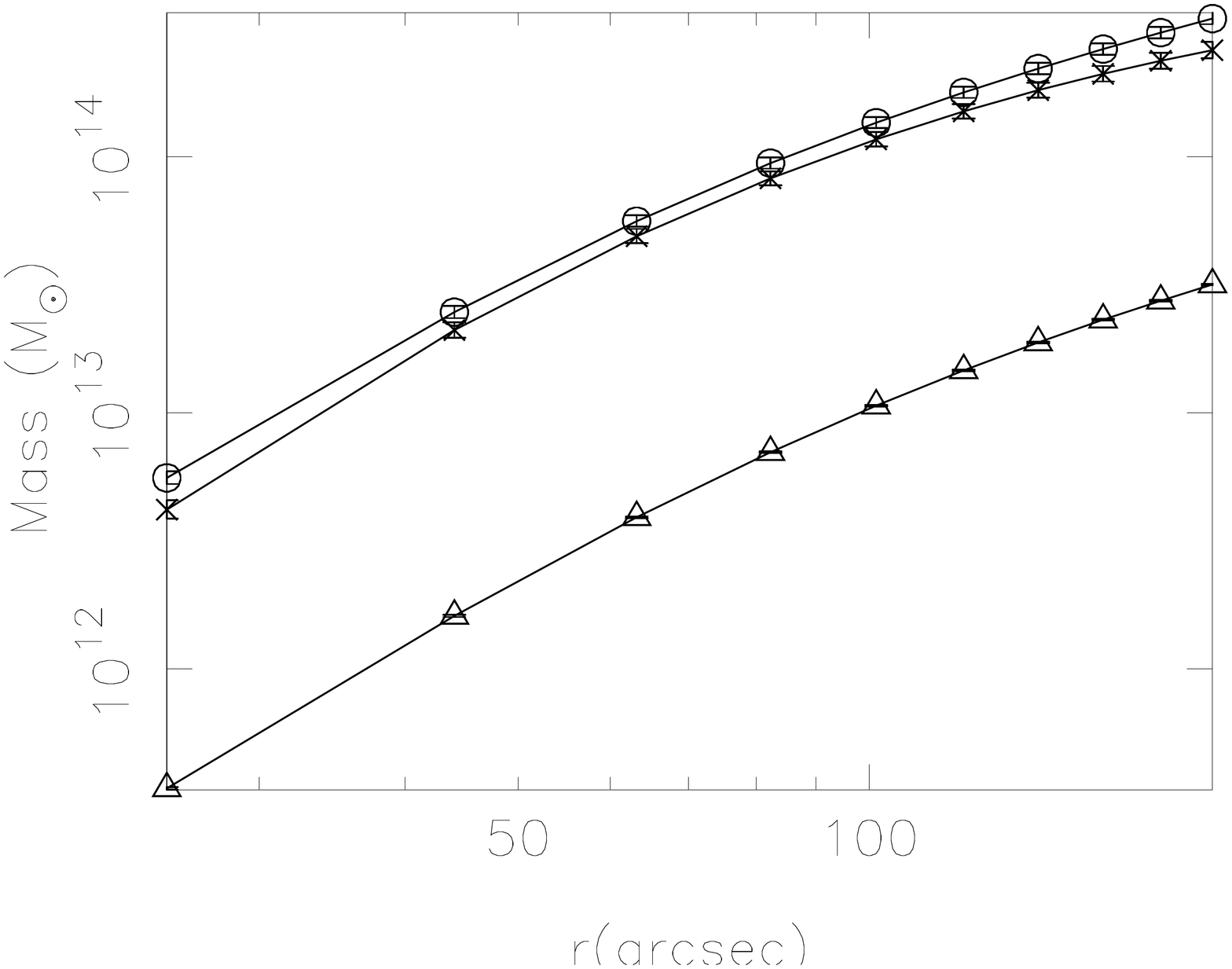}}
\end{figure}
\begin{figure}[p]
\ContinuedFloat
\centering
\subfloat[Abell2259]{\label{Abell2259masses}\includegraphics[width=3.0cm,angle=270]{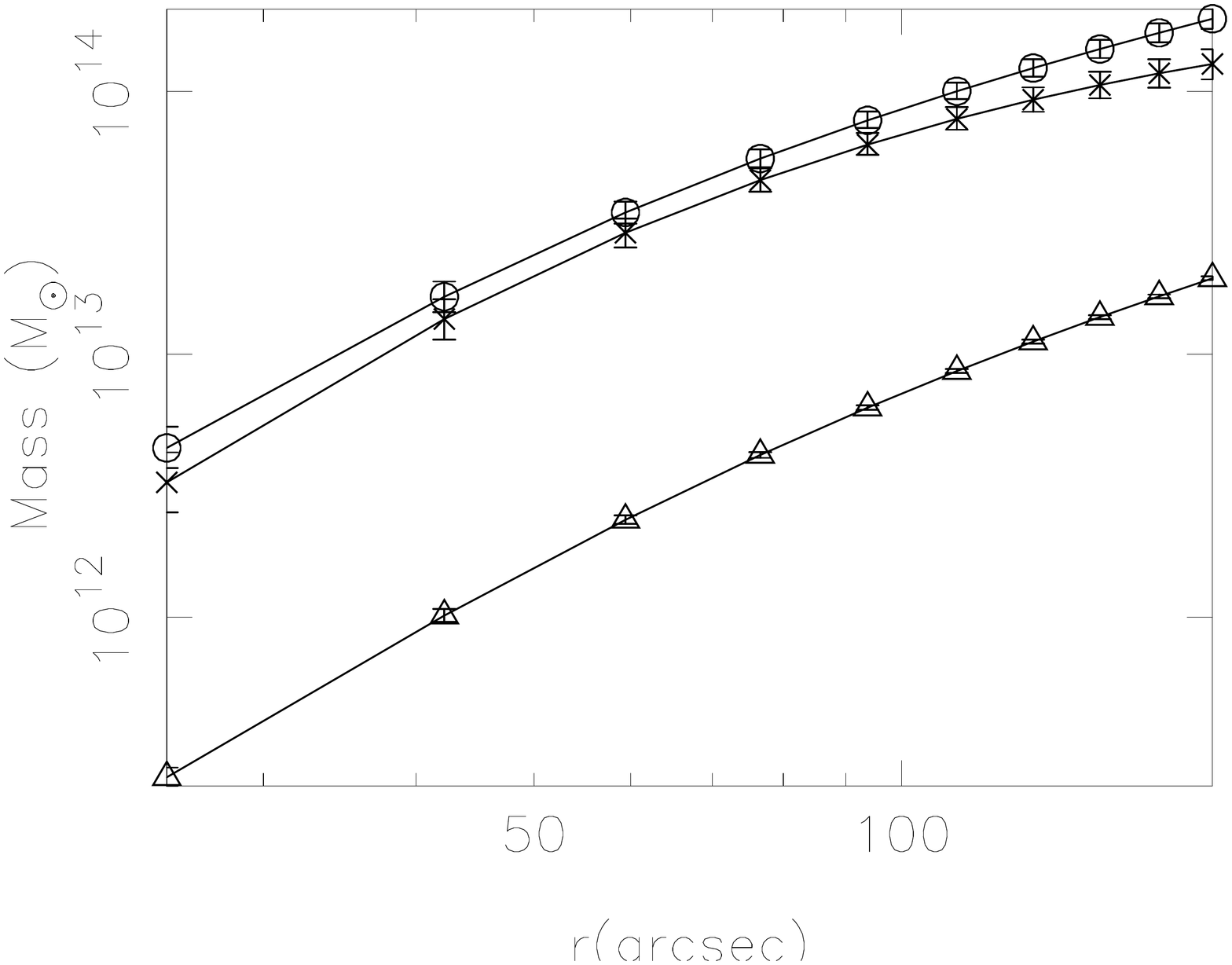}}
\subfloat[Abell2261]{\label{Abell2261masses}\includegraphics[width=3.0cm,angle=270]{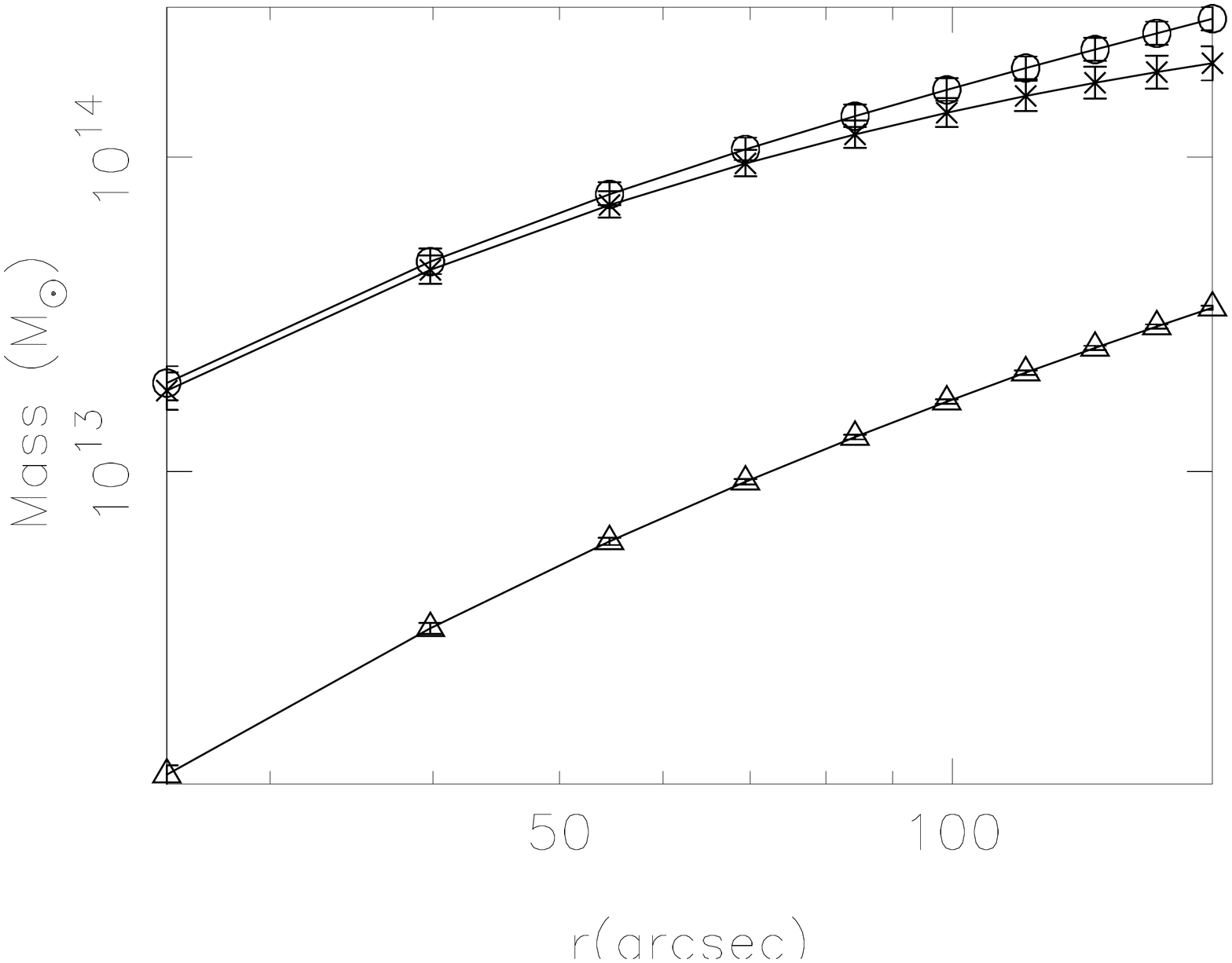}}
\subfloat[Abell267]{\label{Abell267masses}\includegraphics[width=3.0cm,angle=270]{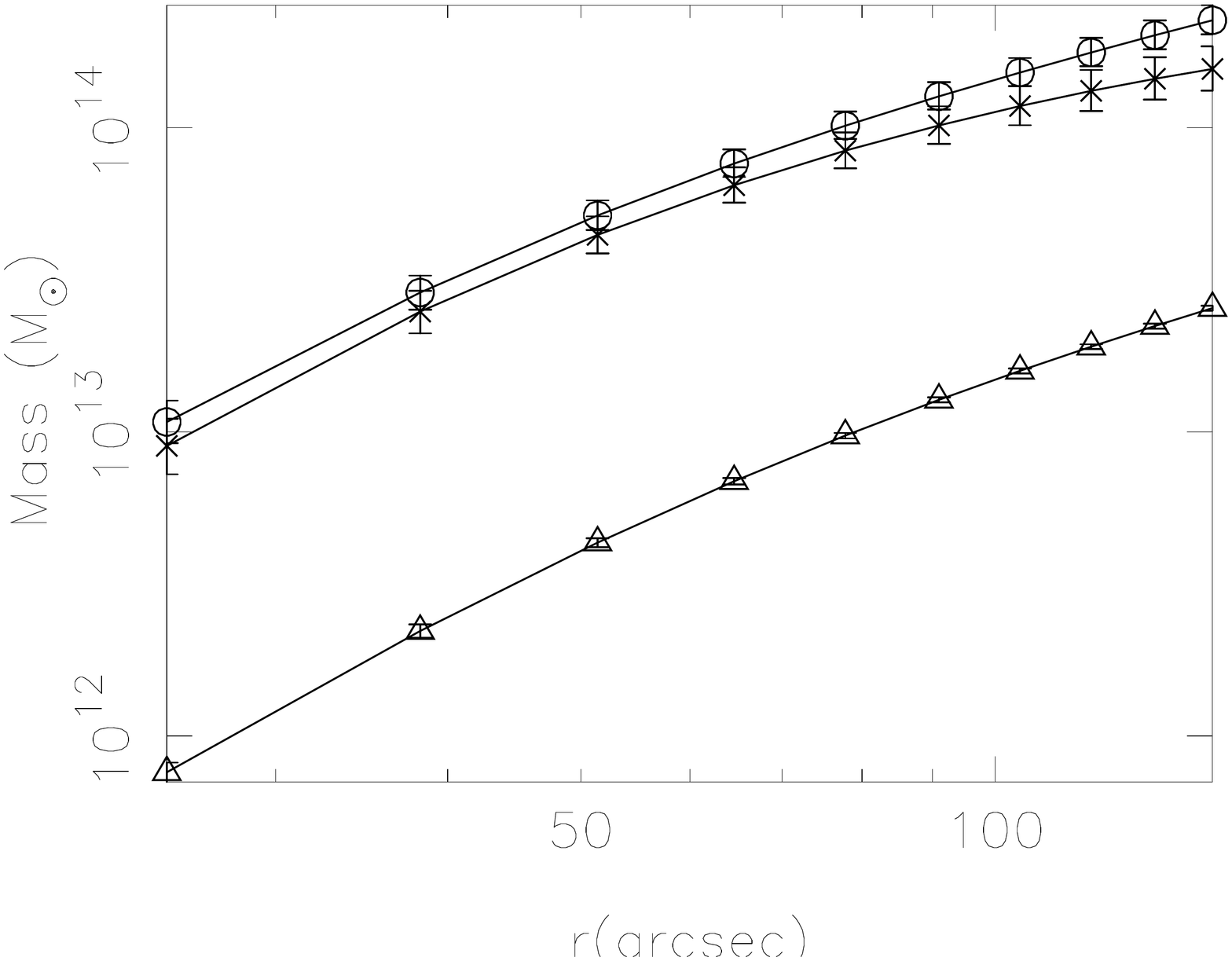}}

\subfloat[Abell370]{\label{Abell370masses}\includegraphics[width=3.0cm,angle=270]{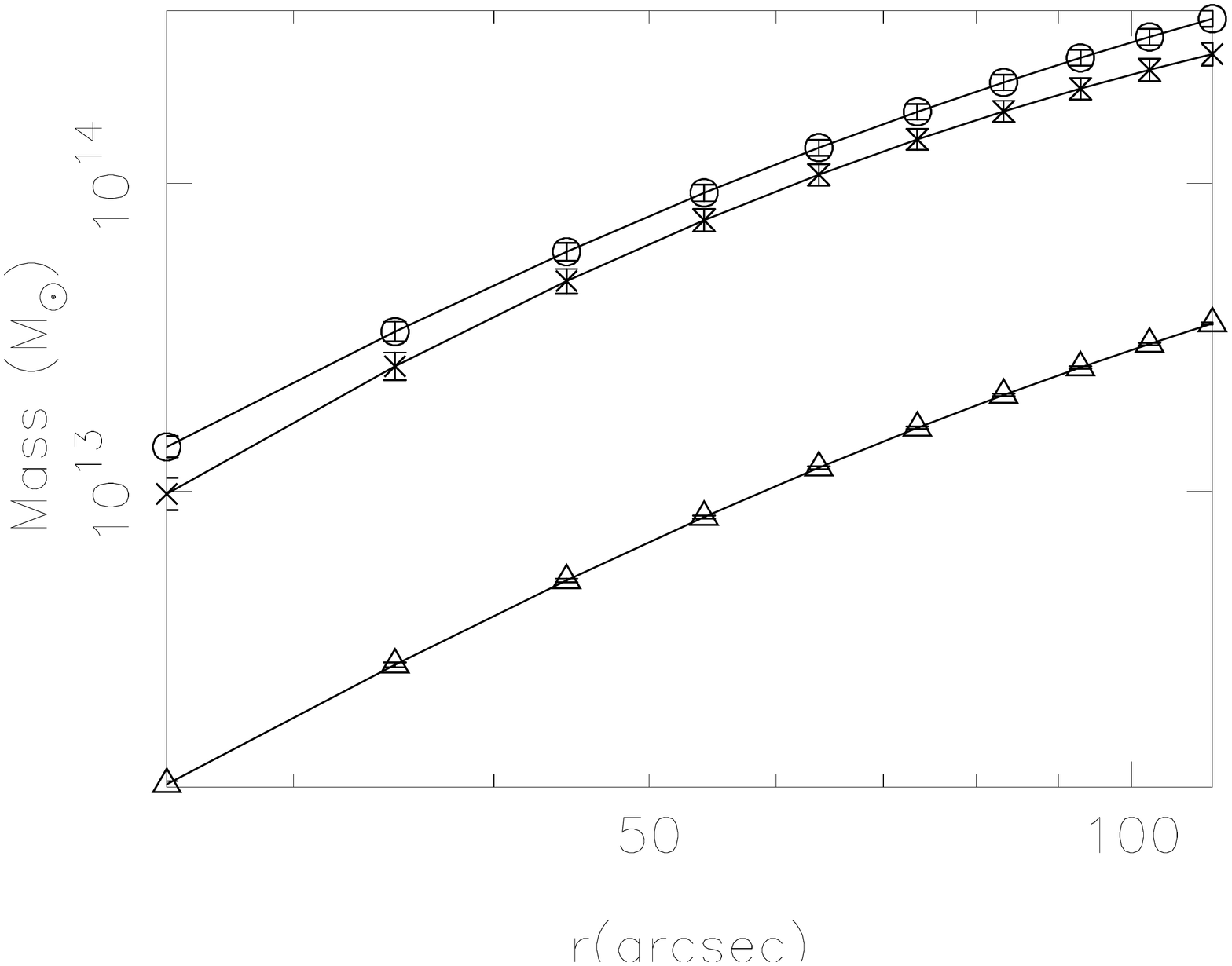}}
\subfloat[Abell586]{\label{Abell586masses}\includegraphics[width=3.0cm,angle=270]{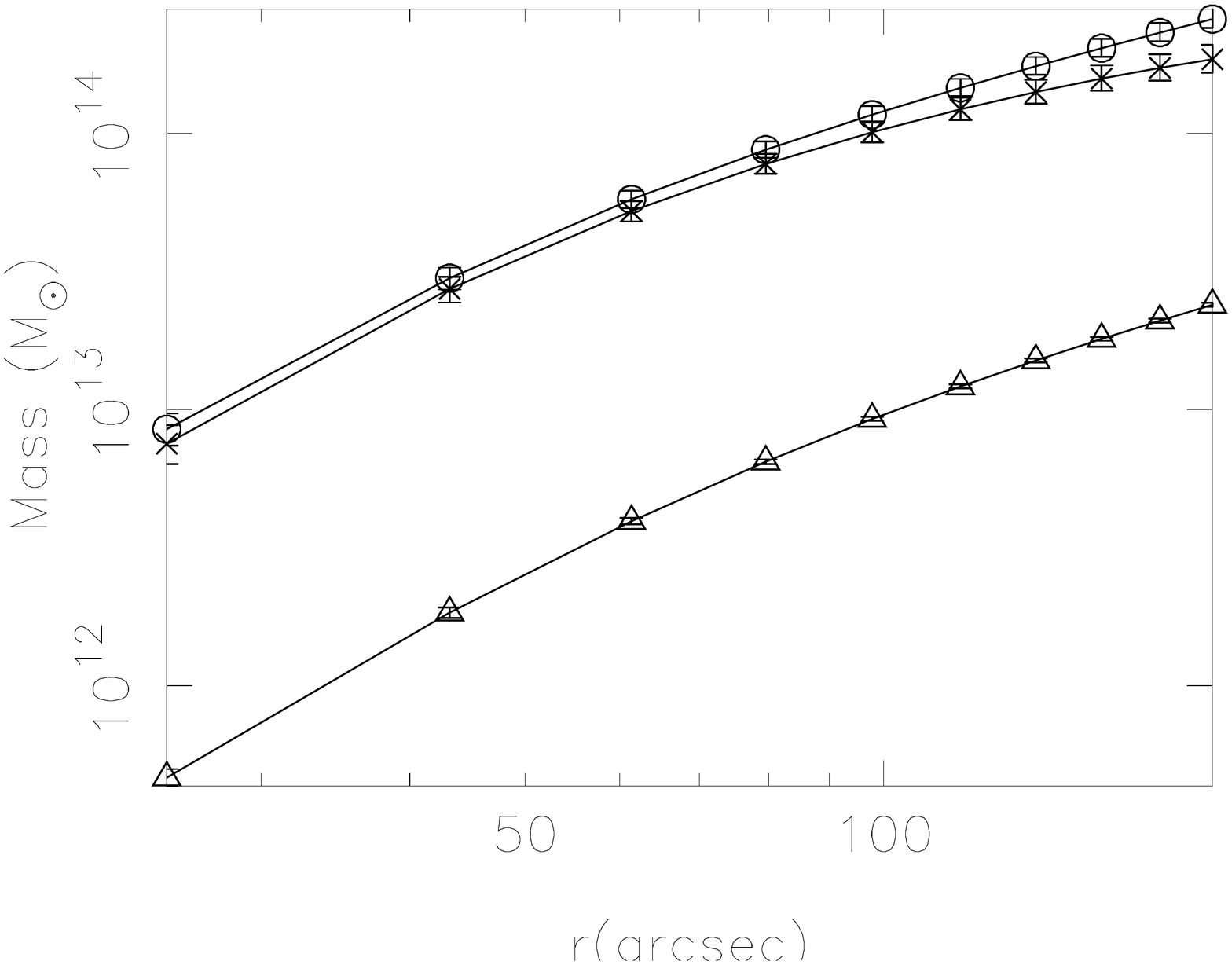}}
\subfloat[Abell611]{\label{Abell611masses}\includegraphics[width=3.0cm,angle=270]{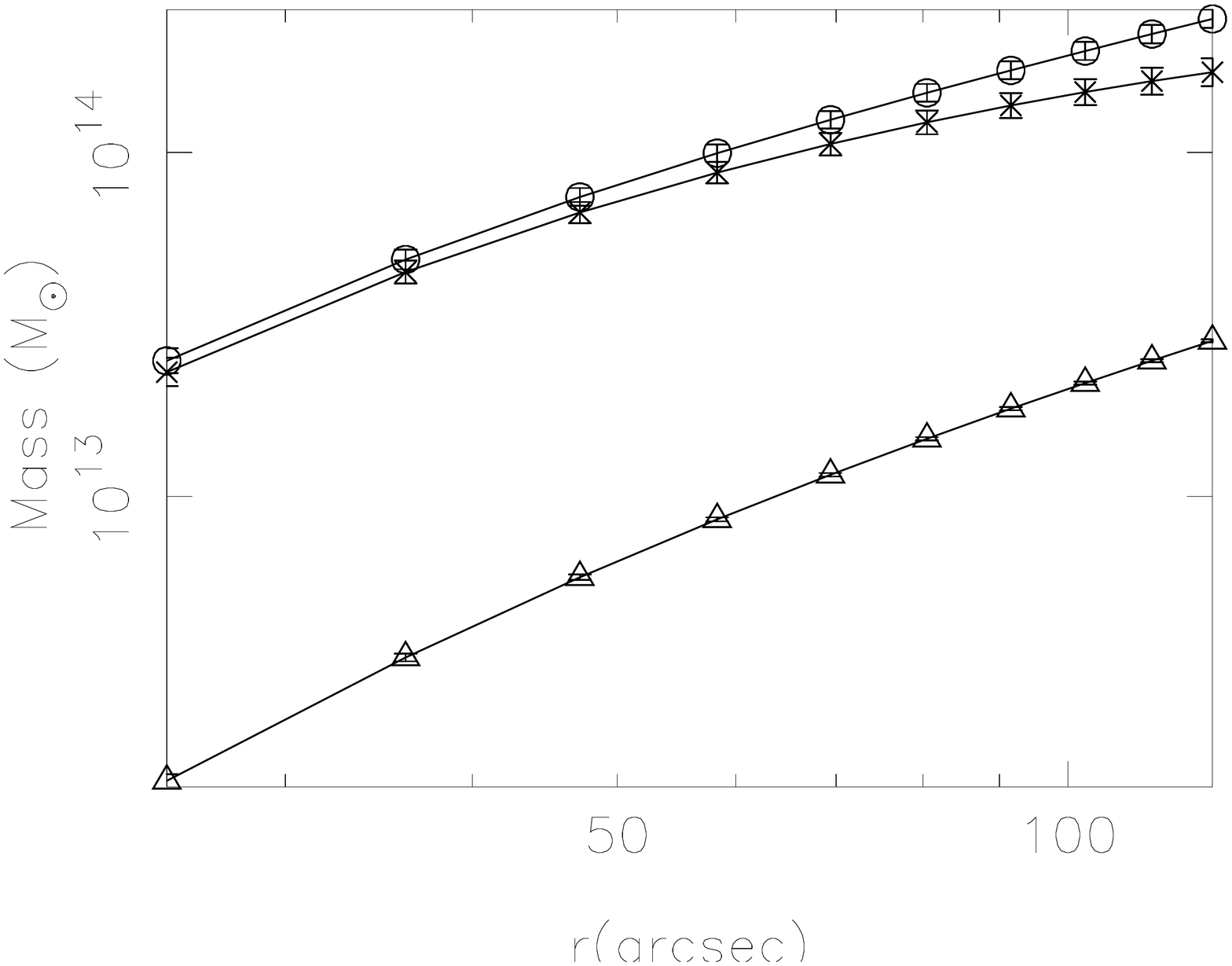}}

\subfloat[Abell665]{\label{Abell665masses}\includegraphics[width=3.0cm,angle=270]{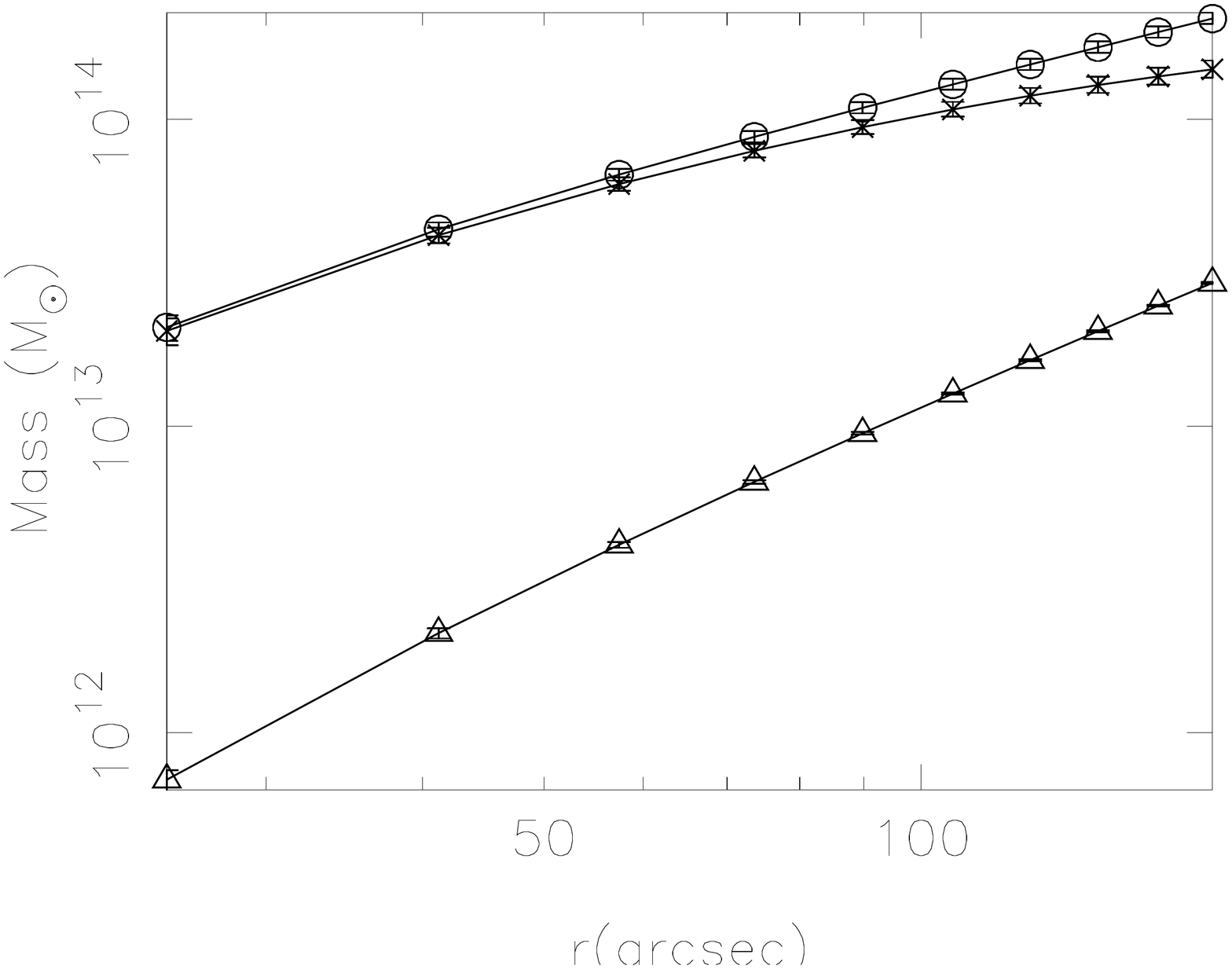}}
\subfloat[Abell68]{\label{Abell68masses}\includegraphics[width=3.0cm,angle=270]{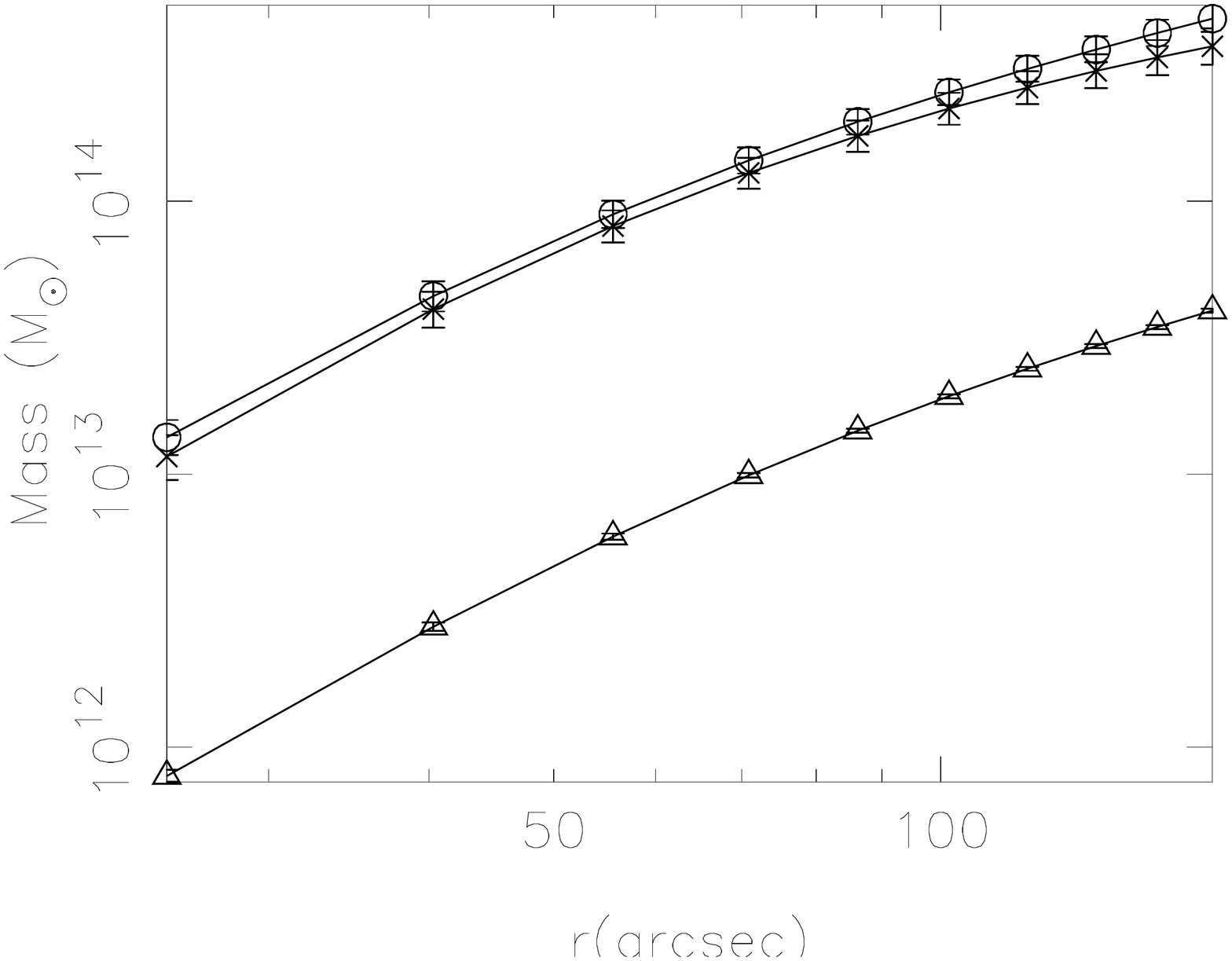}}
\subfloat[Abell697]{\label{Abell697masses}\includegraphics[width=3.0cm,angle=270]{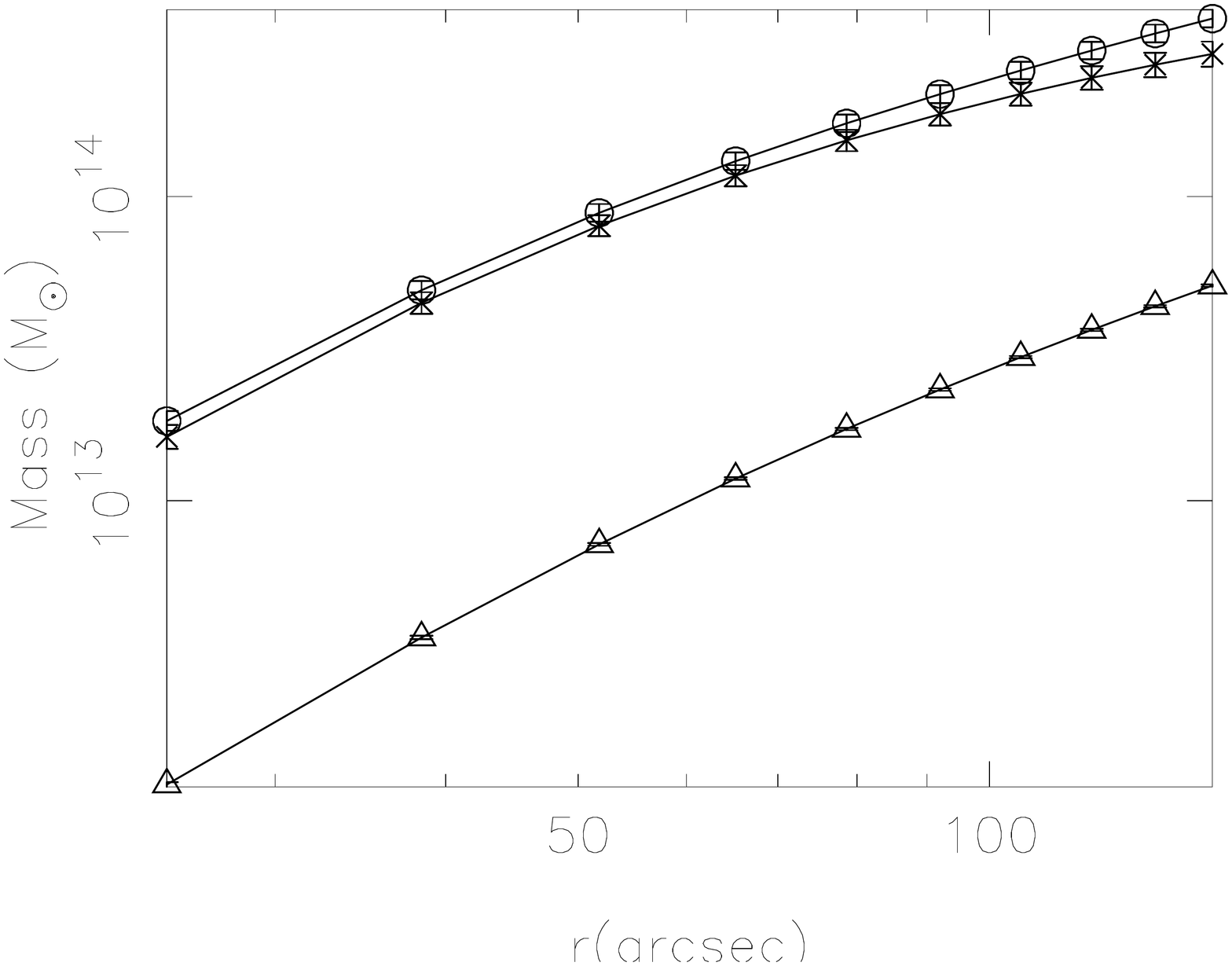}}
\end{figure}
\begin{figure}[p]
\ContinuedFloat
\centering
\subfloat[Abell773]{\label{Abell773masses}\includegraphics[width=3.0cm,angle=270]{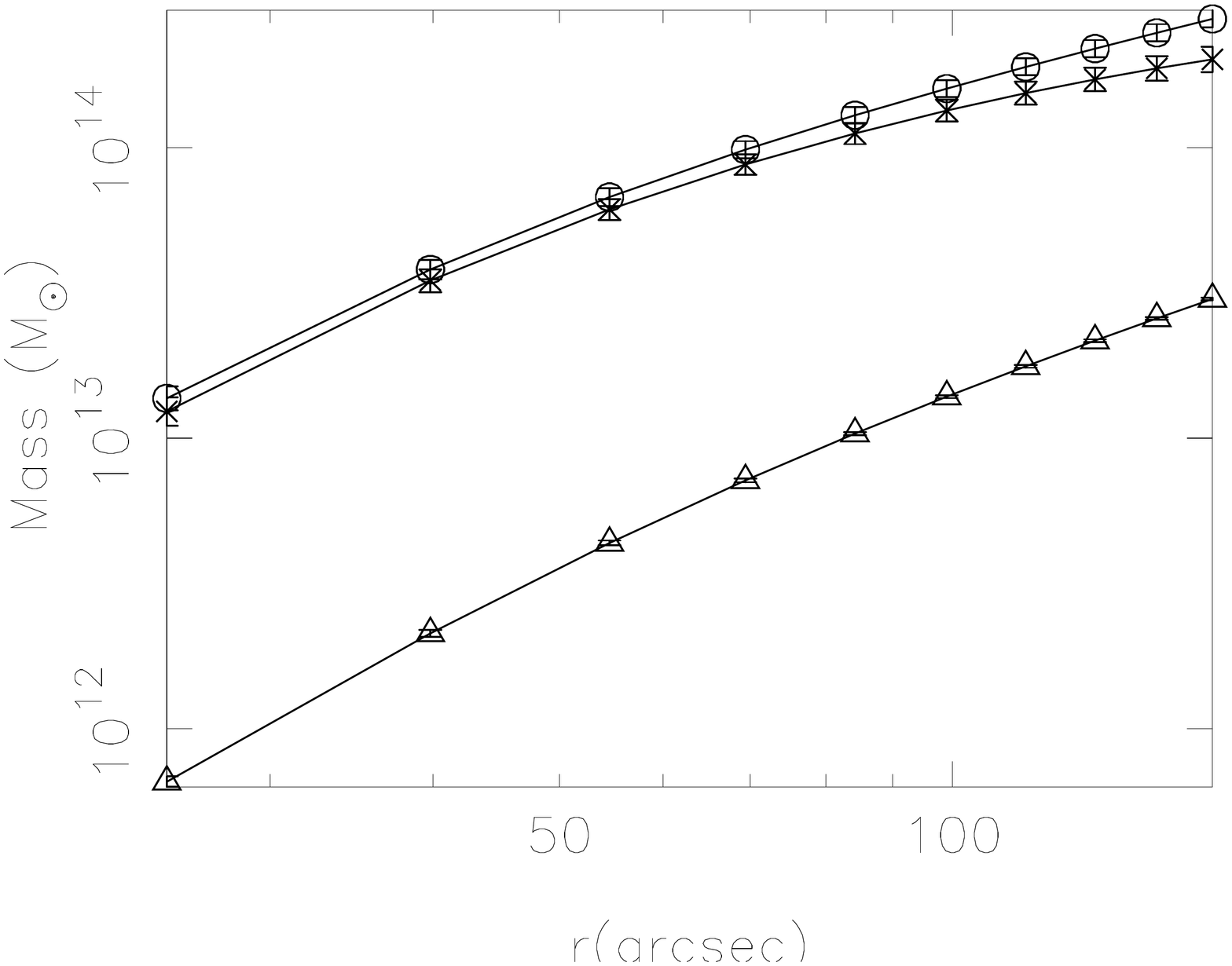}}
\subfloat[CLJ0016+1609]{\label{CLJ0016+1609masses}\includegraphics[width=3.0cm,angle=270]{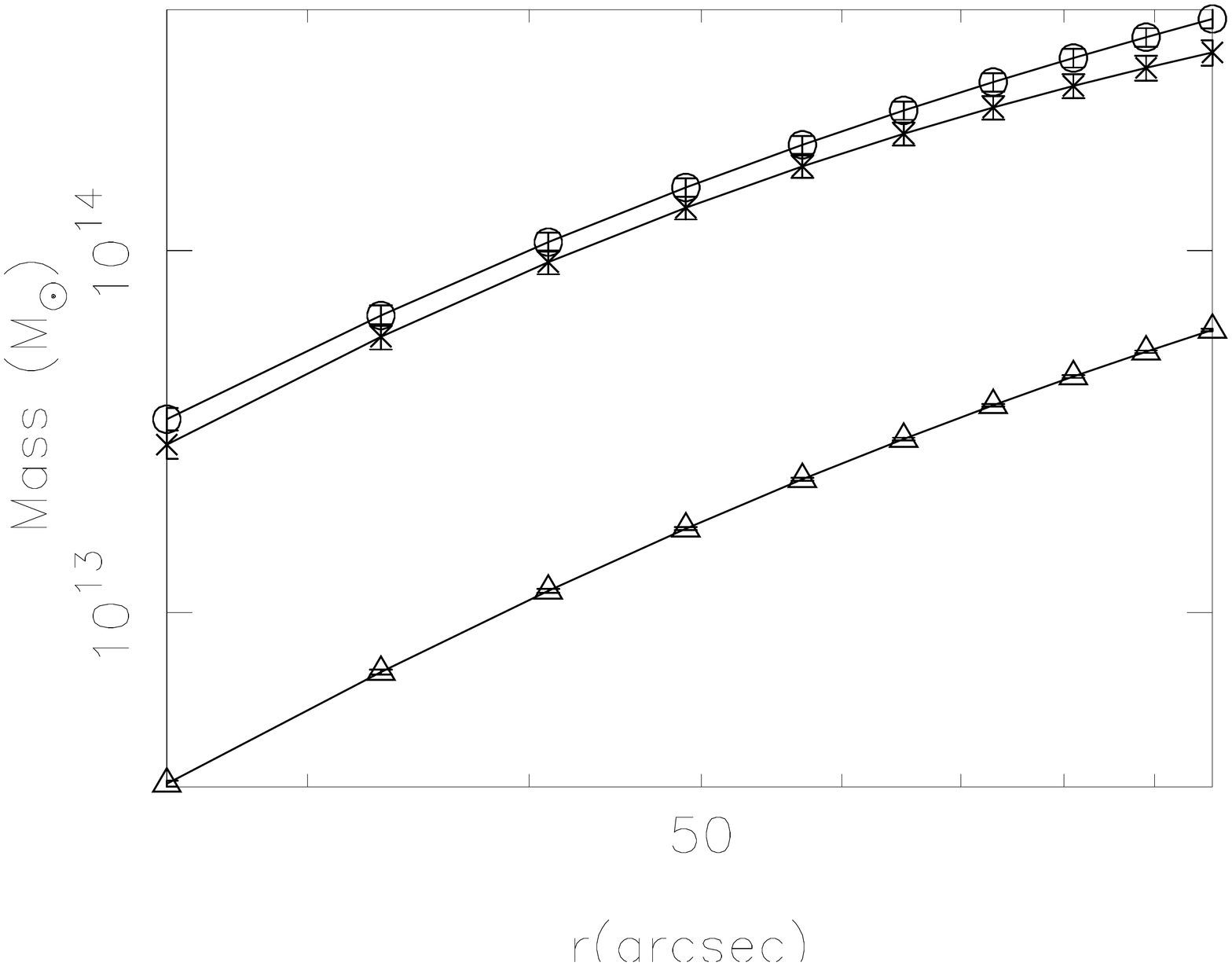}}
\subfloat[CLJ1226+3332]{\label{CLJ1226+3332masses}\includegraphics[width=3.0cm,angle=270]{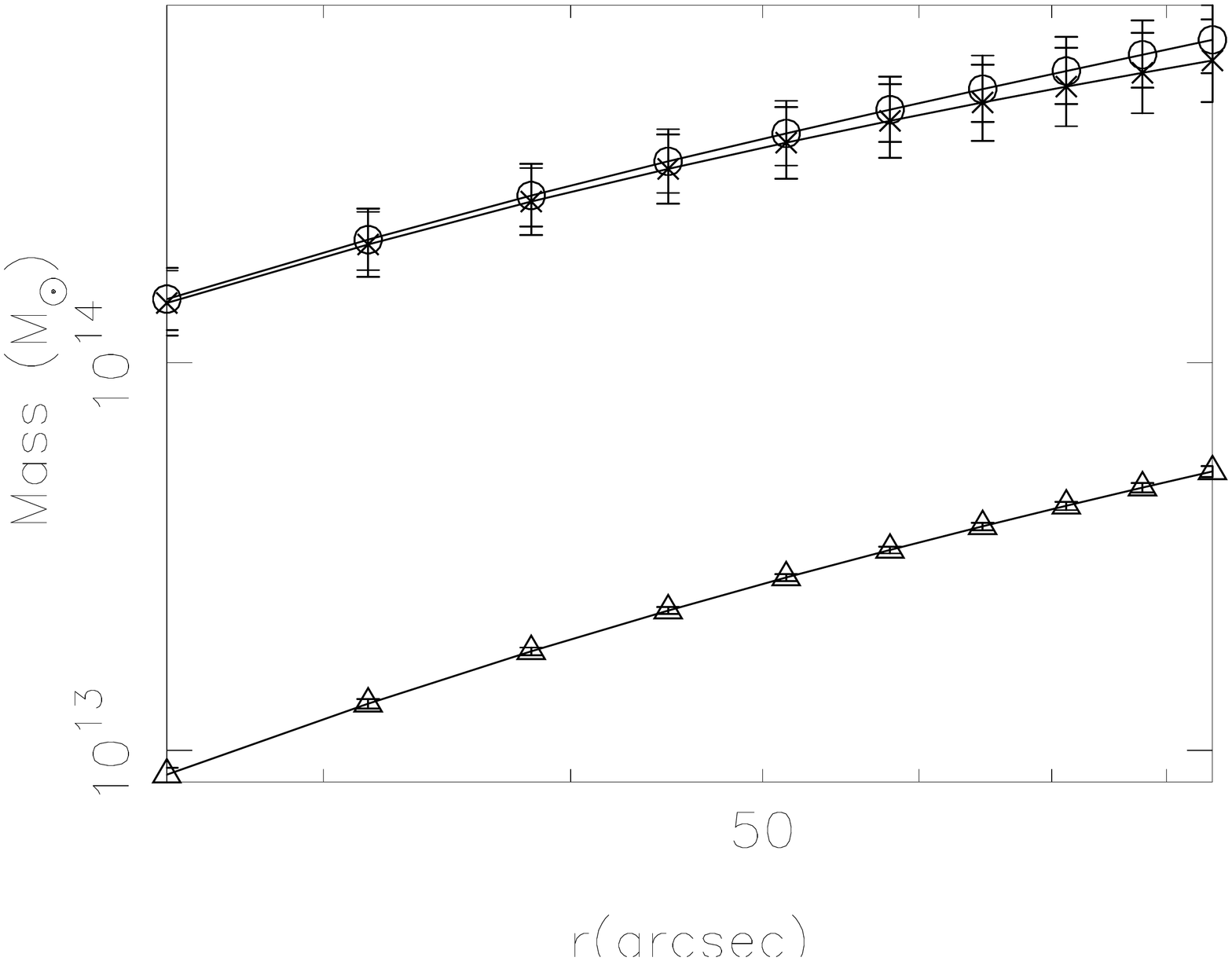}}

\subfloat[MACSJ0647.7+7015]{\label{MACSJ0647.7+7015masses}\includegraphics[width=3.0cm,angle=270]{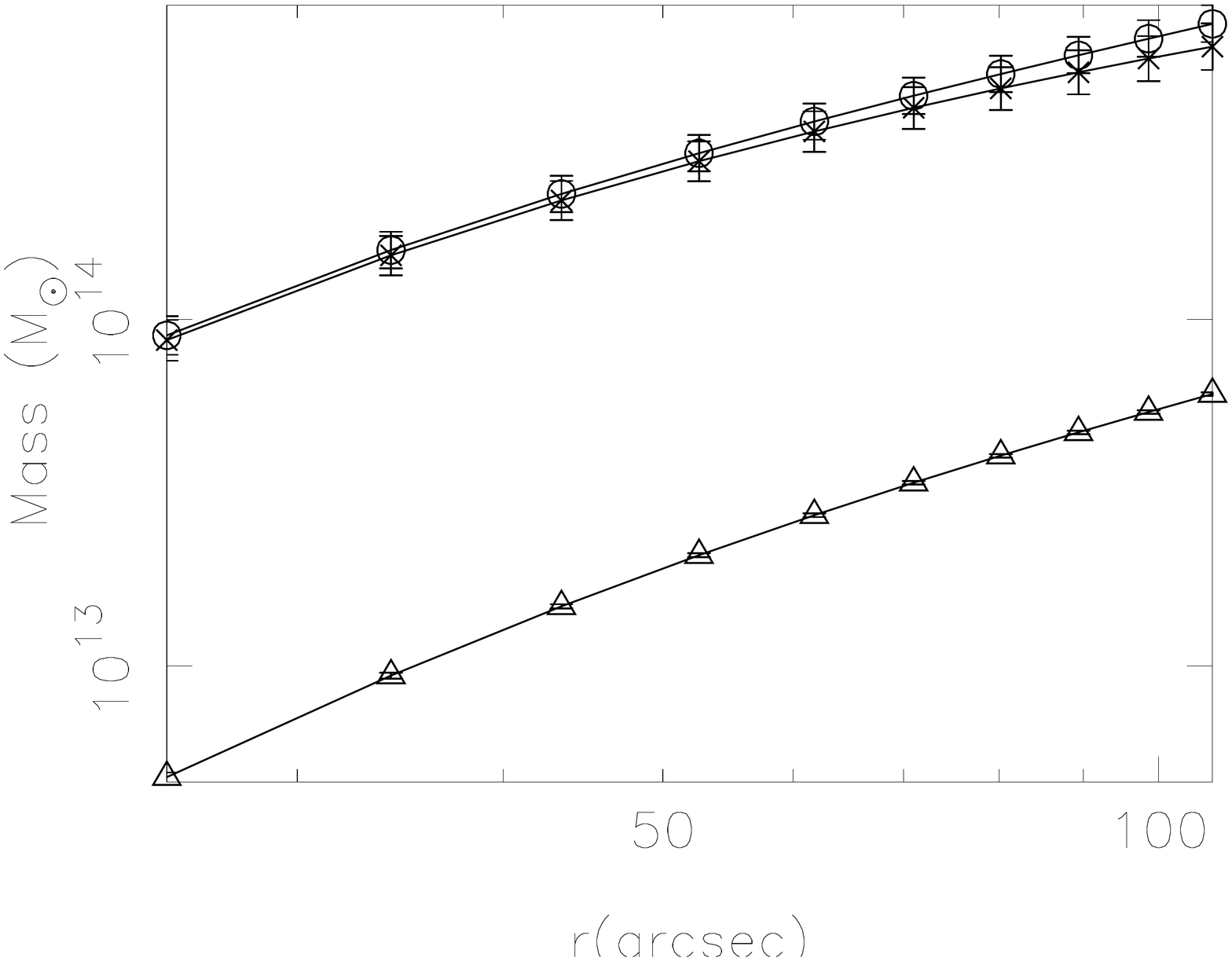}}
\subfloat[MACSJ0744.8+3927]{\label{MACSJ0744.8+3927masses}\includegraphics[width=3.0cm,angle=270]{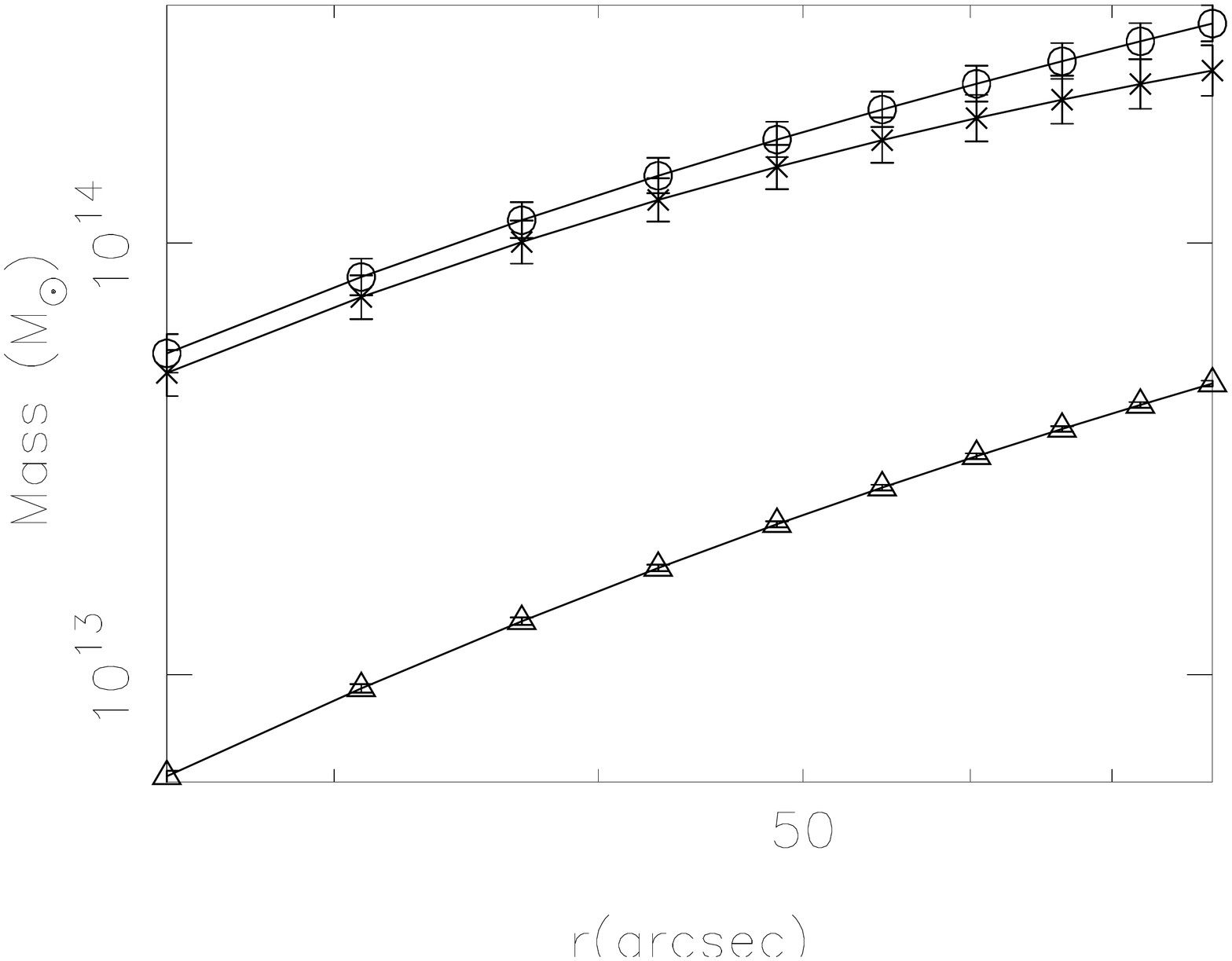}}
\subfloat[MACSJ1149.5+2223]{\label{MACSJ1149.5+2223masses}\includegraphics[width=3.0cm,angle=270]{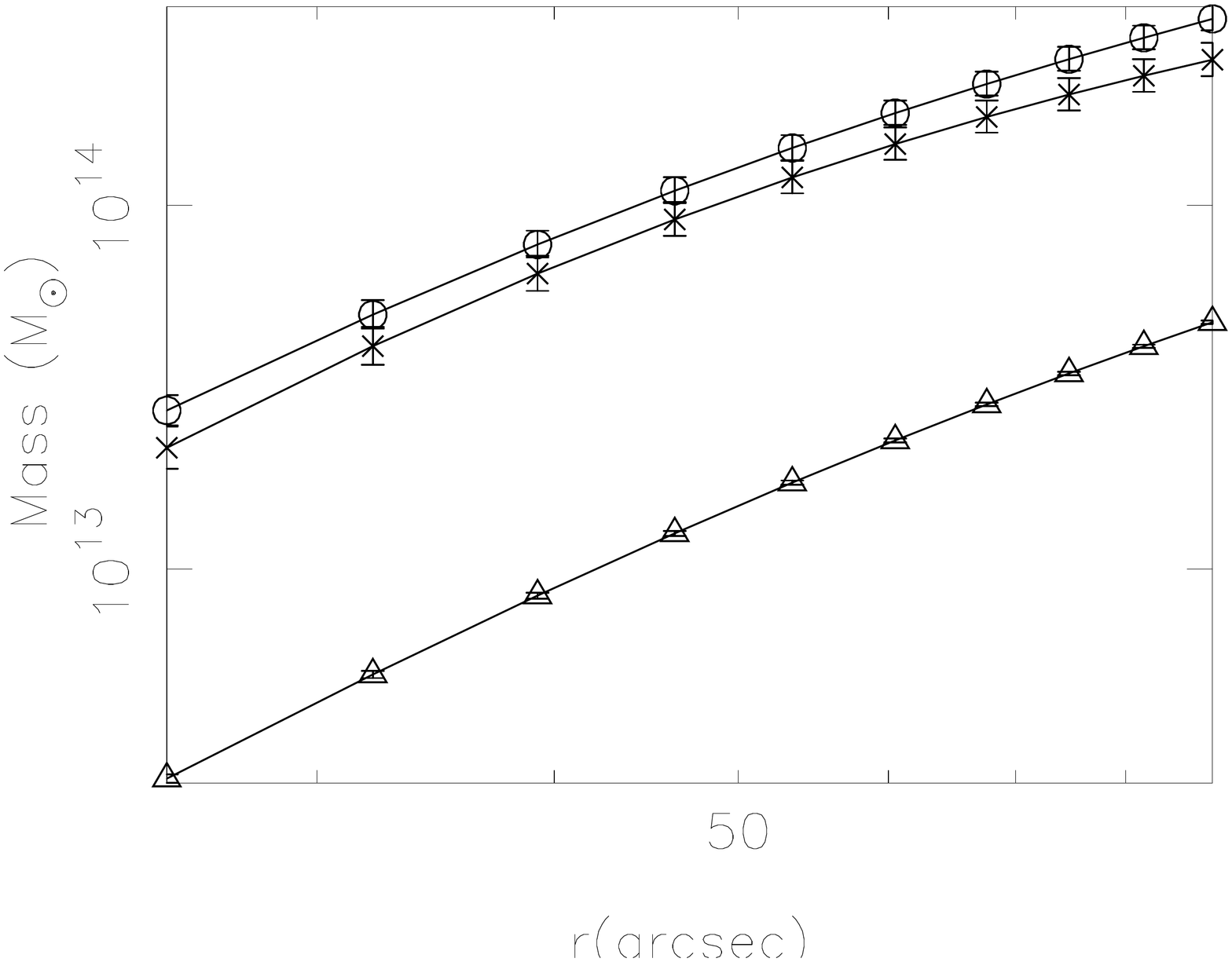}}

\subfloat[MACSJ1311.0-0310]{\label{MACSJ1311.0-0310masses}\includegraphics[width=3.0cm,angle=270]{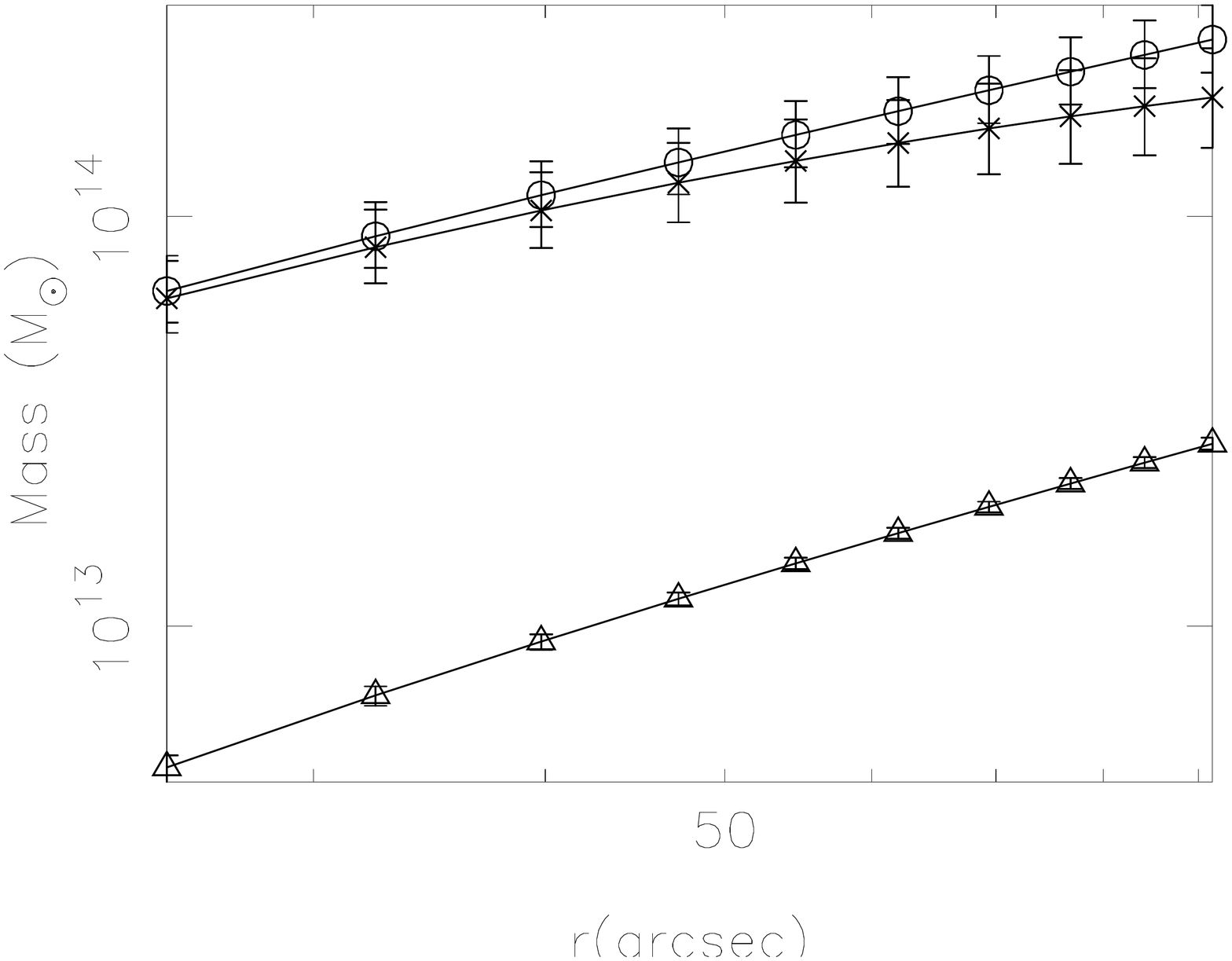}}
\subfloat[MACSJ1423.8+2404]{\label{MACSJ1423.8+2404masses}\includegraphics[width=3.0cm,angle=270]{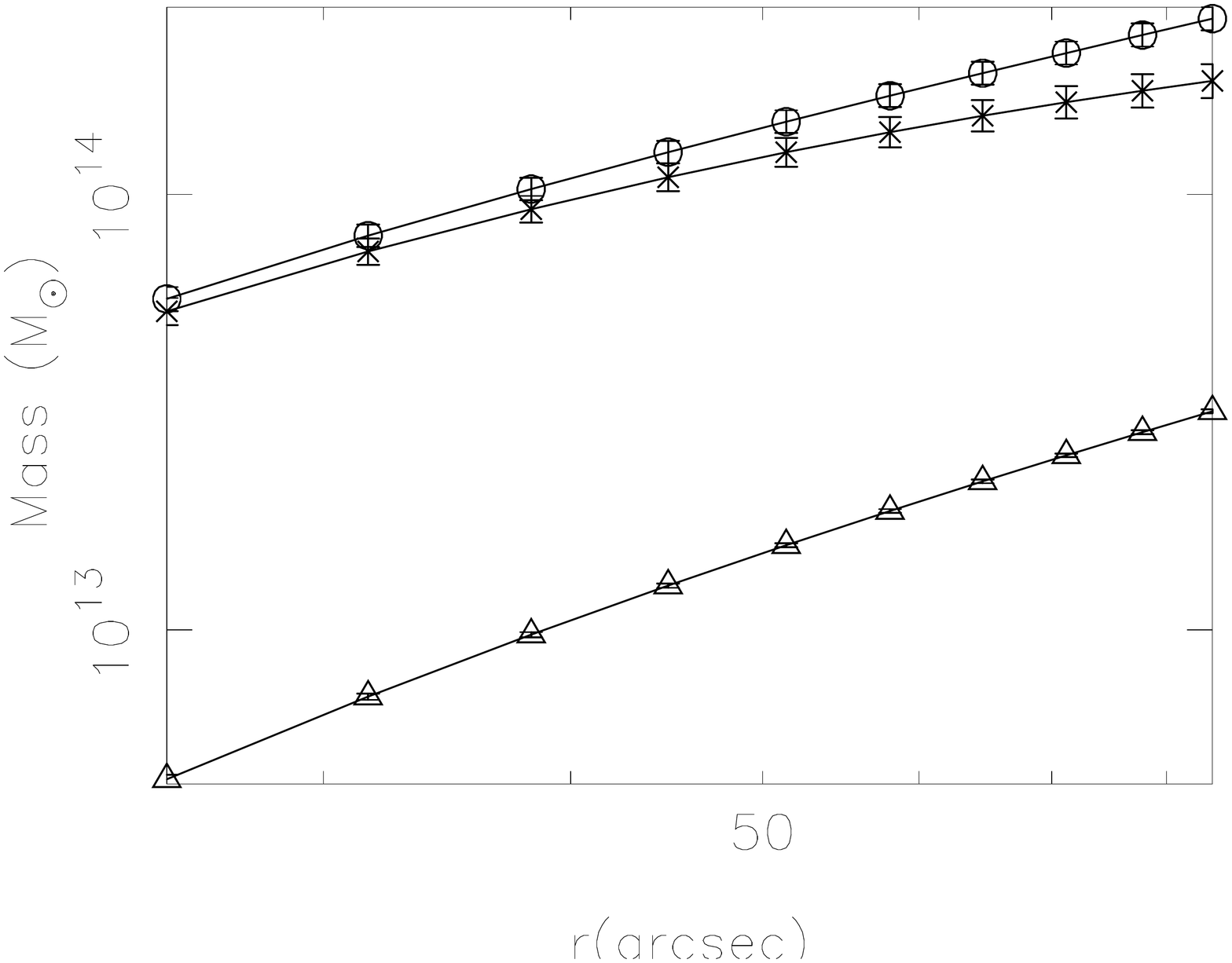}}
\subfloat[MACSJ2129.4-0741]{\label{MACSJ2129.4-0741masses}\includegraphics[width=3.0cm,angle=270]{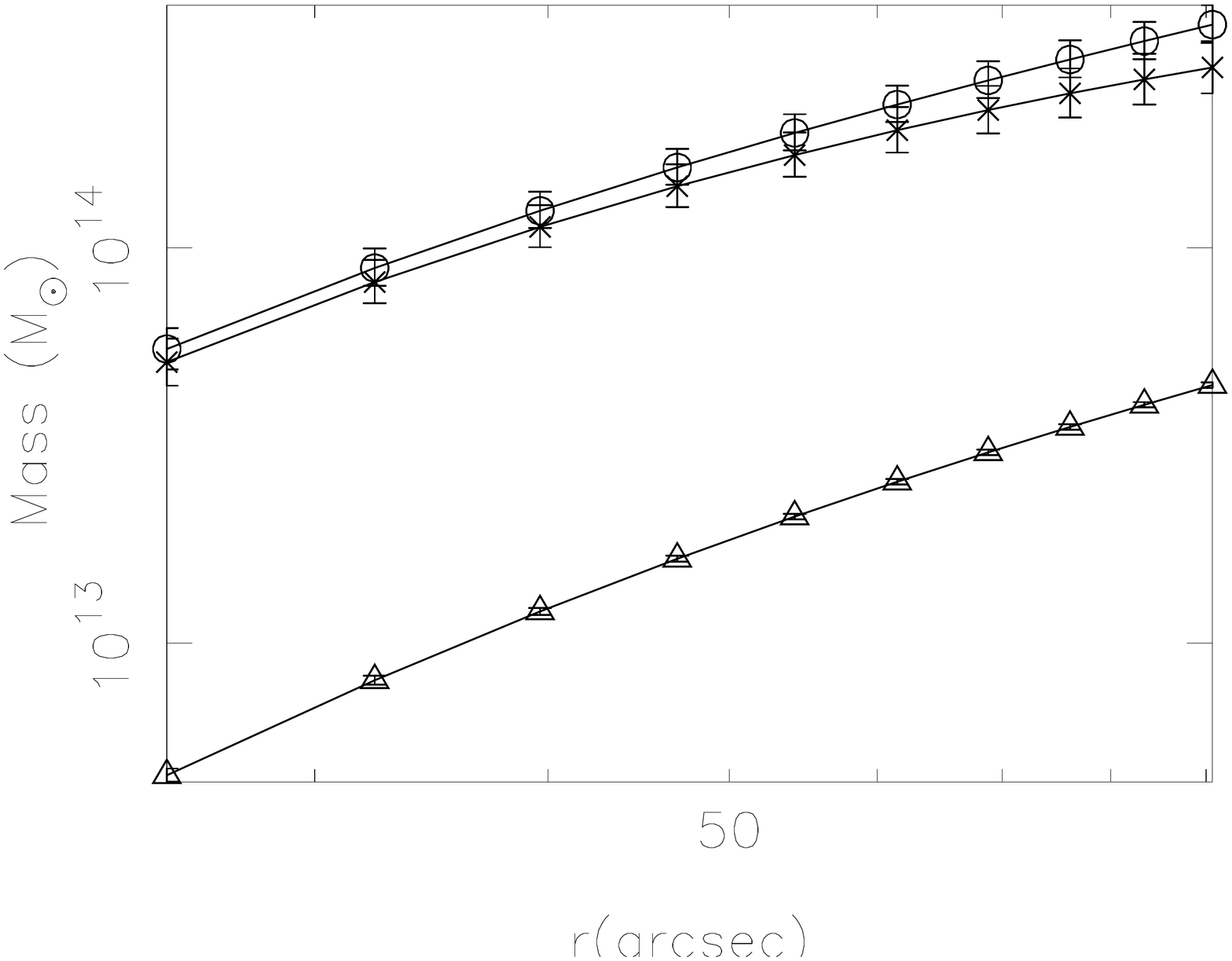}}
\end{figure}
\begin{figure}[p]
\ContinuedFloat
\centering
\subfloat[MACSJ2214.9-1359]{\label{MACSJ2214.9-1359masses}\includegraphics[width=3.0cm,angle=270]{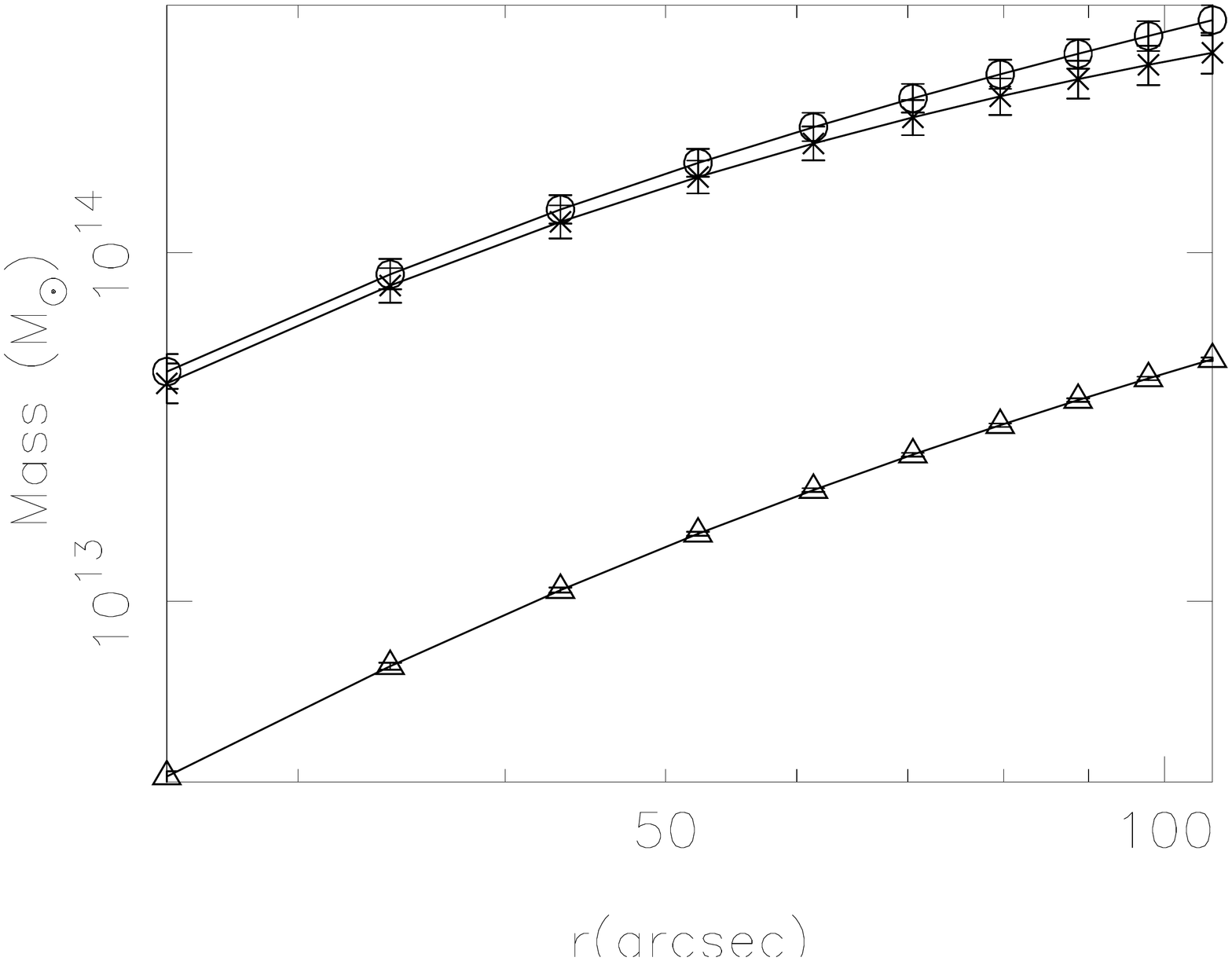}}
\subfloat[MACSJ2228.5+2036]{\label{MACSJ2228.5+2036masses}\includegraphics[width=3.0cm,angle=270]{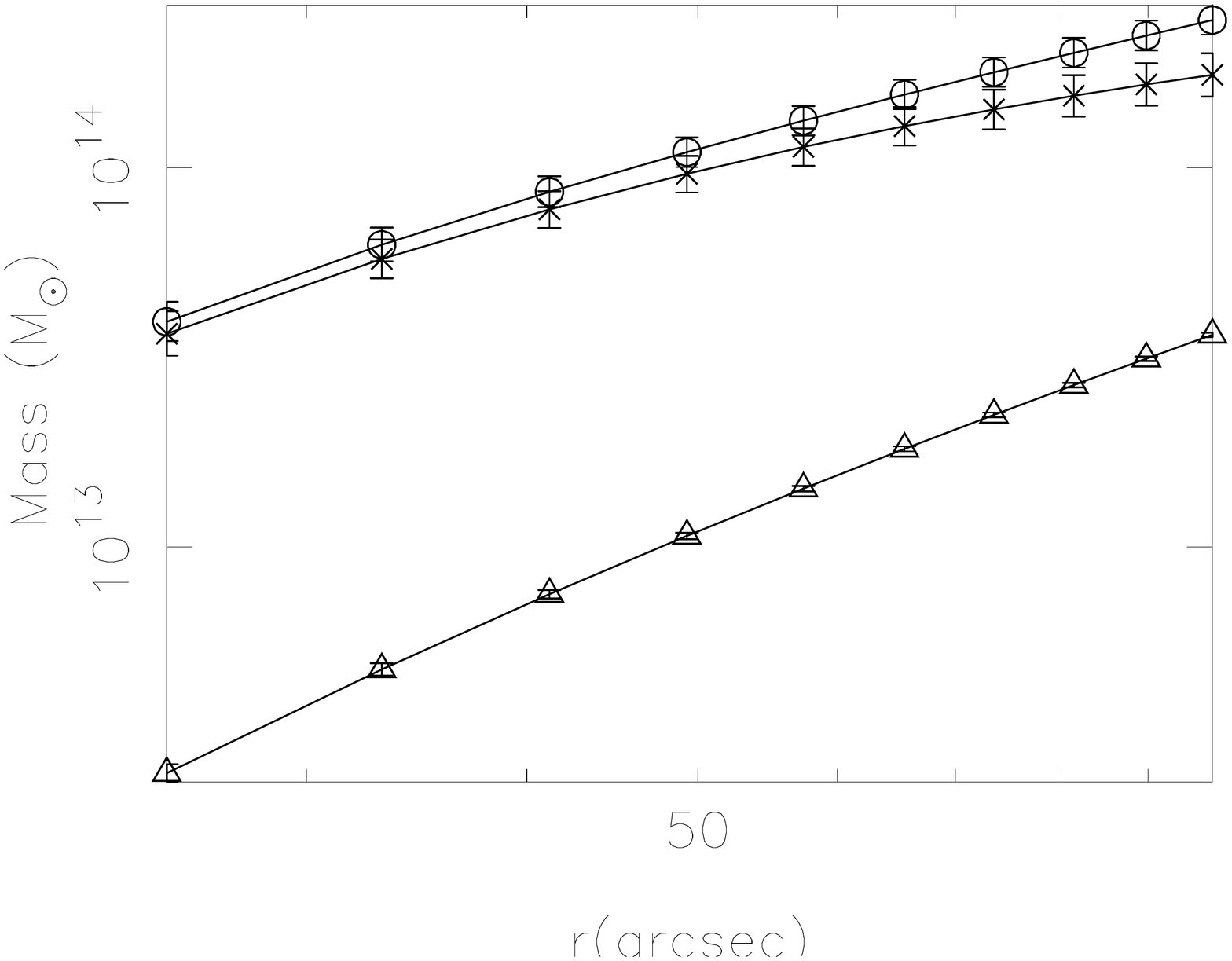}}
\subfloat[MS0451.6-0305]{\label{MS0451.6-0305masses}\includegraphics[width=3.0cm,angle=270]{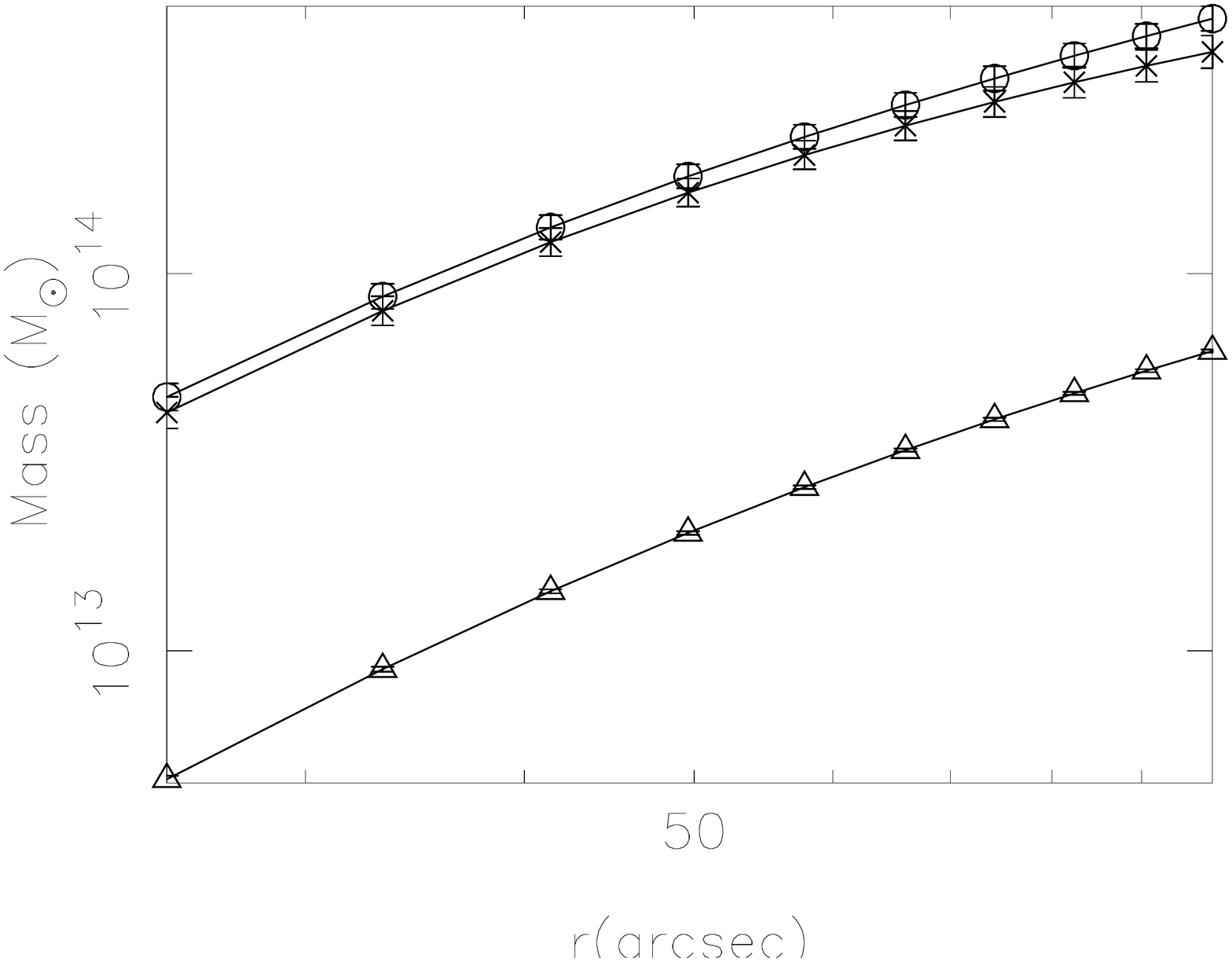}}

\subfloat[MS1054.5-0321]{\label{MS1054.5-0321masses}\includegraphics[width=3.0cm,angle=270]{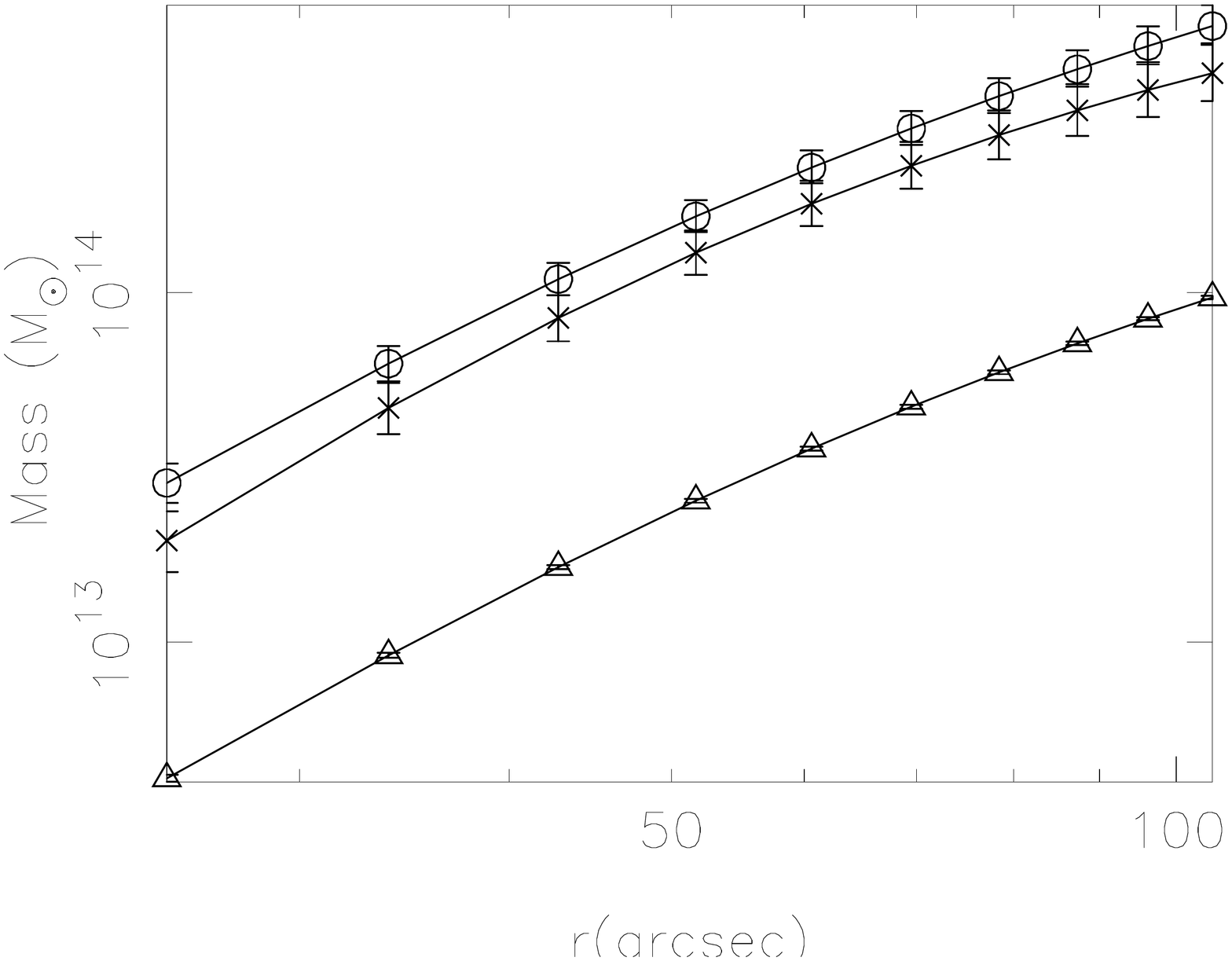}}
\subfloat[MS1137.5+6625]{\label{MS1137.5+6625masses}\includegraphics[width=3.0cm,angle=270]{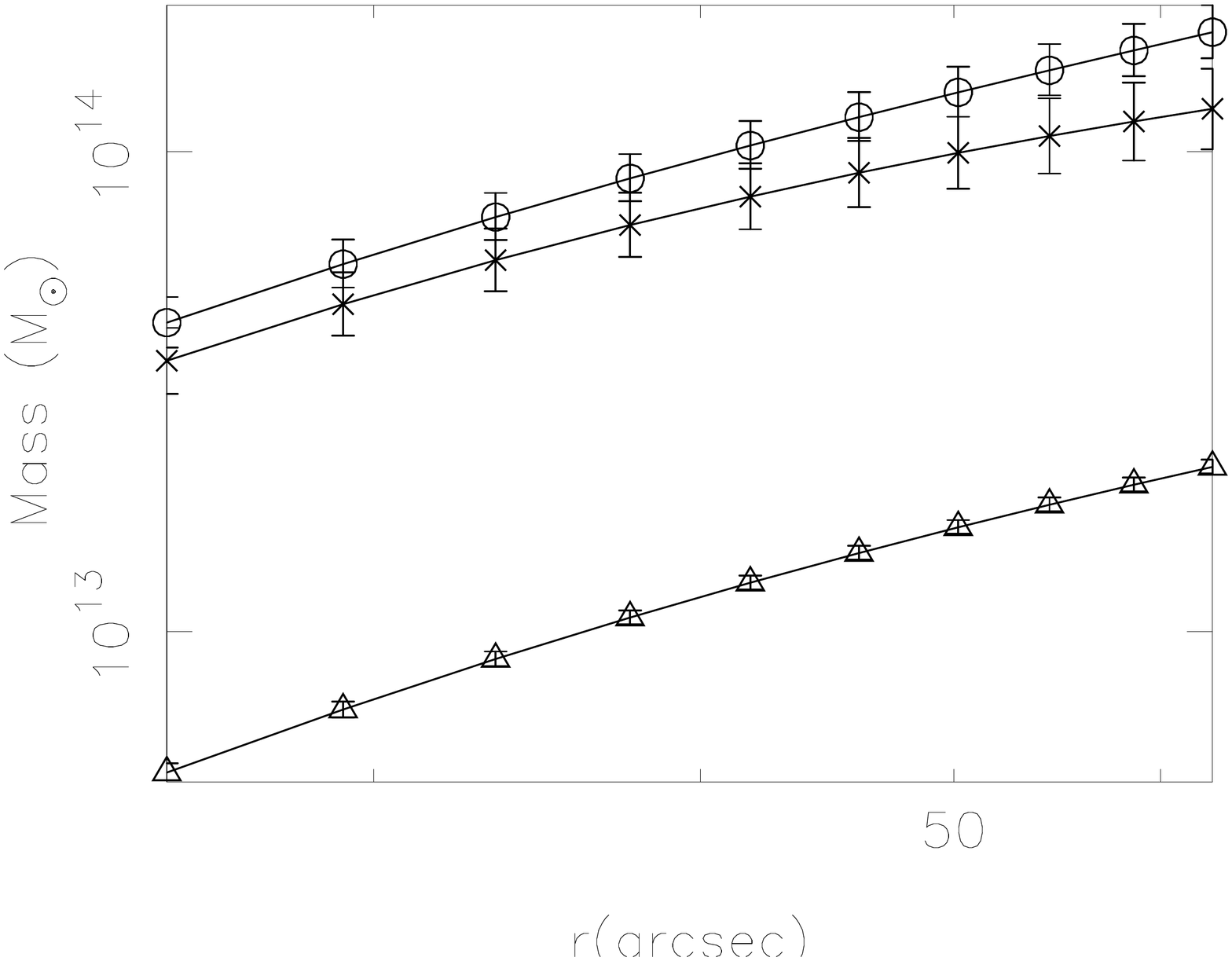}}
\subfloat[MS1358.4+6245]{\label{MS1358.4+6245masses}\includegraphics[width=3.0cm,angle=270]{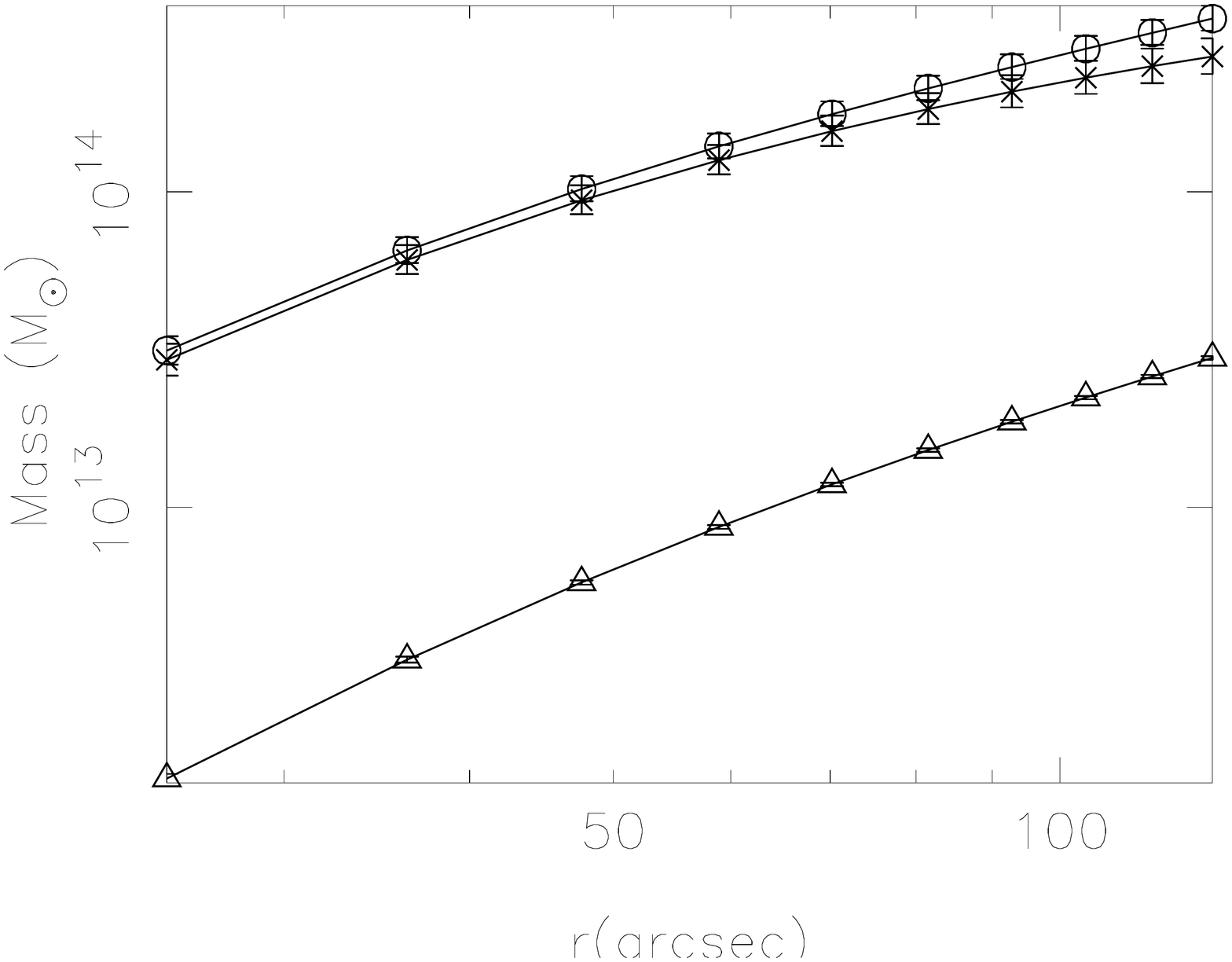}}

\subfloat[MS2053.7-0449]{\label{MS2053.7-0449masses}\includegraphics[width=3.0cm,angle=270]{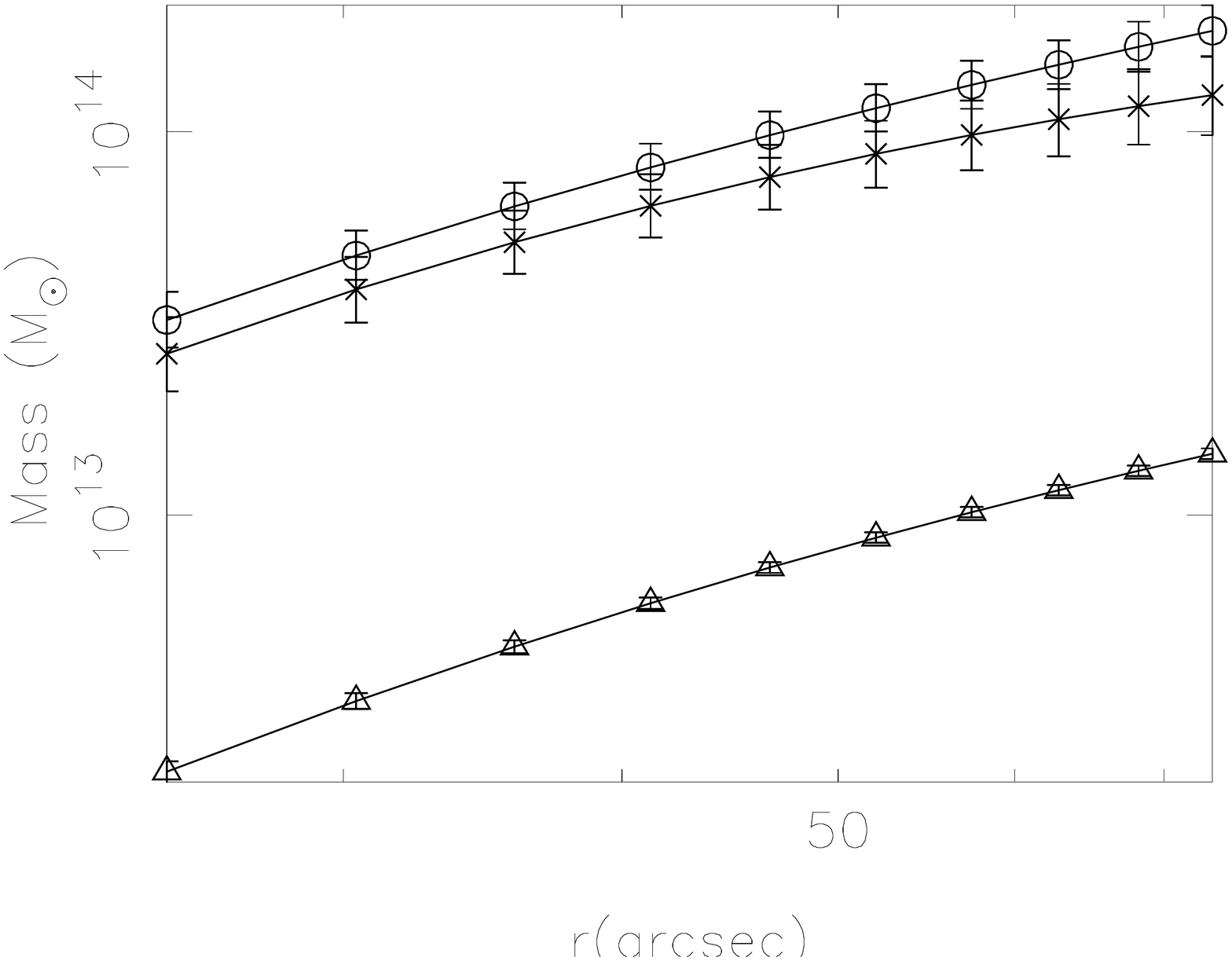}}
\subfloat[RXJ1347.5-1145]{\label{RXJ1347.5-1145masses}\includegraphics[width=3.0cm,angle=270]{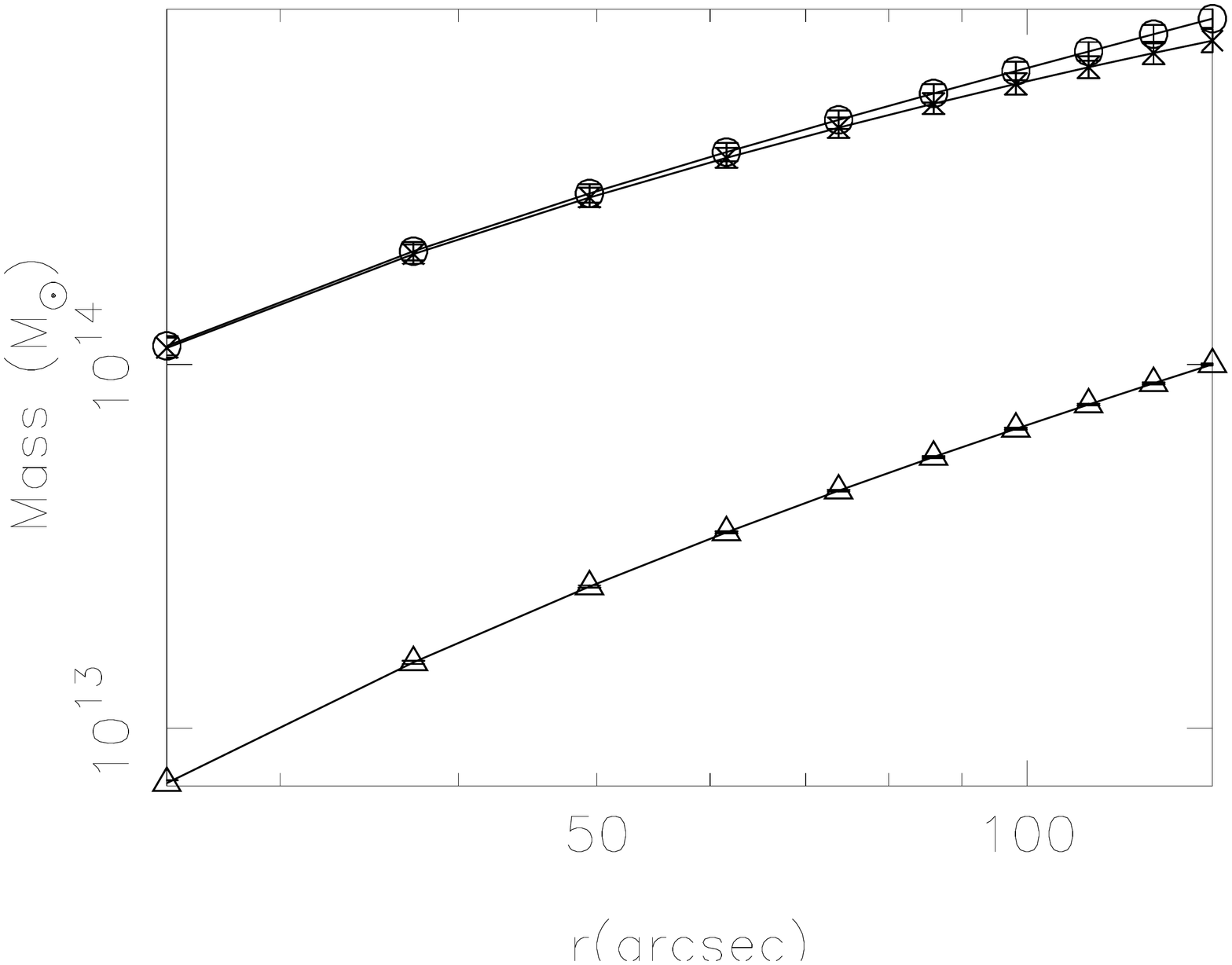}}
\subfloat[RXJ1716.4+6708]{\label{RXJ1716.4+6708masses}\includegraphics[width=3.0cm,angle=270]{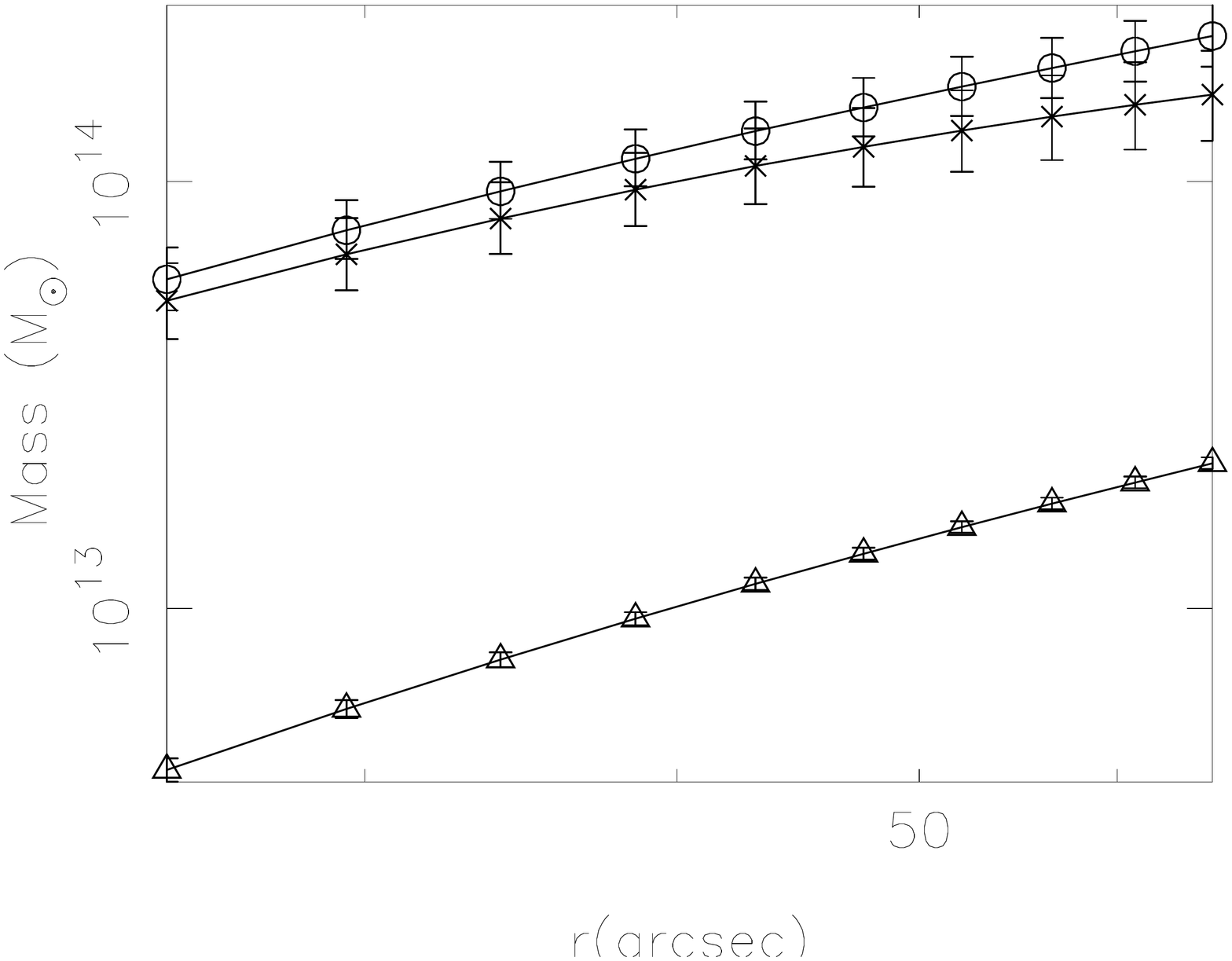}}

\subfloat[RXJ2129.7+0005]{\label{RXJ2129.7+0005masses}\includegraphics[width=3.0cm,angle=270]{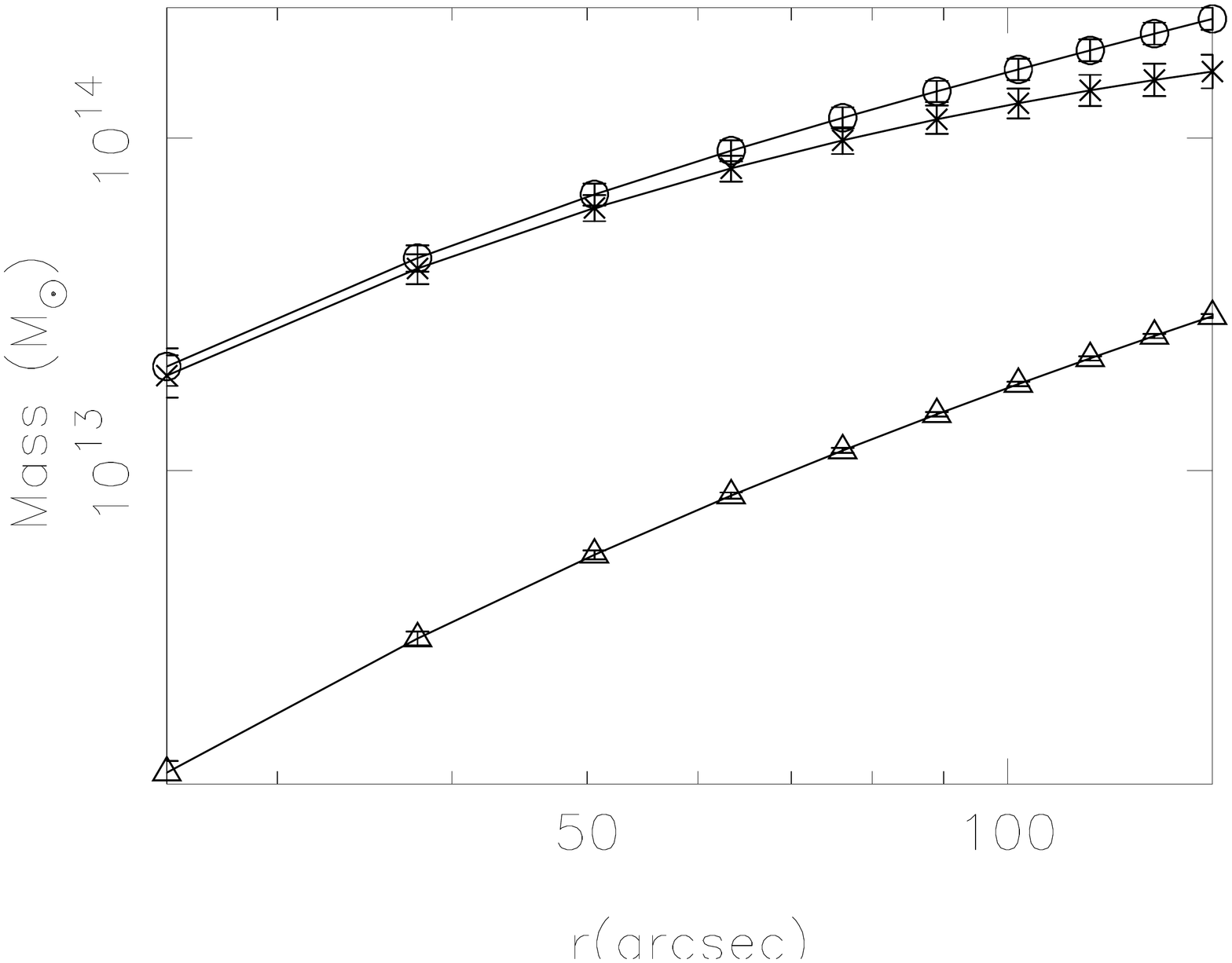}}
\subfloat[ZW3146]{\label{ZW3146masses}\includegraphics[width=3.0cm,angle=270]{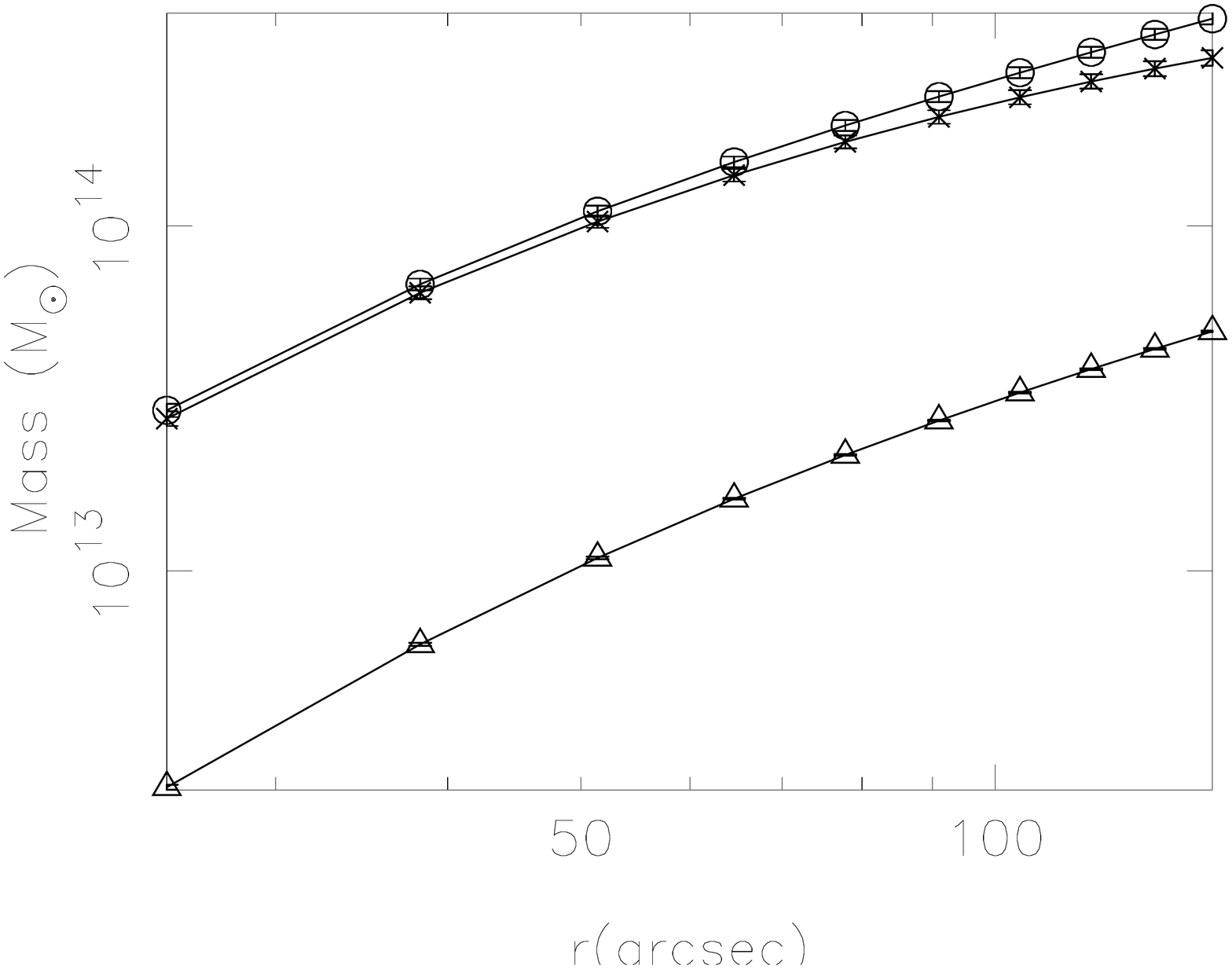}}\end{figure}
\clearpage

\section*{Appendix B}
In this Appendix we provide the radial distribution of the best-fit values of the critical acceleration $a_{0}$, for all clusters. The plot extend between
100~kpc and $r_{2500}$.
\label{appendix2}
\begin{figure}[!h]
\setcounter{figure}{5}
\centering
\caption{\footnotesize The critical acceleration $a_{\circ}$ as a function of radius, under the condition $M_{MOND}=M_{baryon}$ for all 38 clusters}. \label{a0plotappendix}
\subfloat[Abell1413]{\label{Abell1413a0}\includegraphics[width=3.0cm,angle=270]{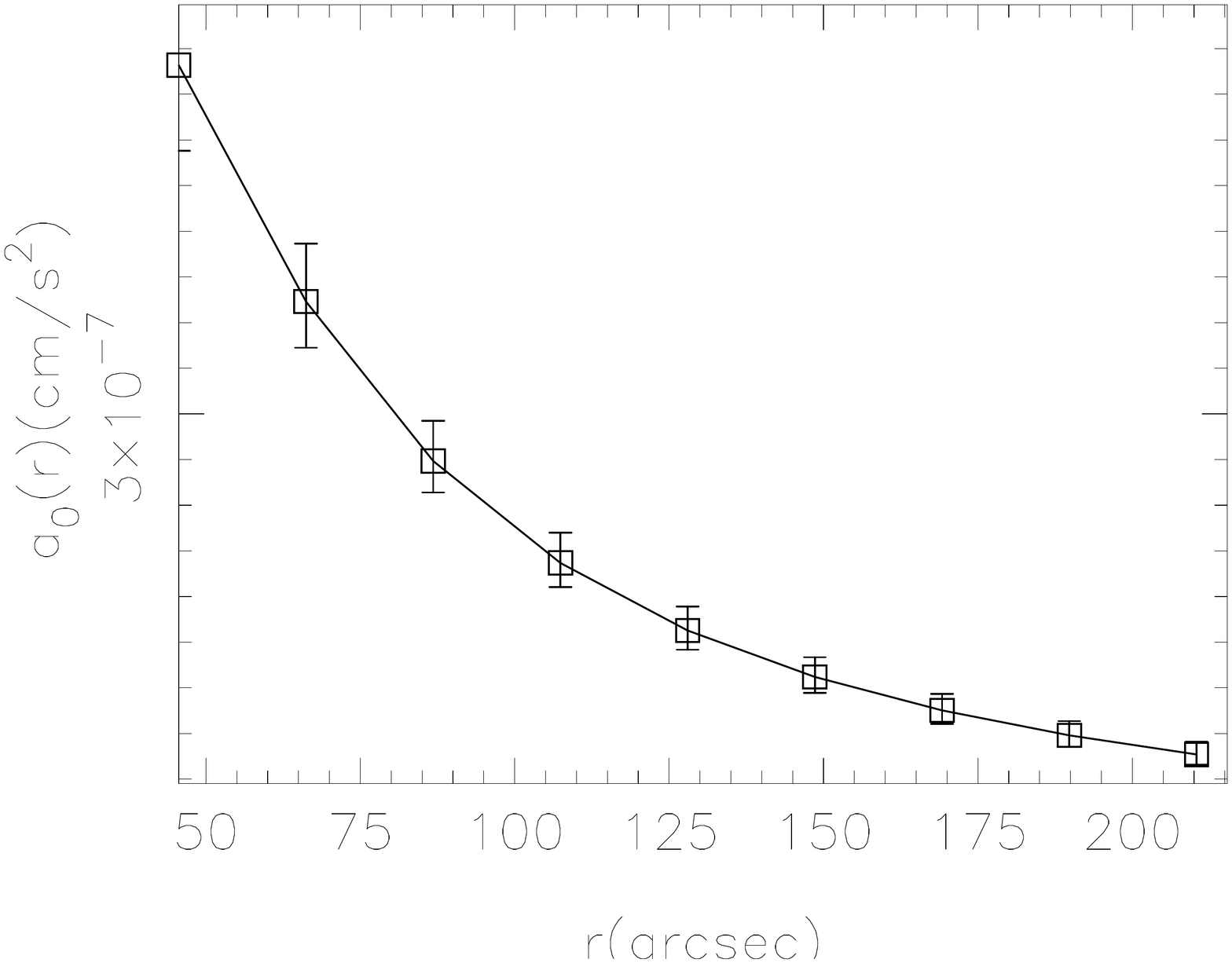}}
\subfloat[Abell1689]{\label{Abell1689a0}\includegraphics[width=3.0cm,angle=270]{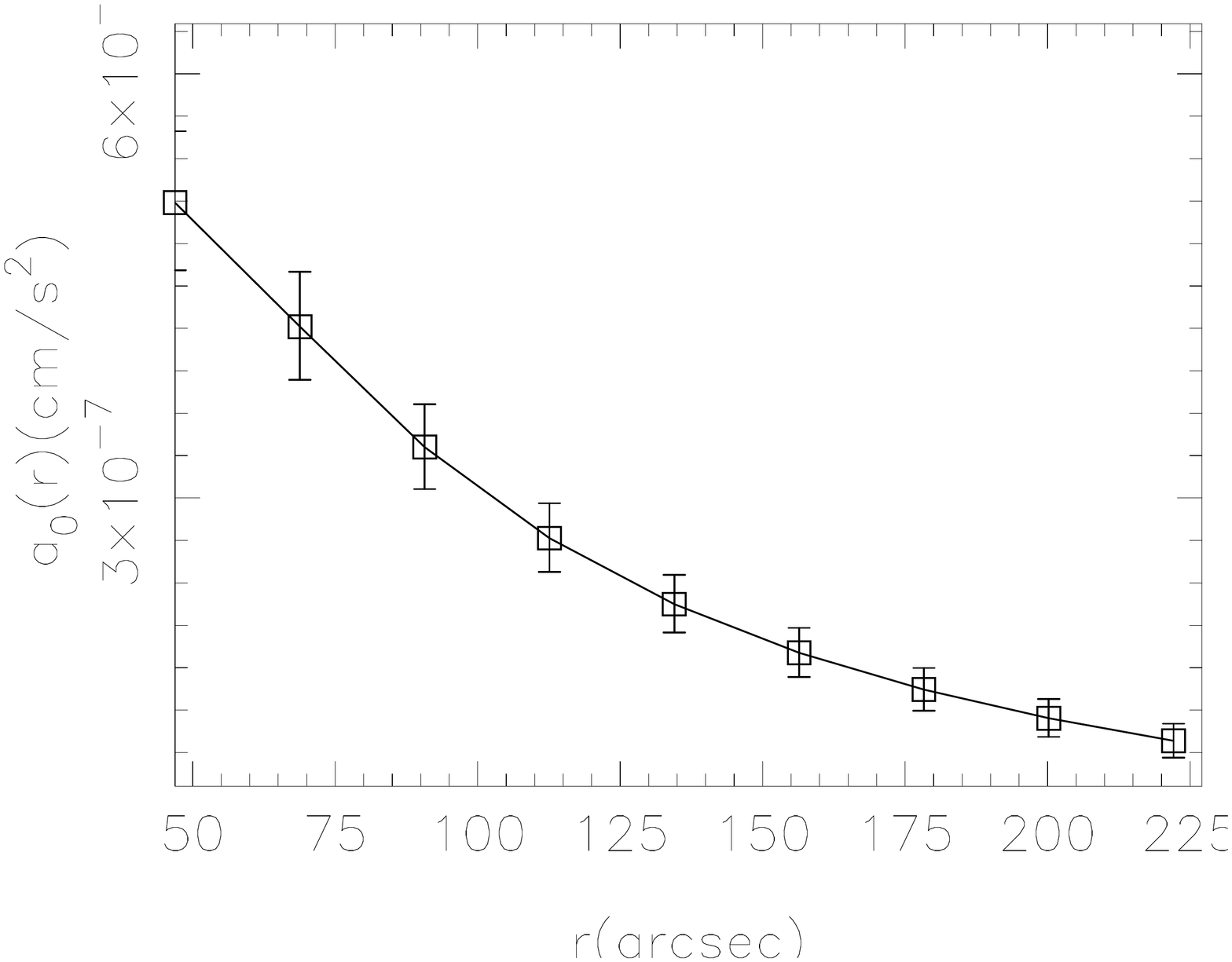}}
\subfloat[Abell1835]{\label{Abell1835a0}\includegraphics[width=3.0cm,angle=270]{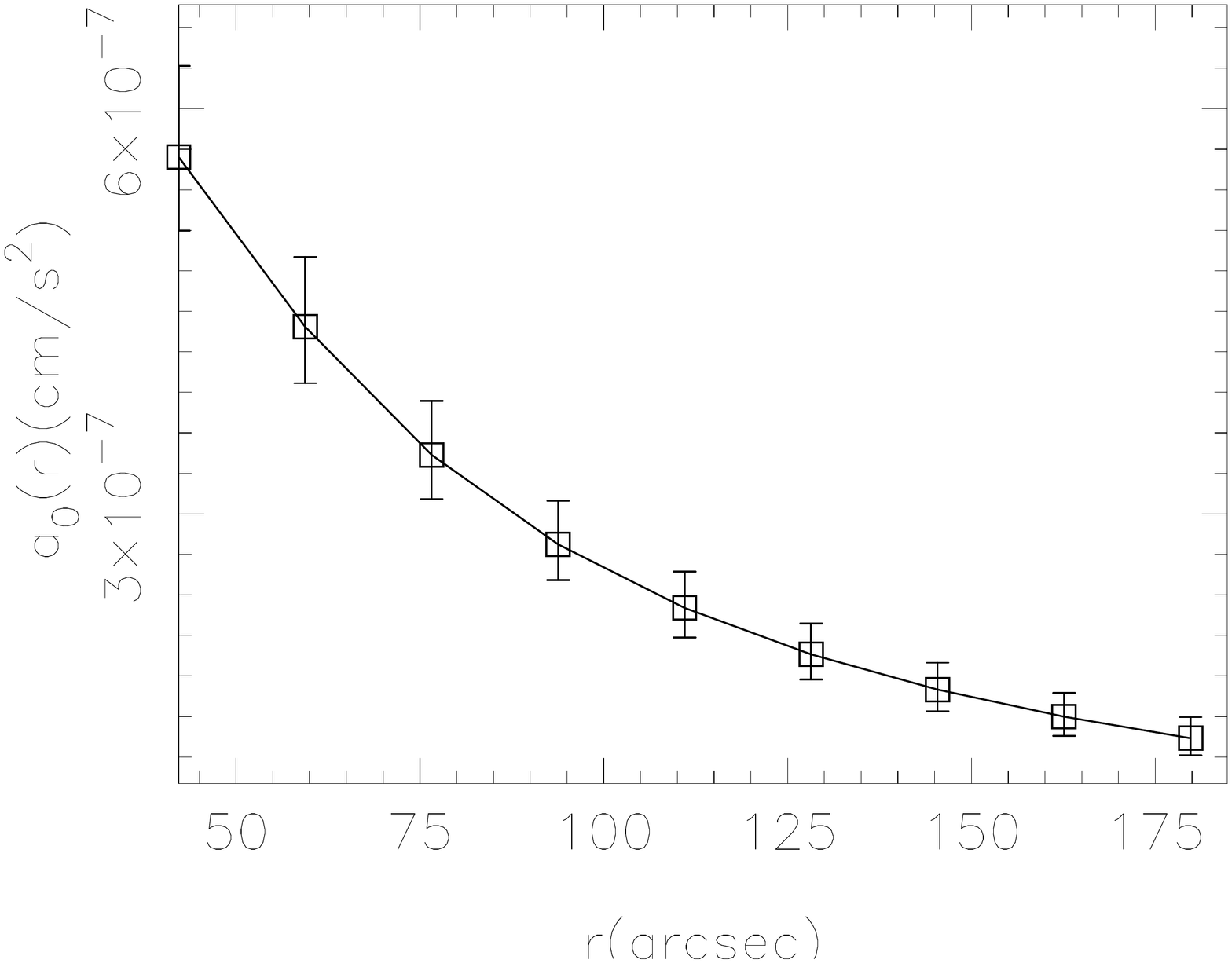}}

\subfloat[Abell1914]{\label{Abell1914a0}\includegraphics[width=3.0cm,angle=270]{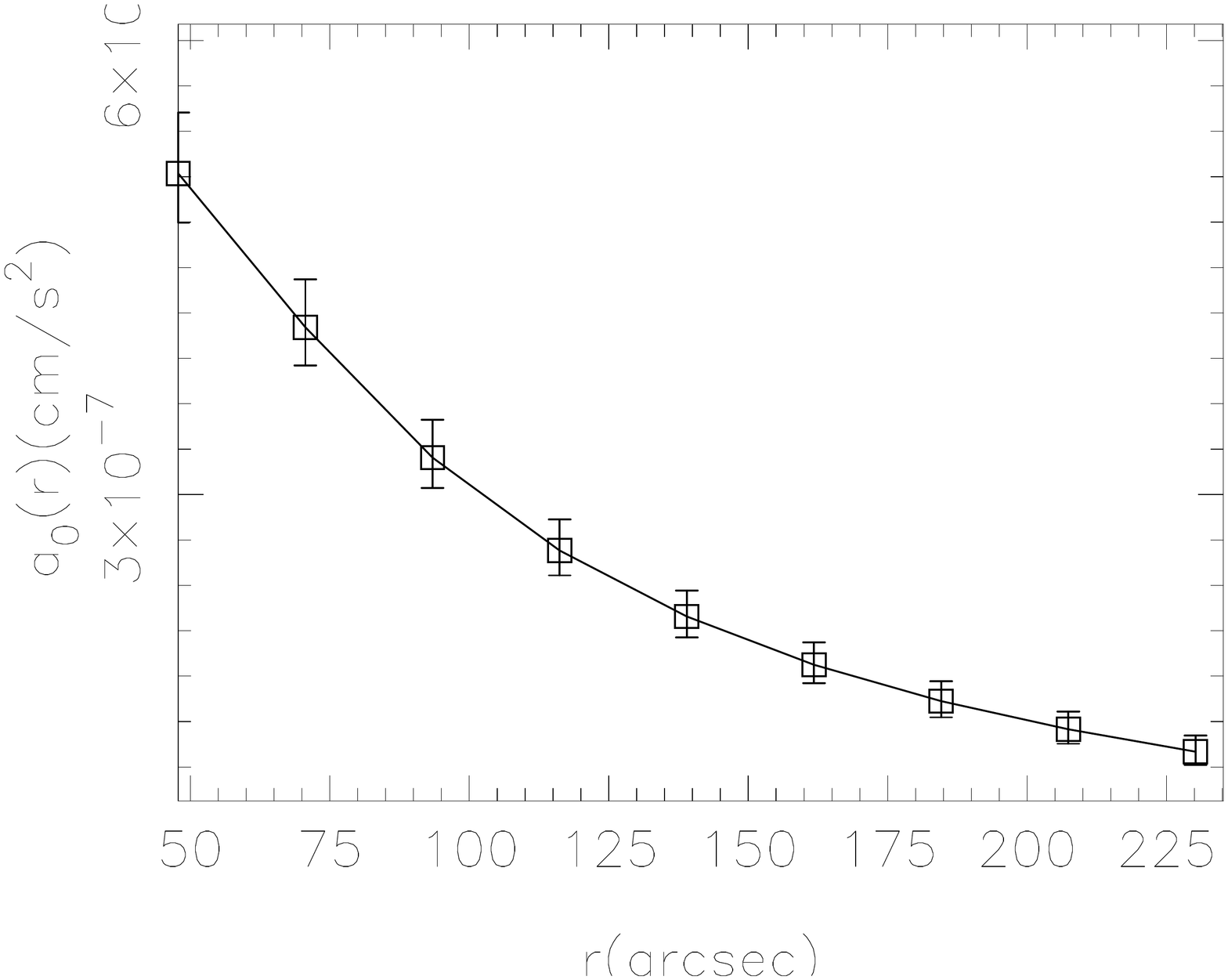}}
\subfloat[Abell1995]{\label{Abell1995a0}\includegraphics[width=3.0cm,angle=270]{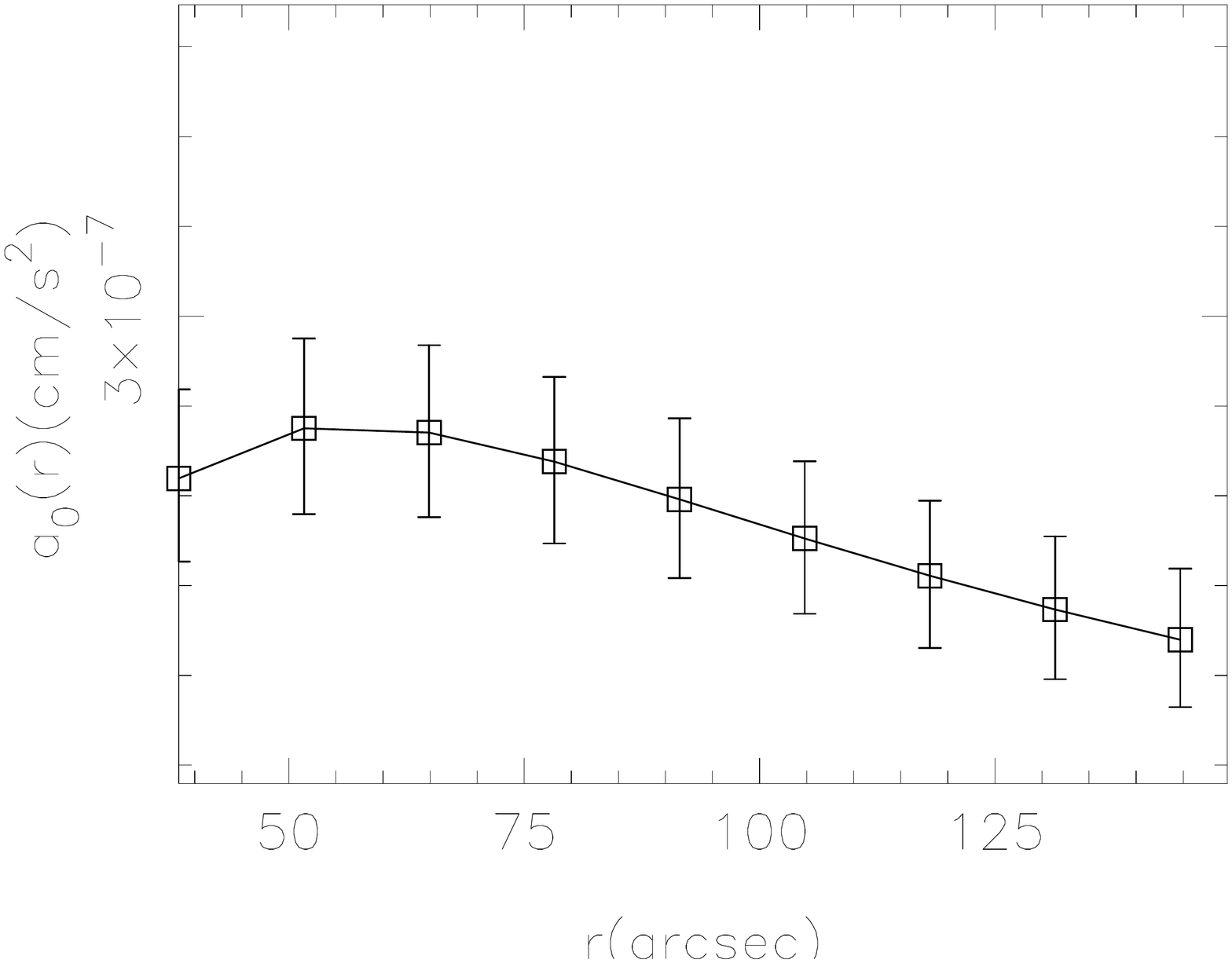}}
\subfloat[Abell2111]{\label{Abell2111a0}\includegraphics[width=3.0cm,angle=270]{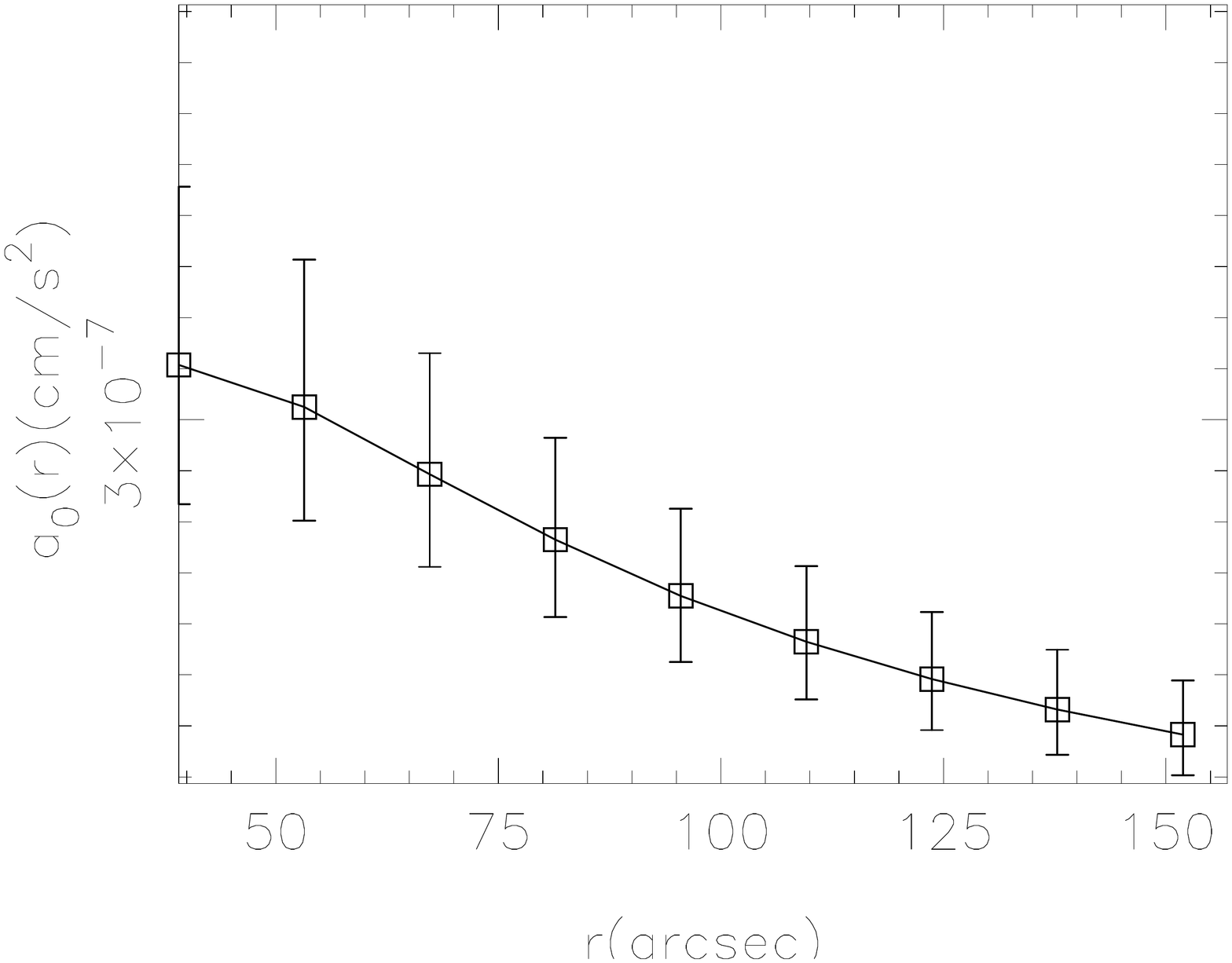}}

\subfloat[Abell2163]{\label{Abell2163a0}\includegraphics[width=3.0cm,angle=270]{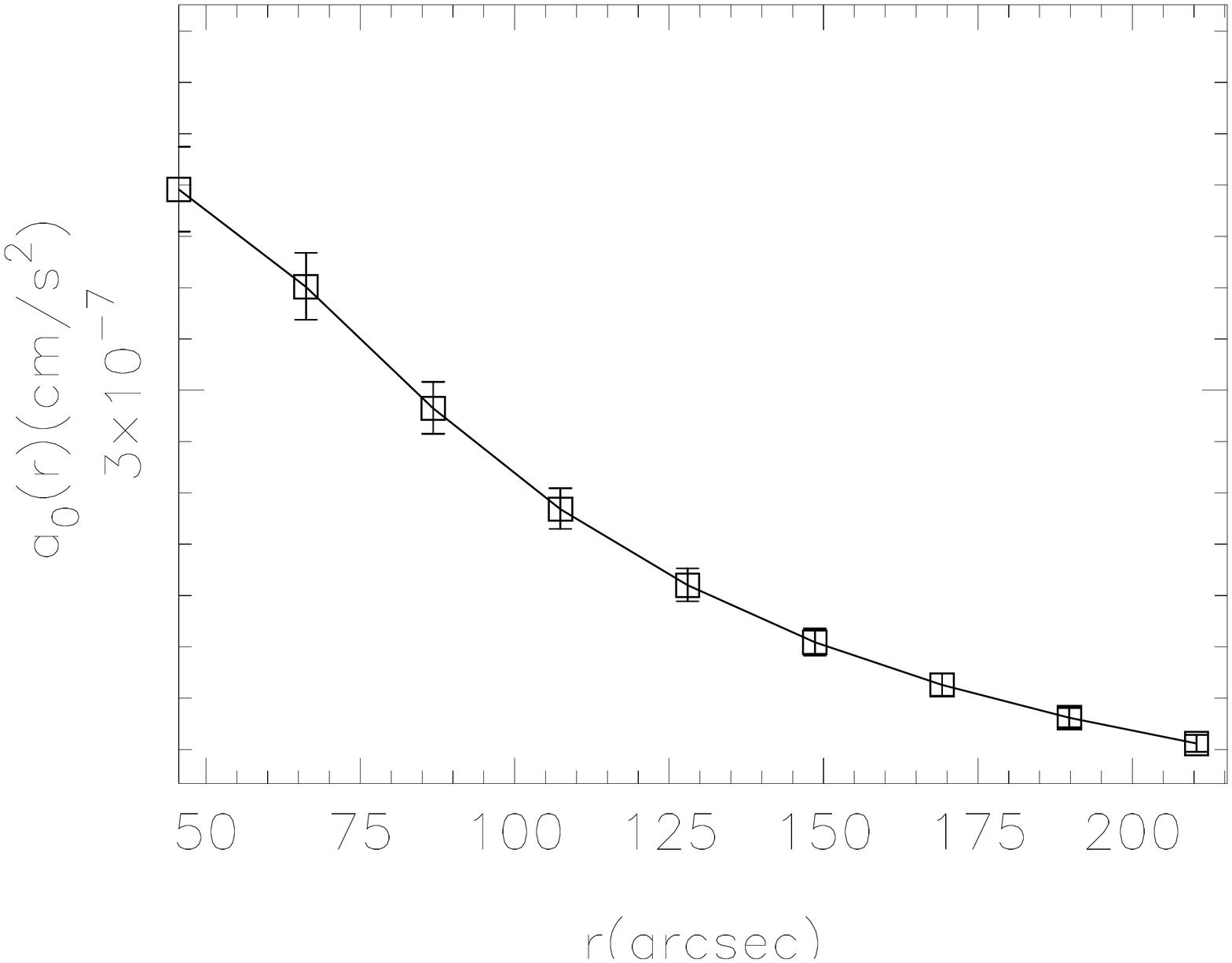}}
\subfloat[Abell2204]{\label{Abell2204a0}\includegraphics[width=3.0cm,angle=270]{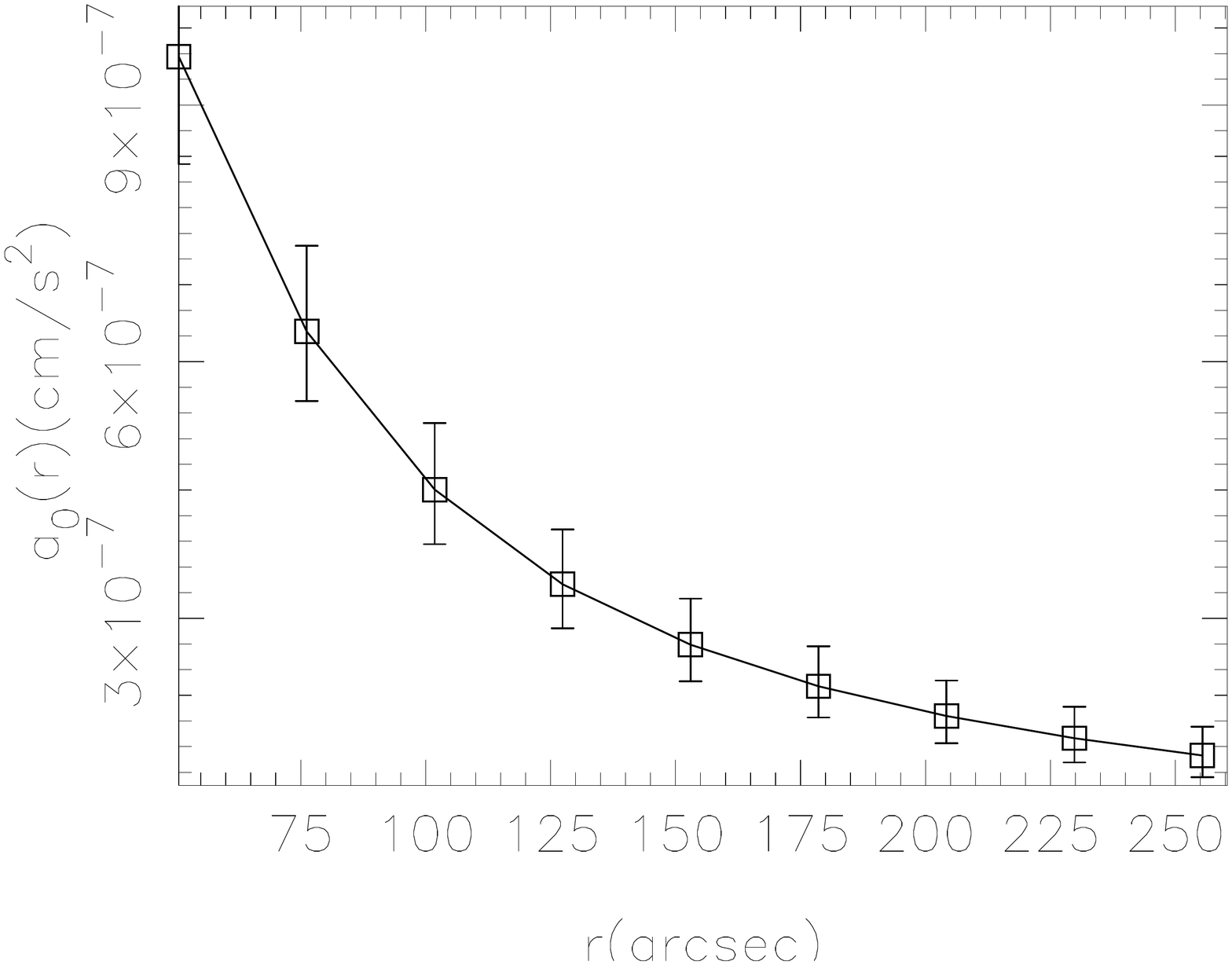}}
\subfloat[Abell2218]{\label{Abell2218a0}\includegraphics[width=3.0cm,angle=270]{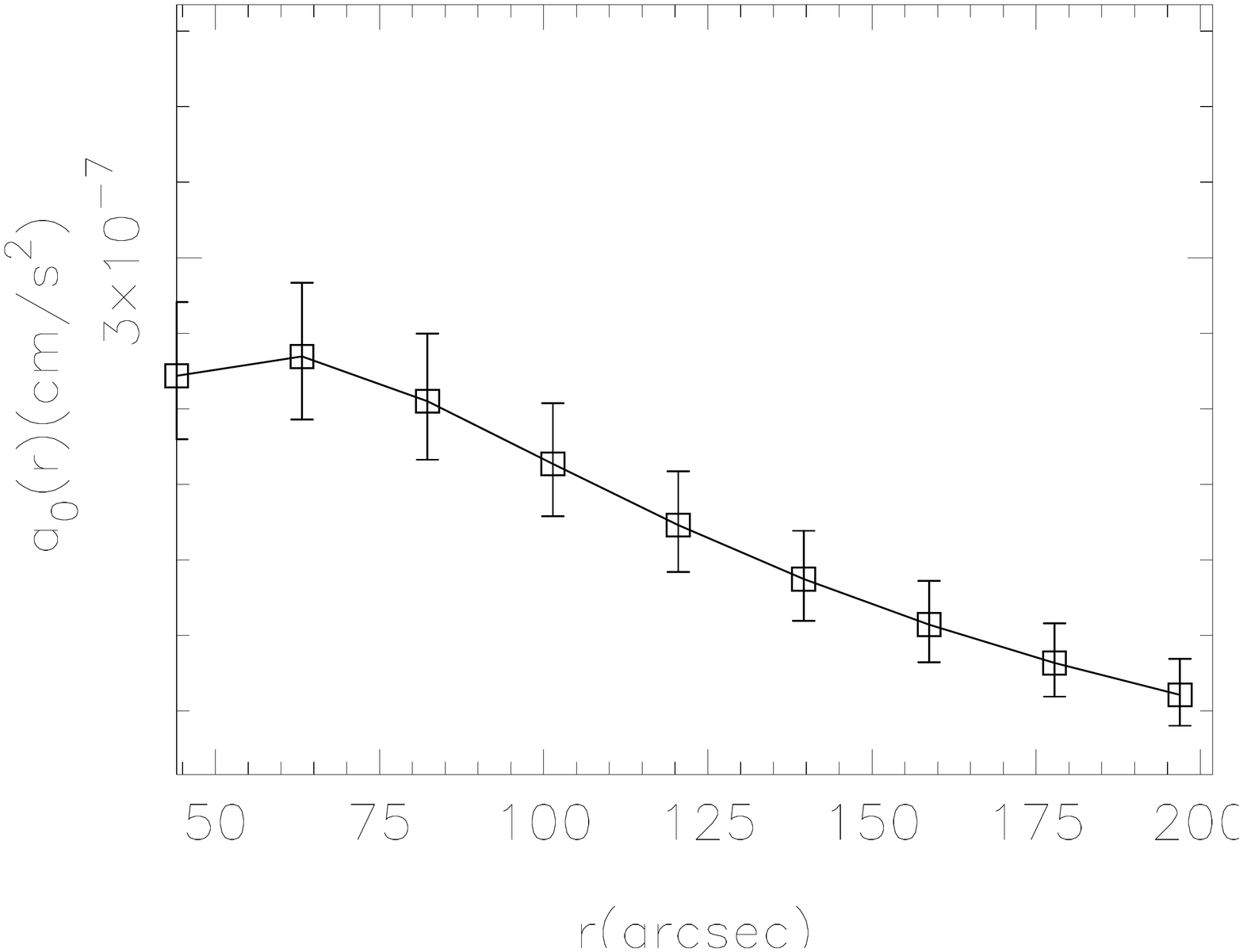}}

\end{figure}
\begin{figure}[p]
\ContinuedFloat
\centering
\subfloat[Abell2259]{\label{Abell2259a0}\includegraphics[width=3.0cm,angle=270]{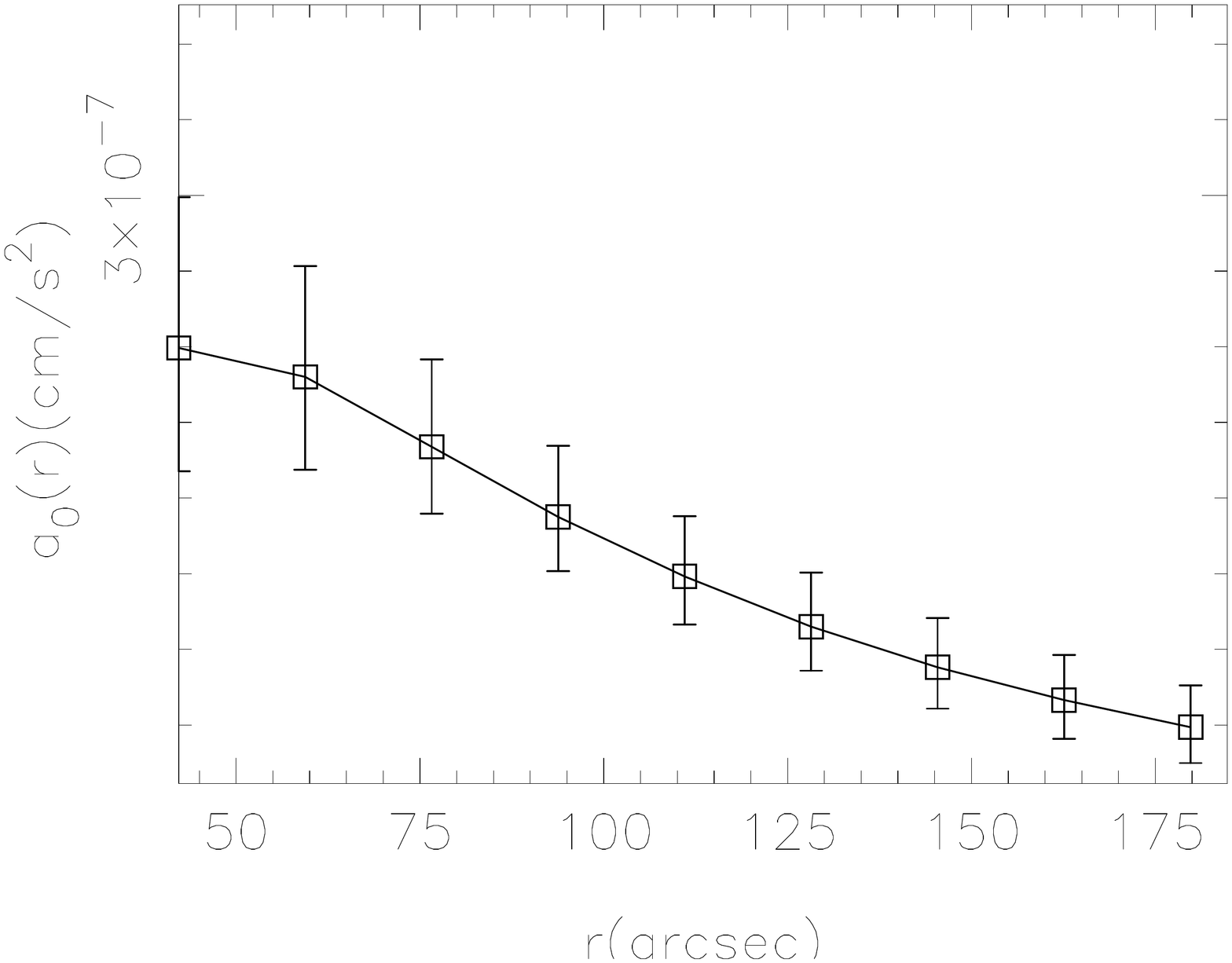}}
\subfloat[Abell2261]{\label{Abell2261a0}\includegraphics[width=3.0cm,angle=270]{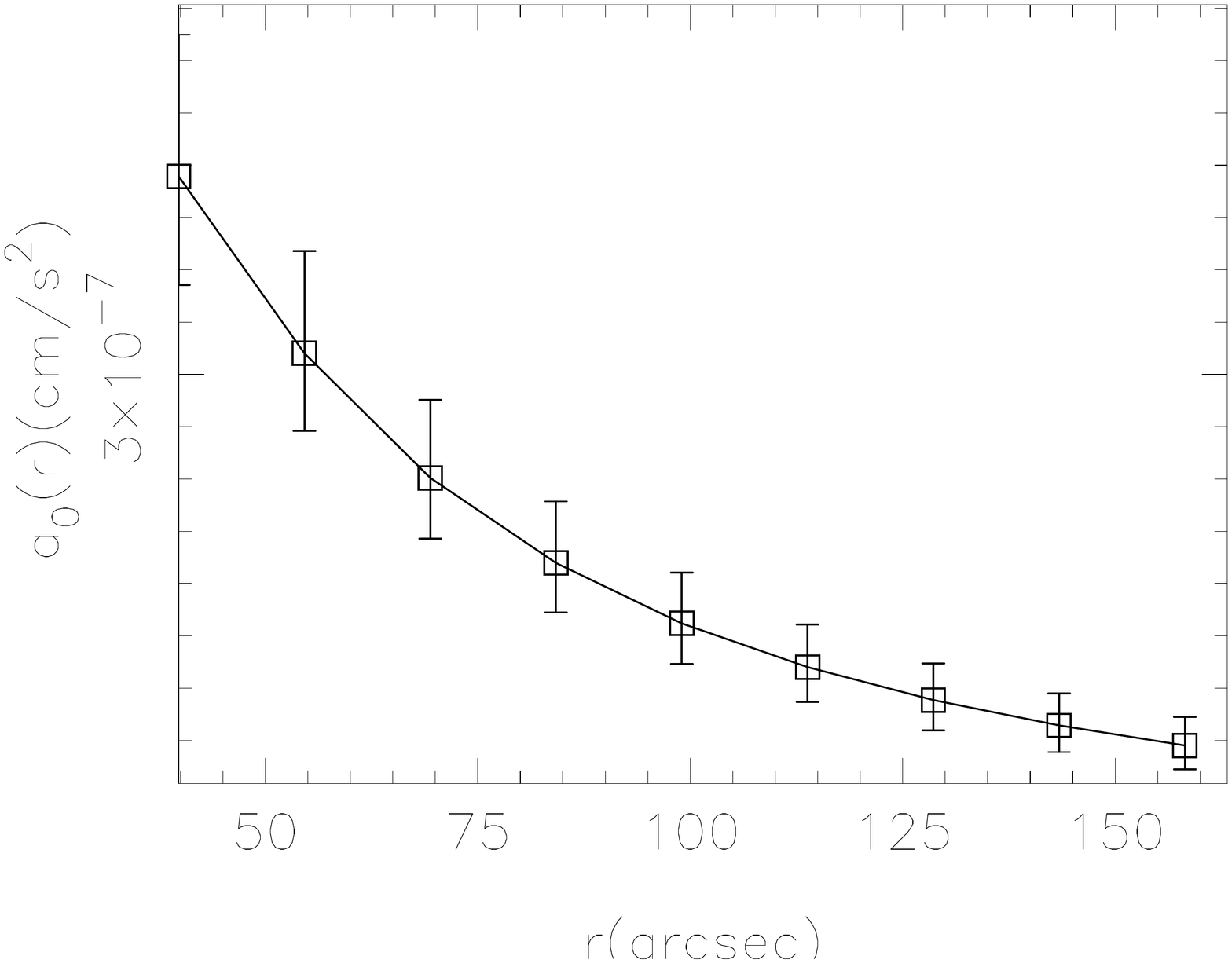}}
\subfloat[Abell267]{\label{Abell267a0}\includegraphics[width=3.0cm,angle=270]{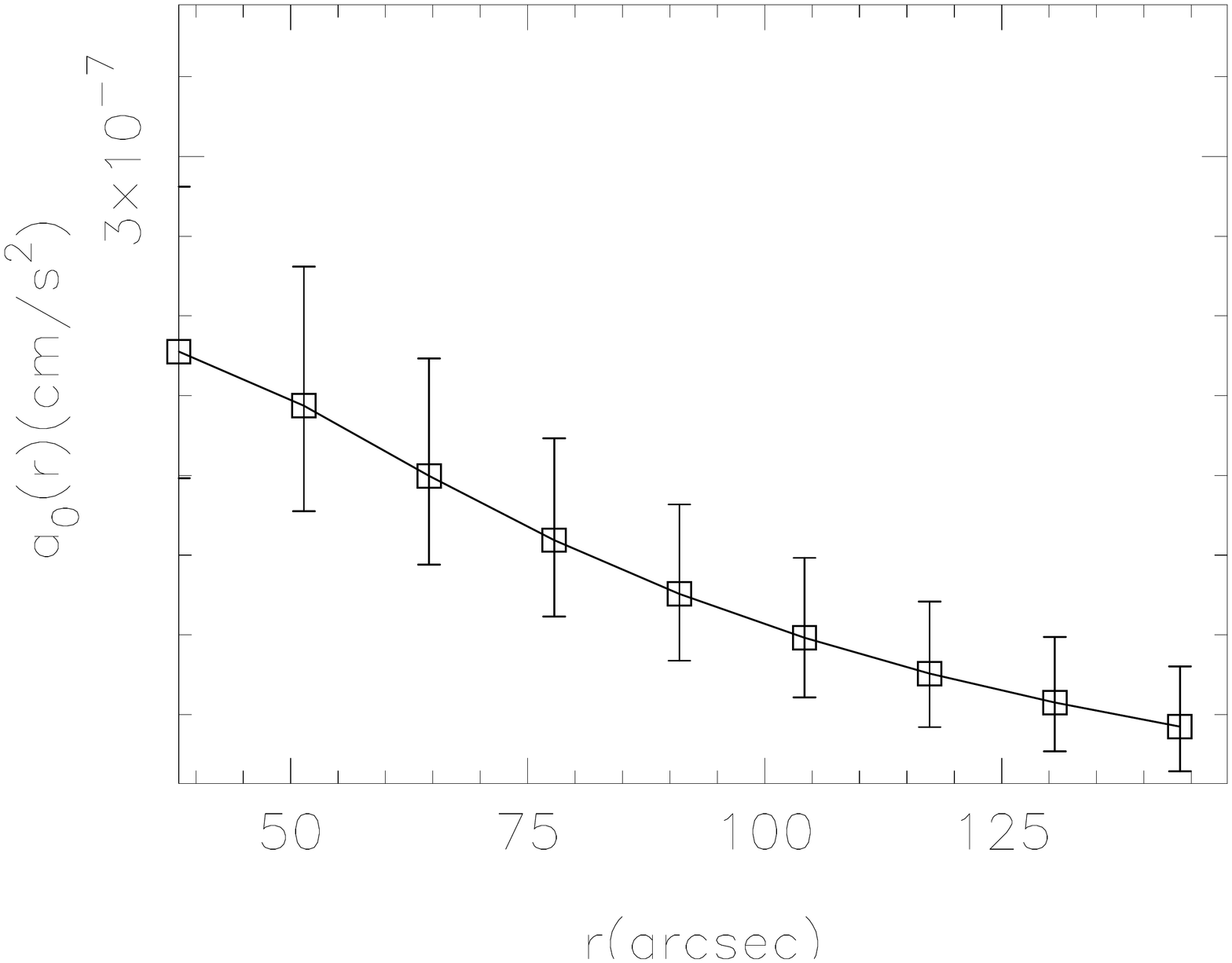}}

\subfloat[Abell370]{\label{Abell370a0}\includegraphics[width=3.0cm,angle=270]{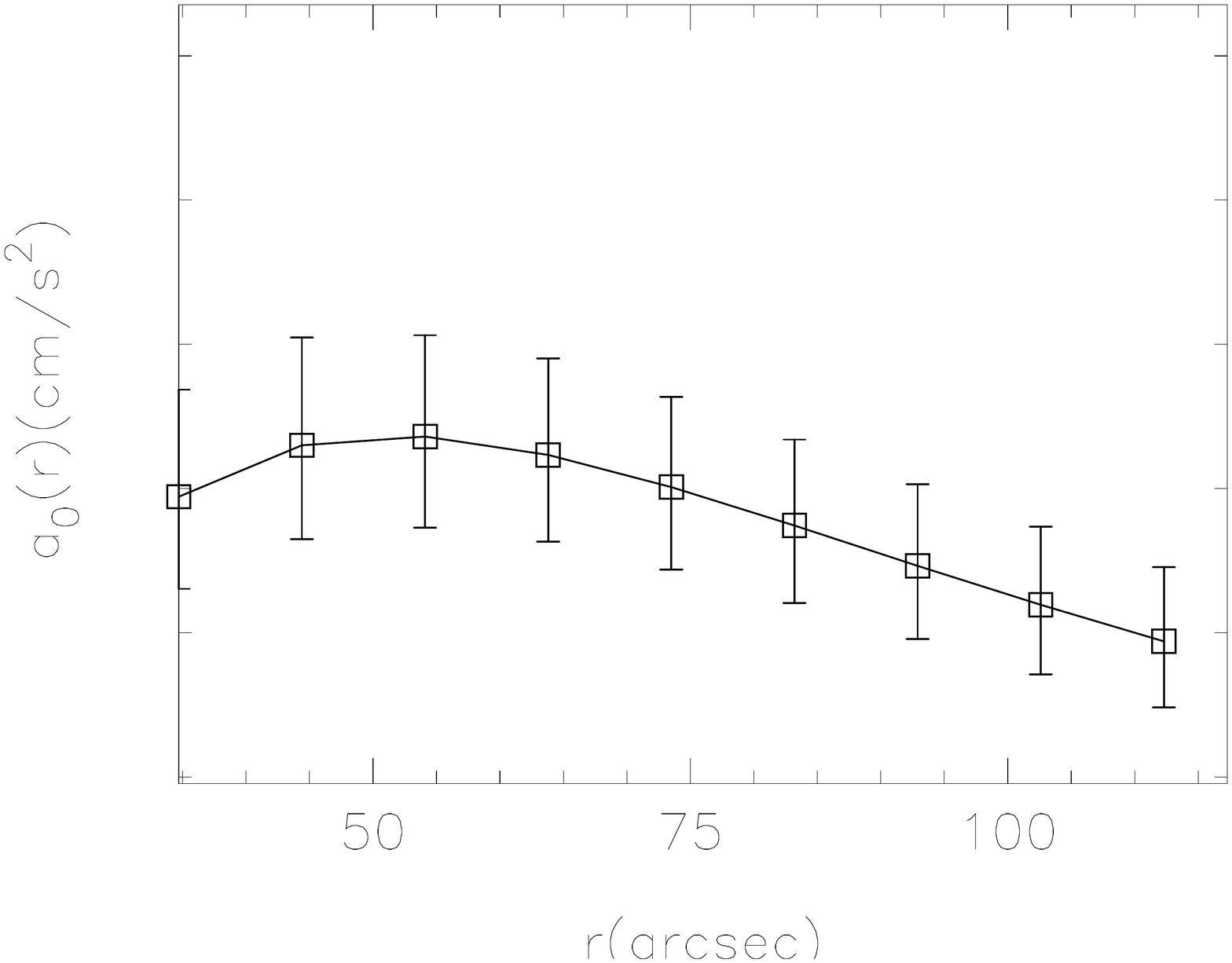}}
\subfloat[Abell586]{\label{Abell586a0}\includegraphics[width=3.0cm,angle=270]{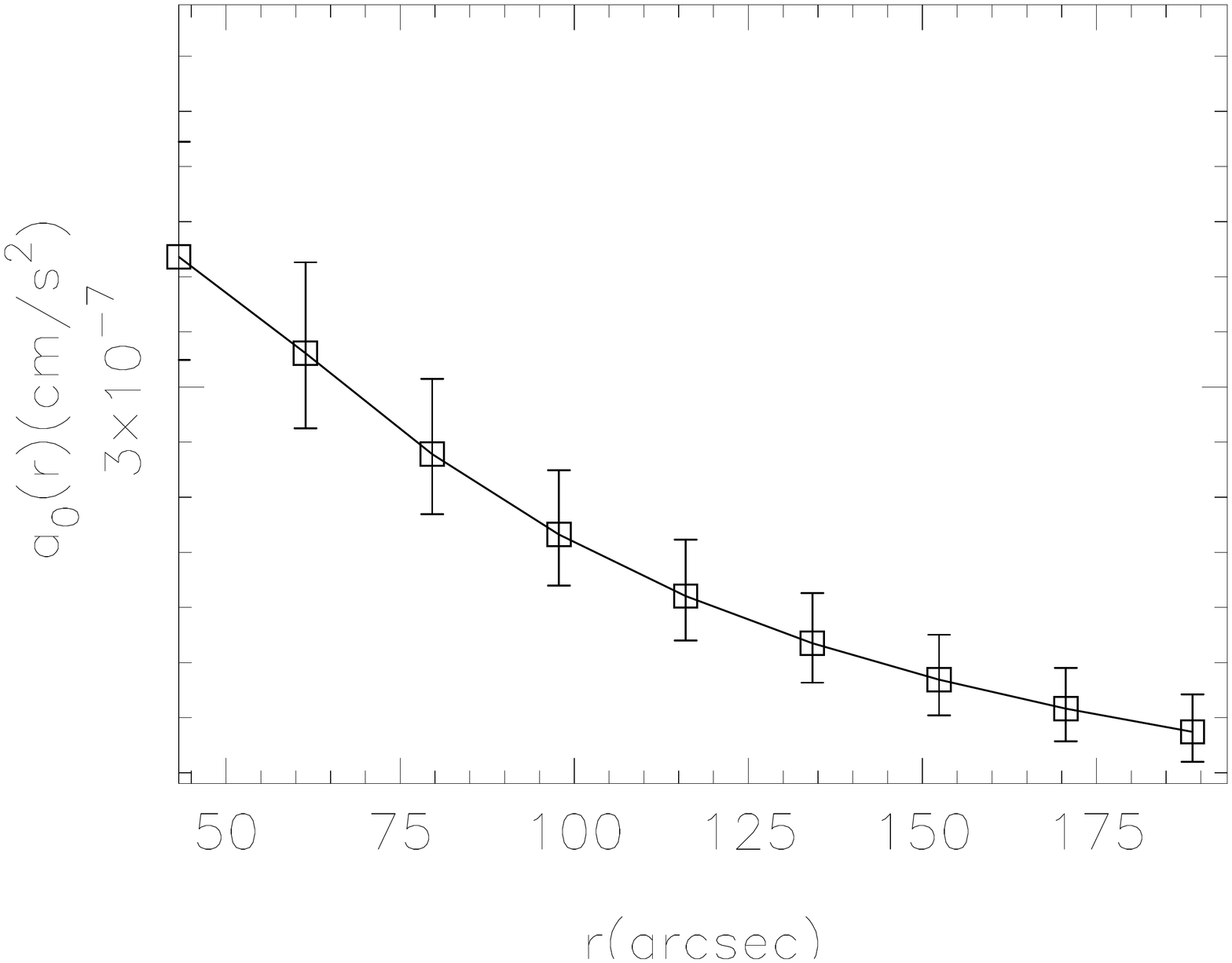}}
\subfloat[Abell611]{\label{Abell611a0}\includegraphics[width=3.0cm,angle=270]{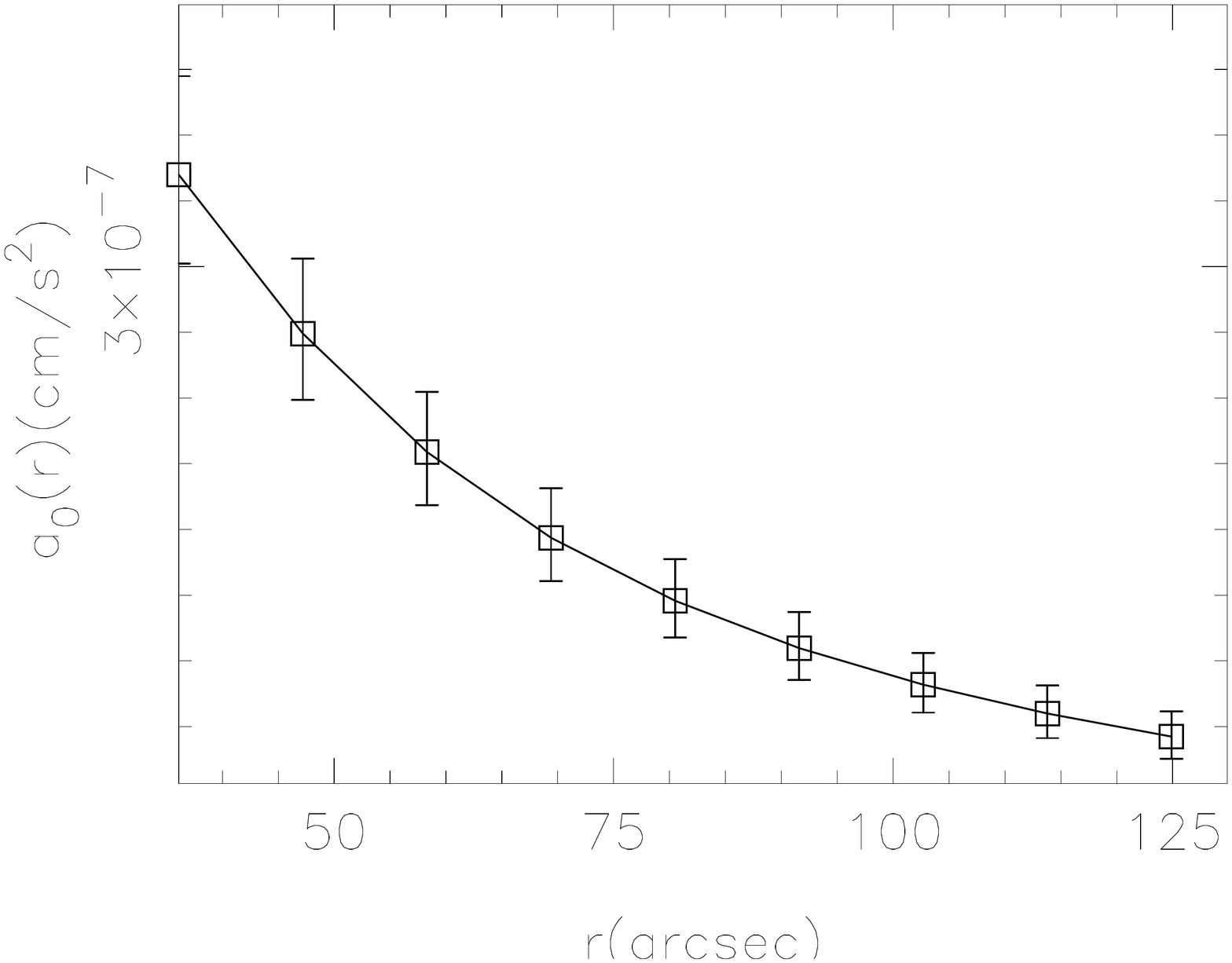}}

\subfloat[Abell665]{\label{Abell665a0}\includegraphics[width=3.0cm,angle=270]{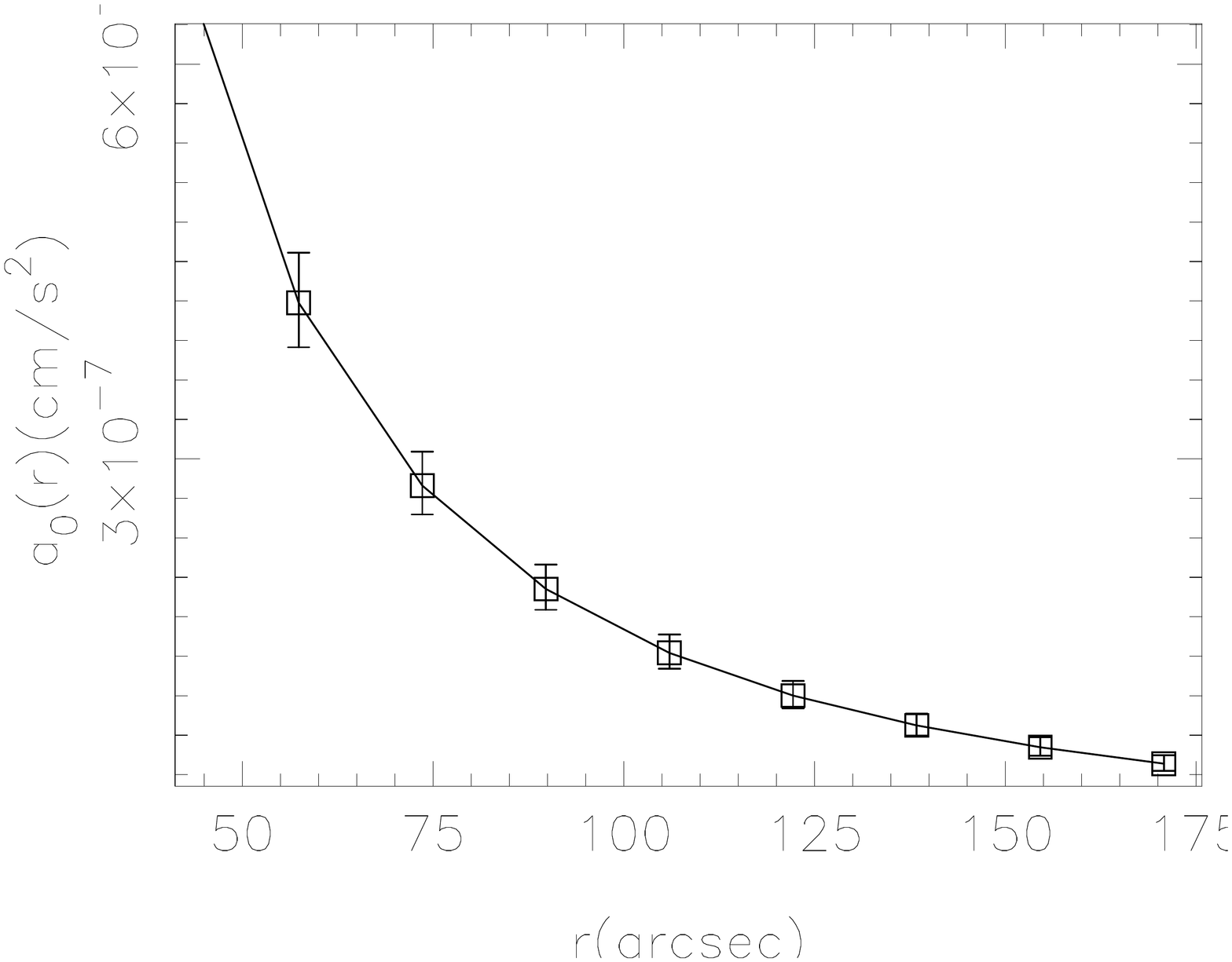}}
\subfloat[Abell68]{\label{Abell68a0}\includegraphics[width=3.0cm,angle=270]{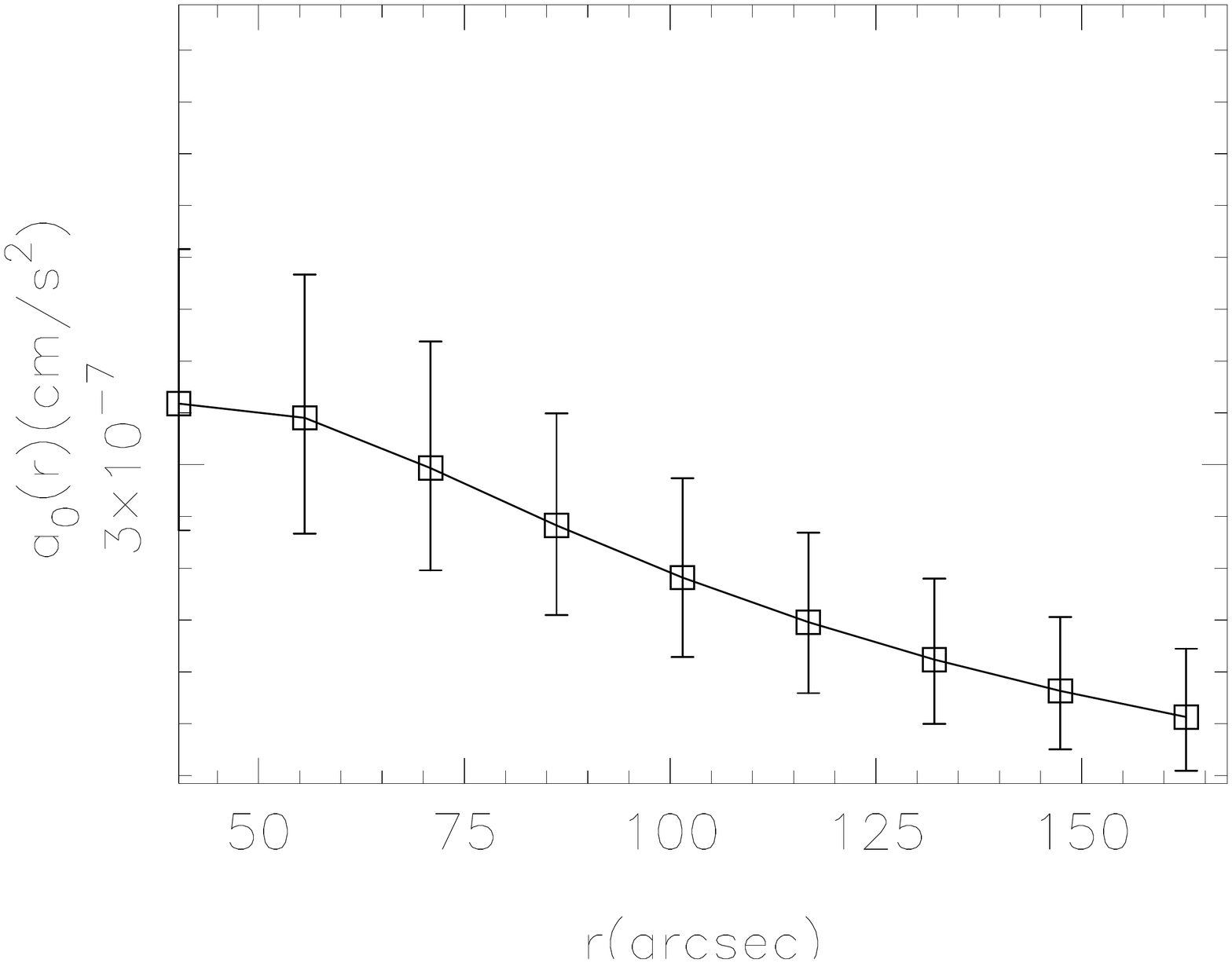}}
\subfloat[Abell697]{\label{Abell697a0}\includegraphics[width=3.0cm,angle=270]{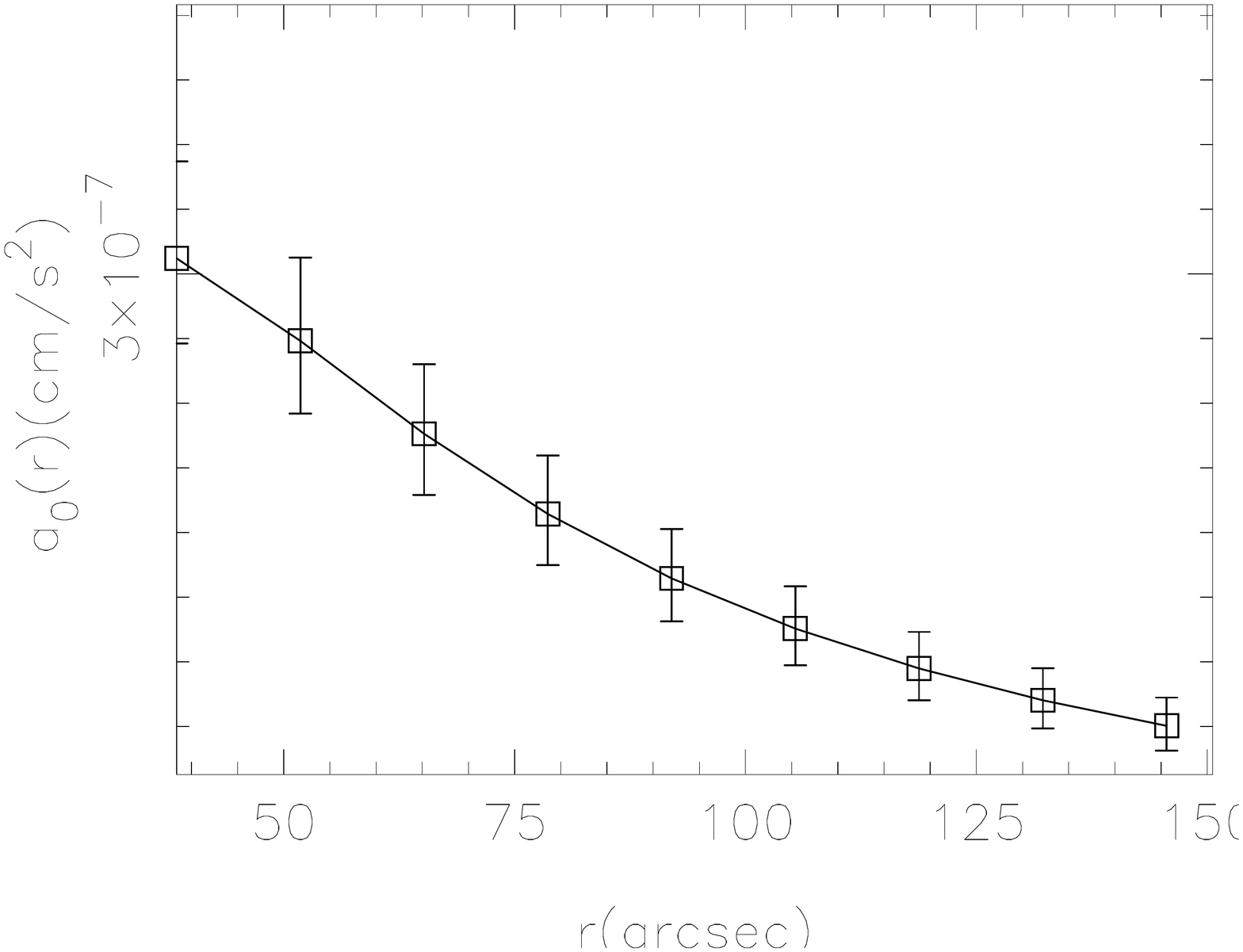}}
\end{figure}

\begin{figure}[p]
\ContinuedFloat
\centering
\subfloat[Abell773]{\label{Abell773a0}\includegraphics[width=3.0cm,angle=270]{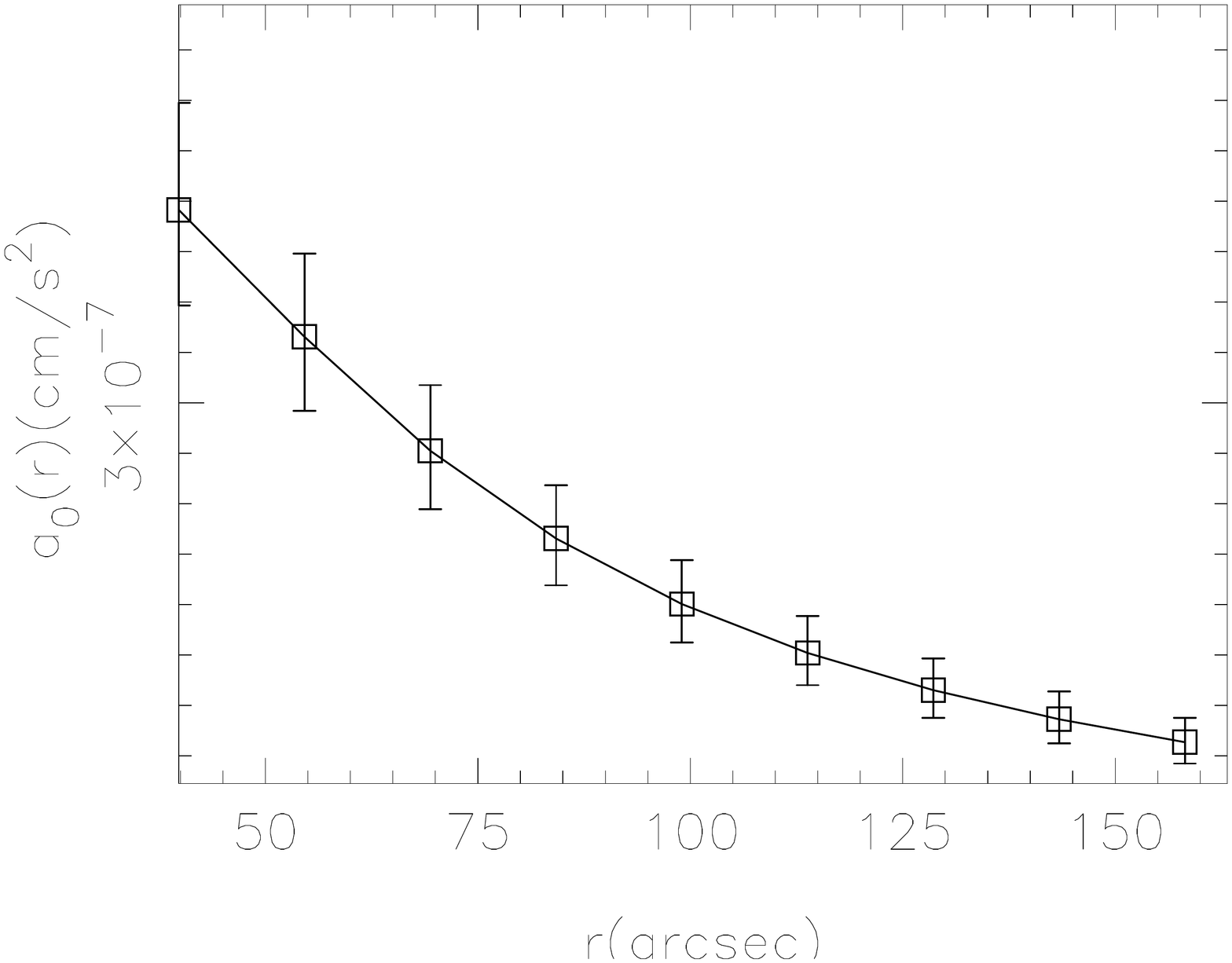}}
\subfloat[CLJ0016+1609]{\label{CLJ0016+1609a0}\includegraphics[width=3.0cm,angle=270]{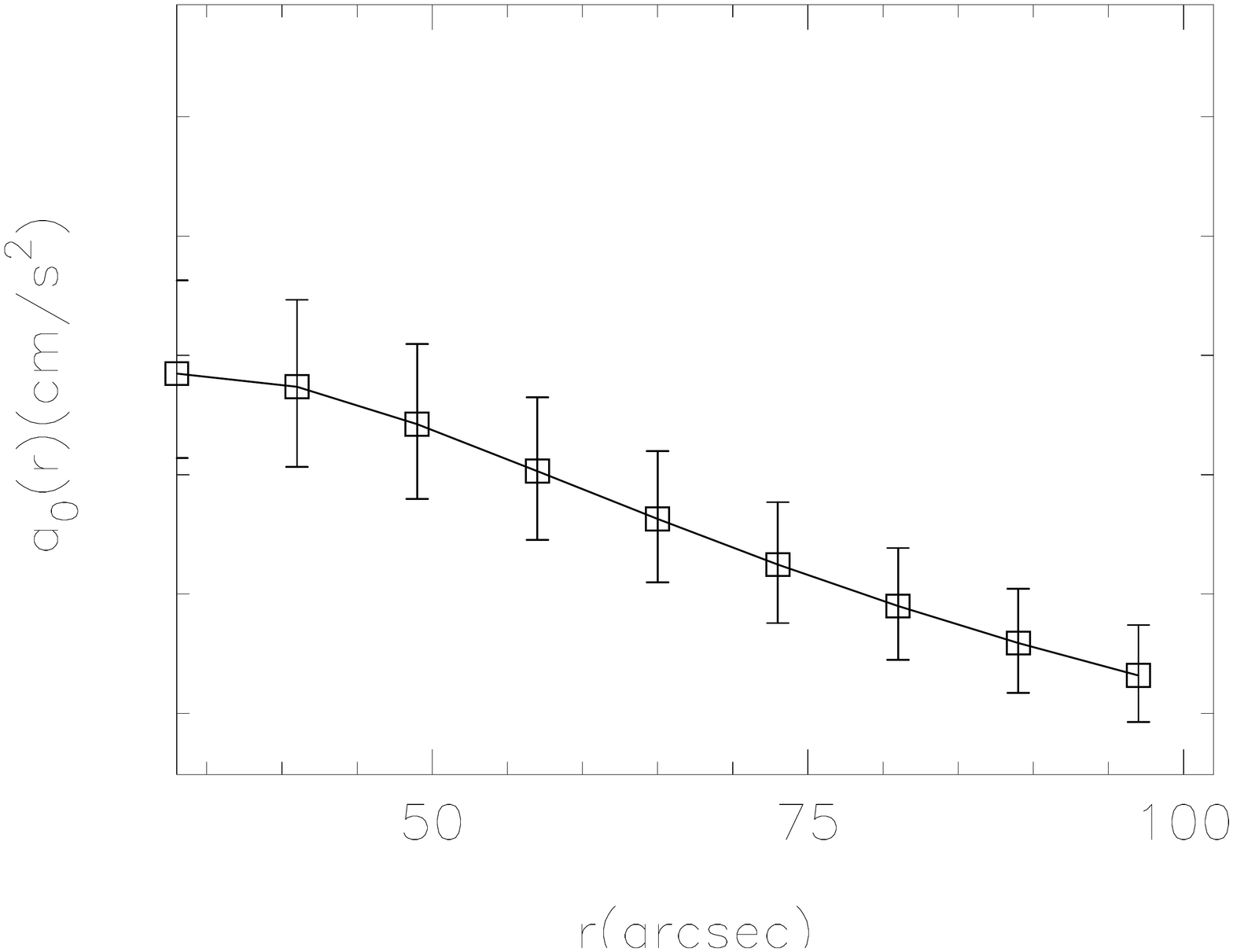}}
\subfloat[CLJ1226+3332]{\label{CLJ1226+3332a0}\includegraphics[width=3.0cm,angle=270]{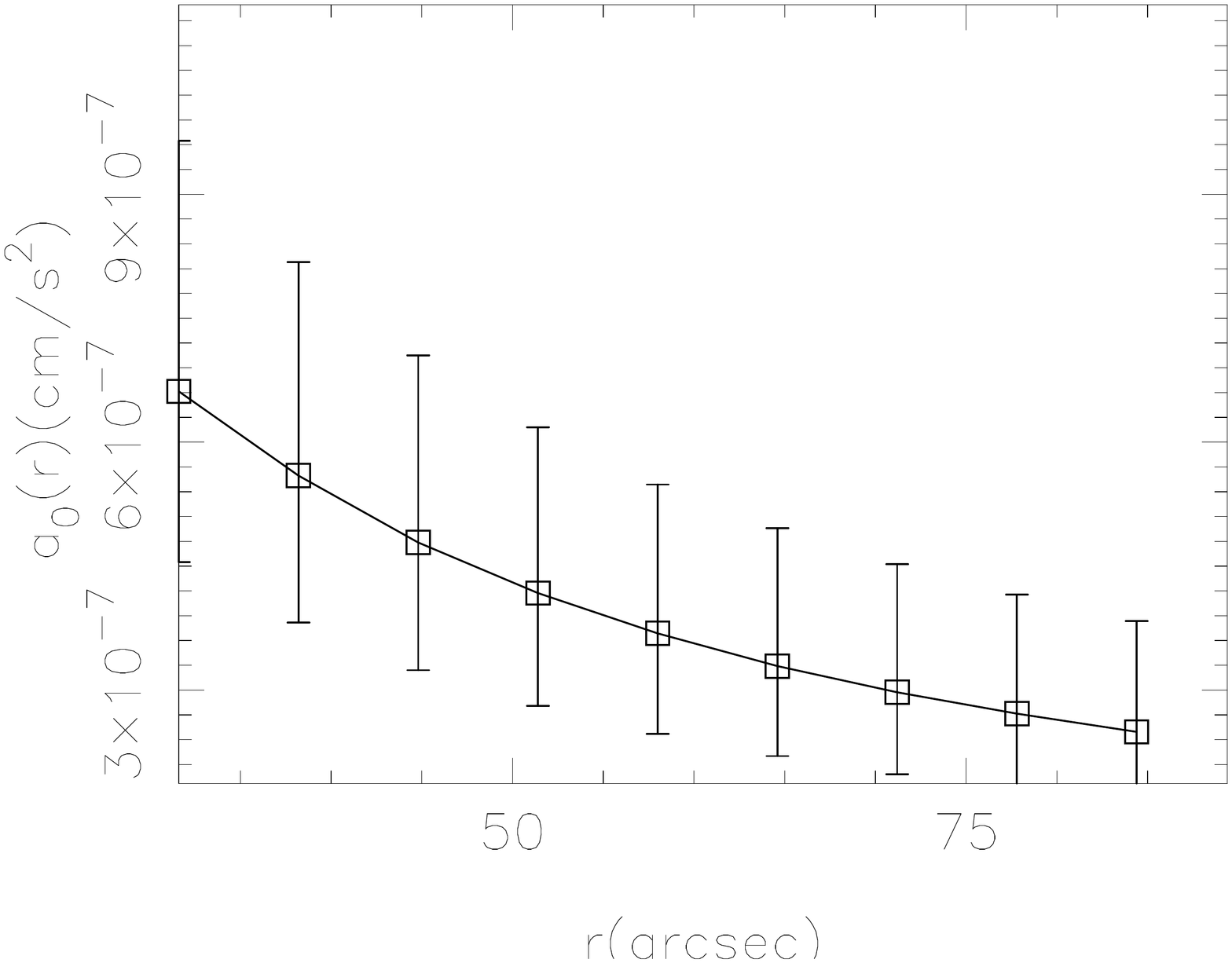}}

\subfloat[MACSJ0647.7+7015]{\label{MACSJ0647.7+7015a0}\includegraphics[width=3.0cm,angle=270]{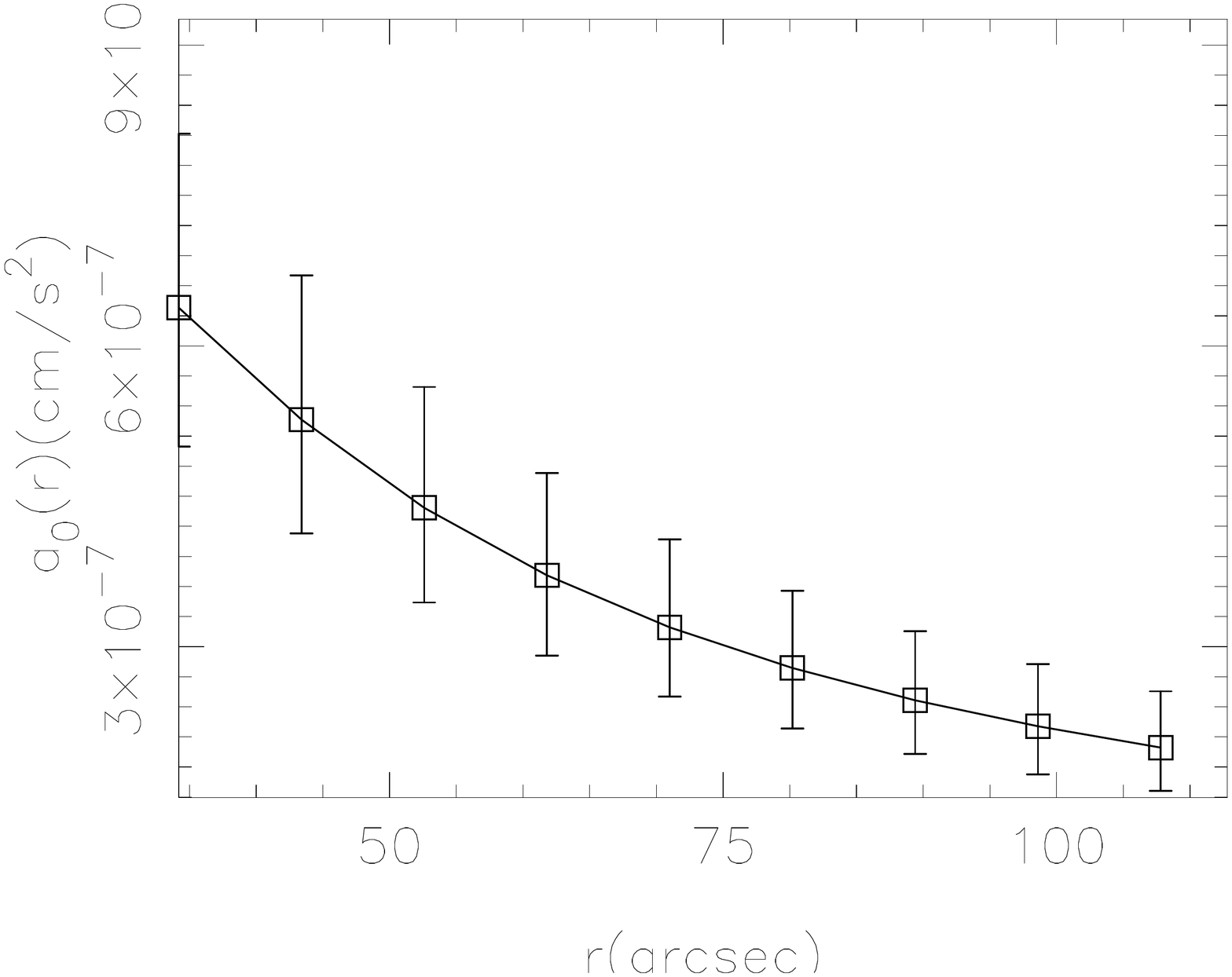}}
\subfloat[MACSJ0744.8+3927]{\label{MACSJ0744.8+3927a0}\includegraphics[width=3.0cm,angle=270]{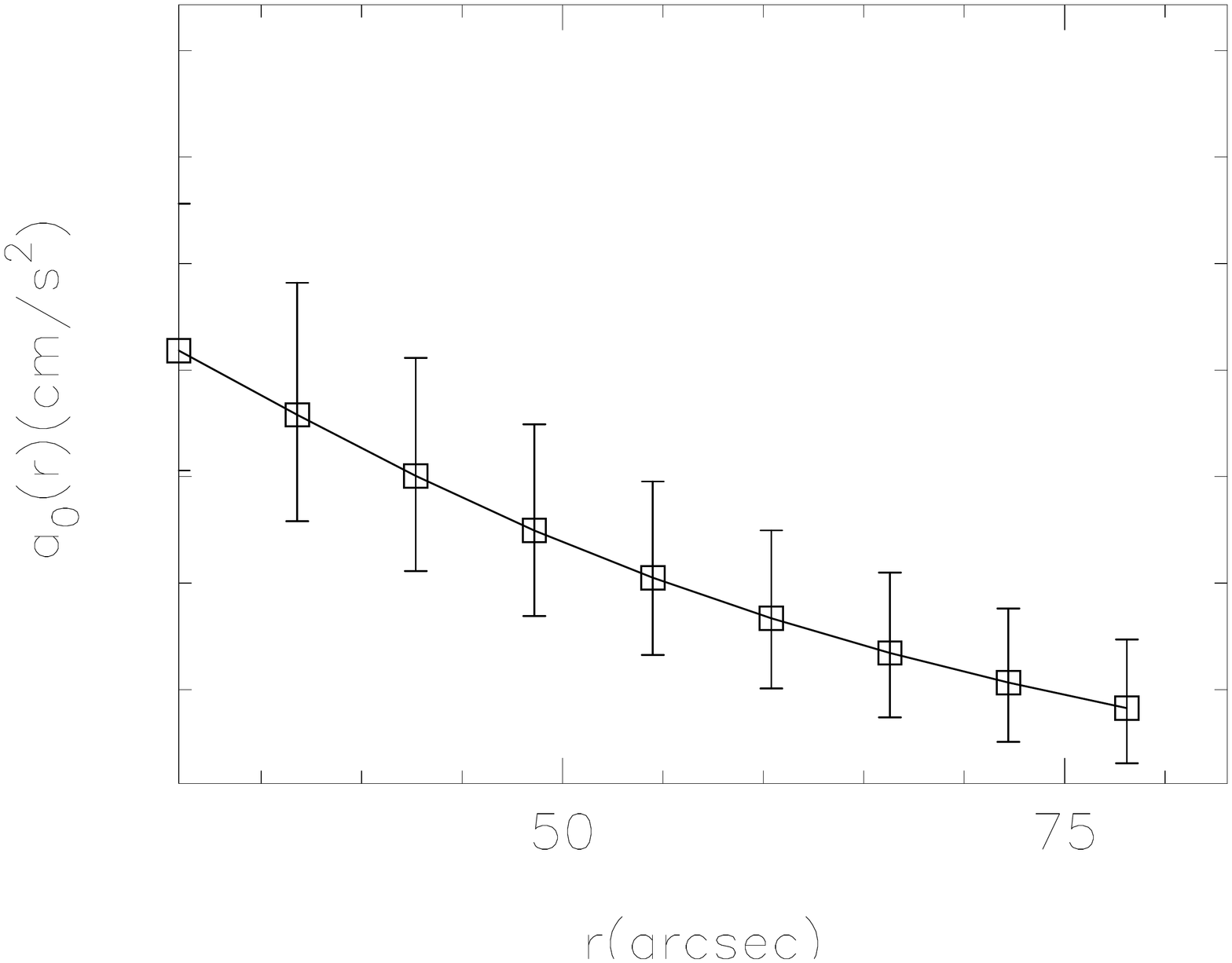}}
\subfloat[MACSJ1149.5+2223]{\label{MACSJ1149.5+2223a0}\includegraphics[width=3.0cm,angle=270]{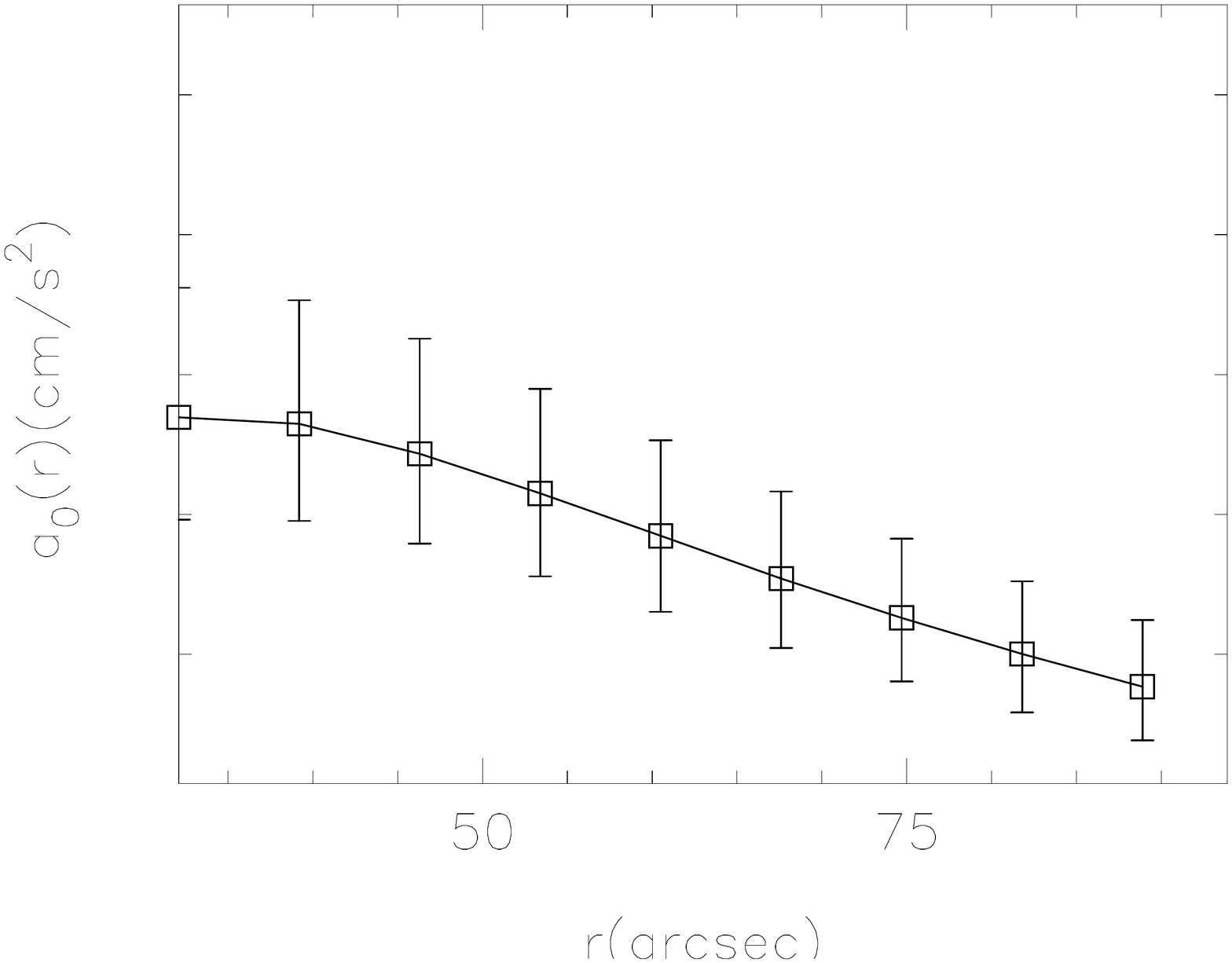}}

\subfloat[MACSJ1311.0-0310]{\label{MACSJ1311.0-0310a0}\includegraphics[width=3.0cm,angle=270]{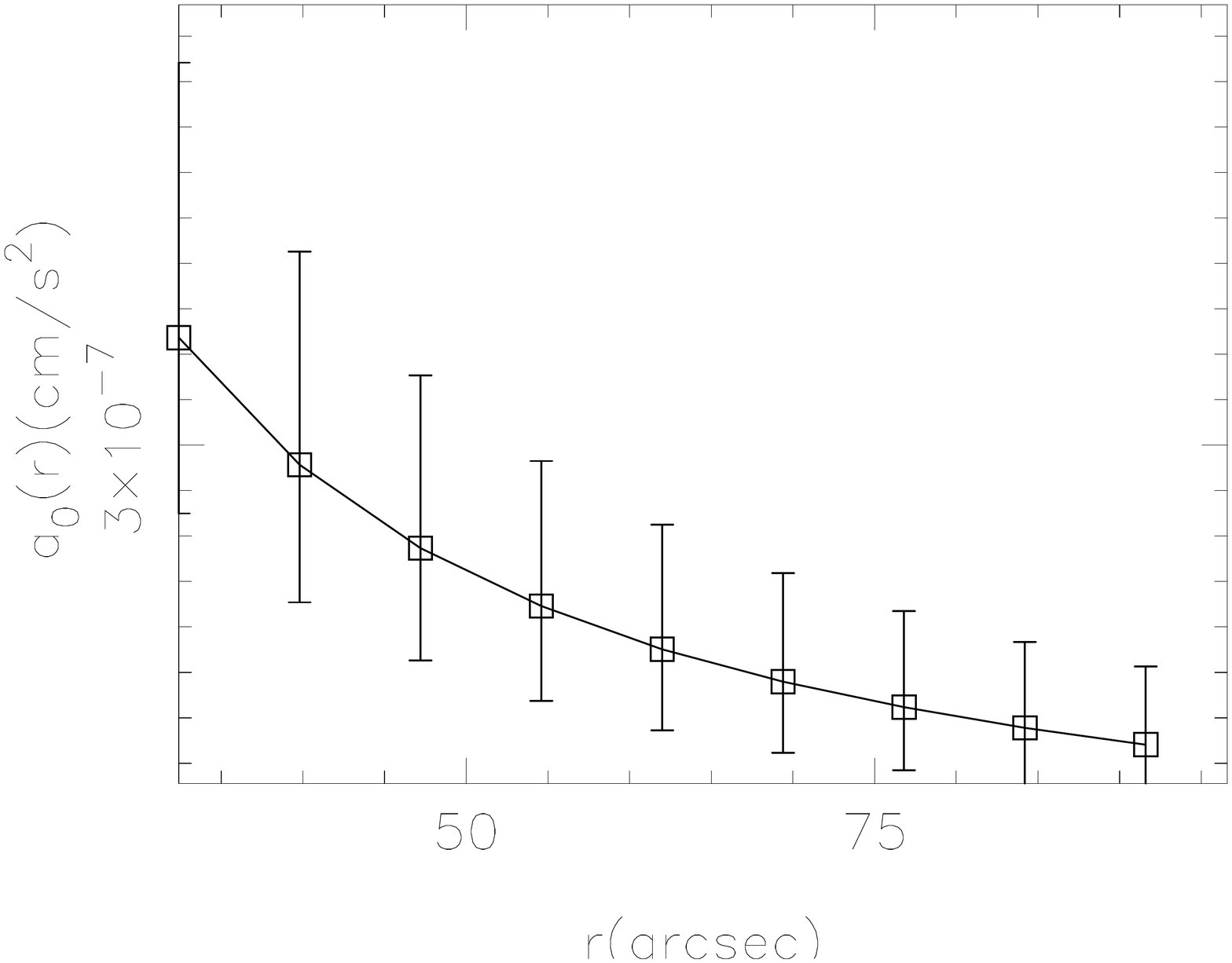}}
\subfloat[MACSJ1423.8+2404]{\label{MACSJ1423.8+2404a0}\includegraphics[width=3.0cm,angle=270]{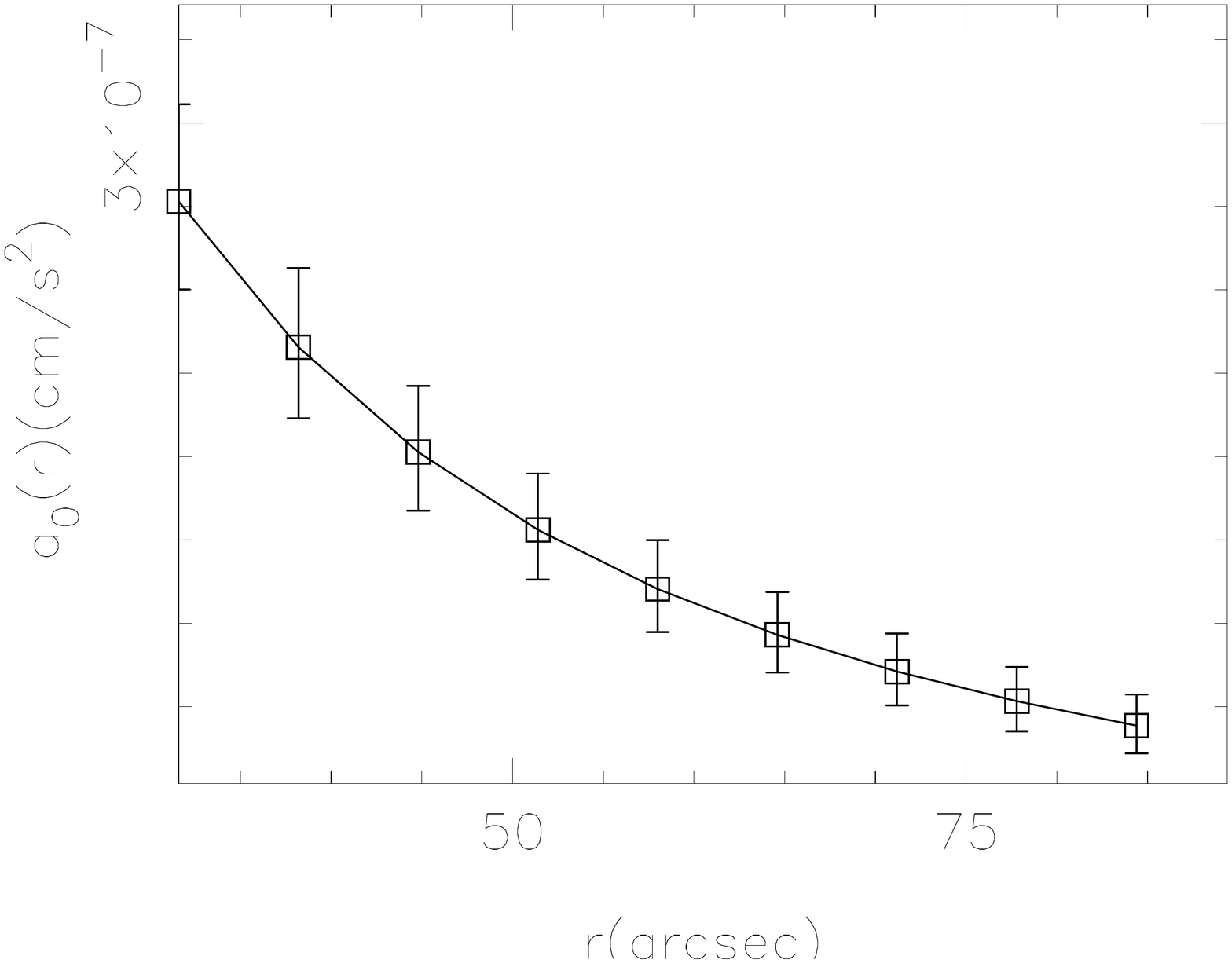}}
\subfloat[MACSJ2129.4-0741]{\label{MACSJ2129.4-0741a0}\includegraphics[width=3.0cm,angle=270]{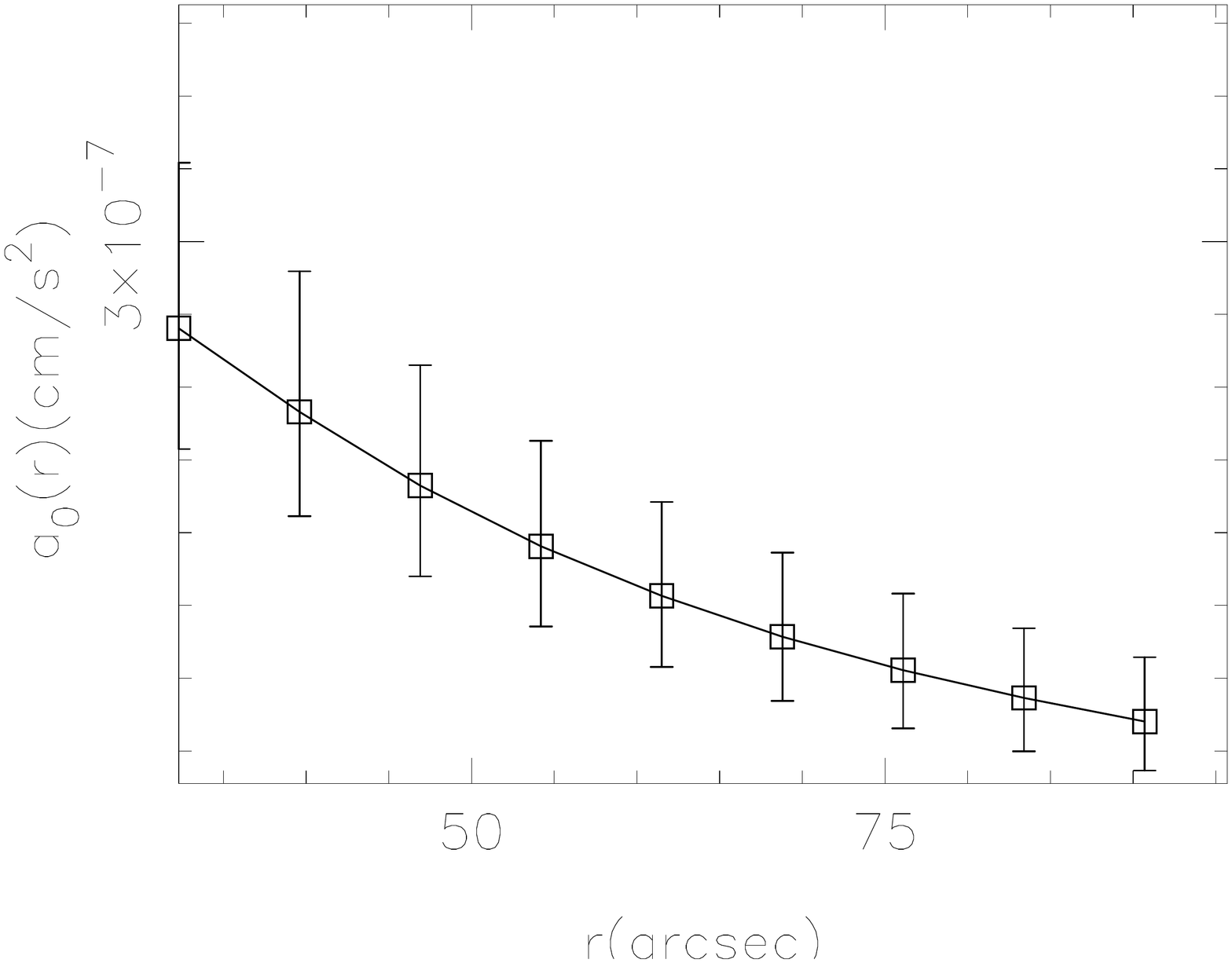}}

\end{figure}
\begin{figure}[p]
\ContinuedFloat
\centering
\subfloat[MACSJ2214.9-1359]{\label{MACSJ2214.9-1359a0}\includegraphics[width=3.0cm,angle=270]{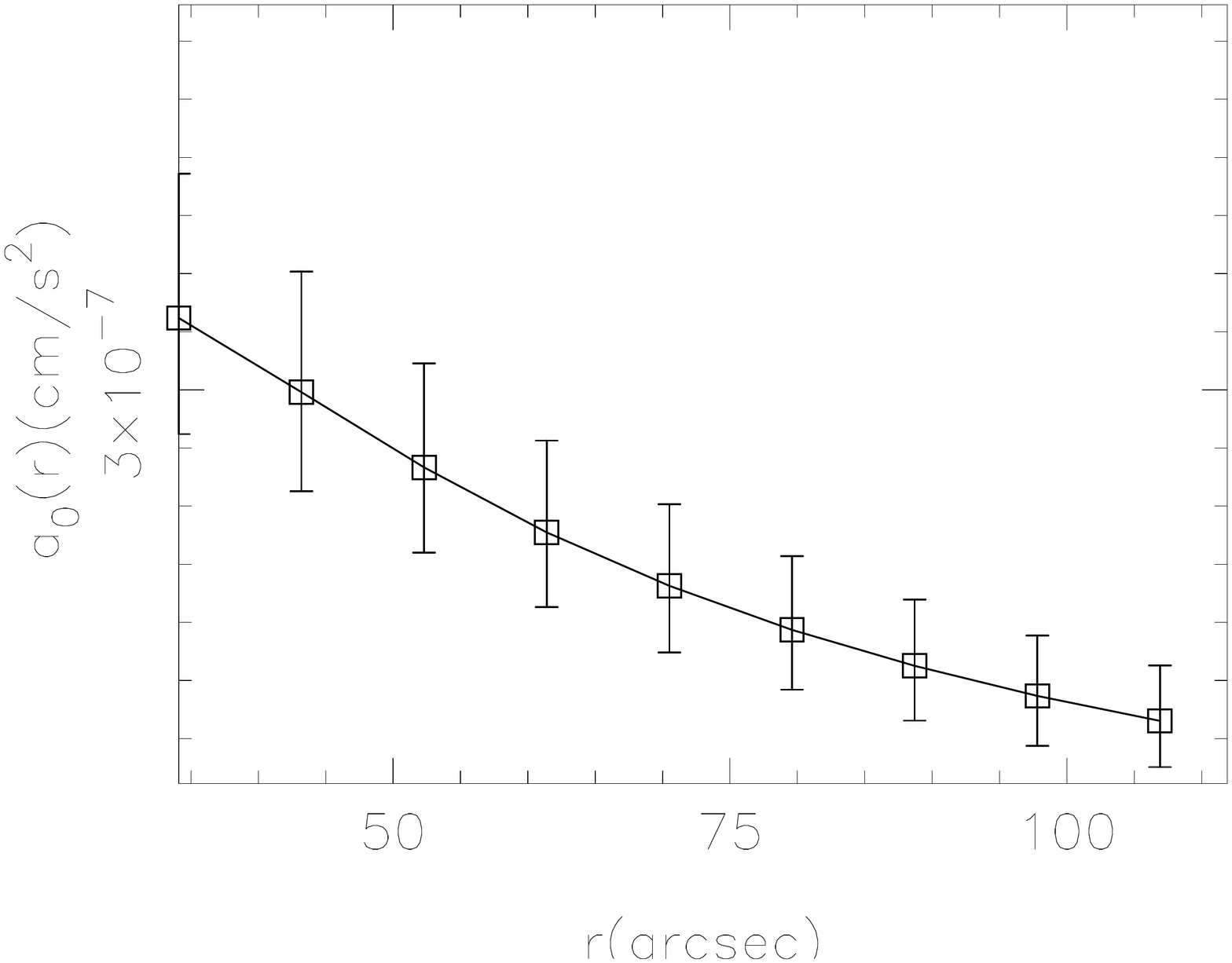}}
\subfloat[MACSJ2228.5+2036]{\label{MACSJ2228.5+2036a0}\includegraphics[width=3.0cm,angle=270]{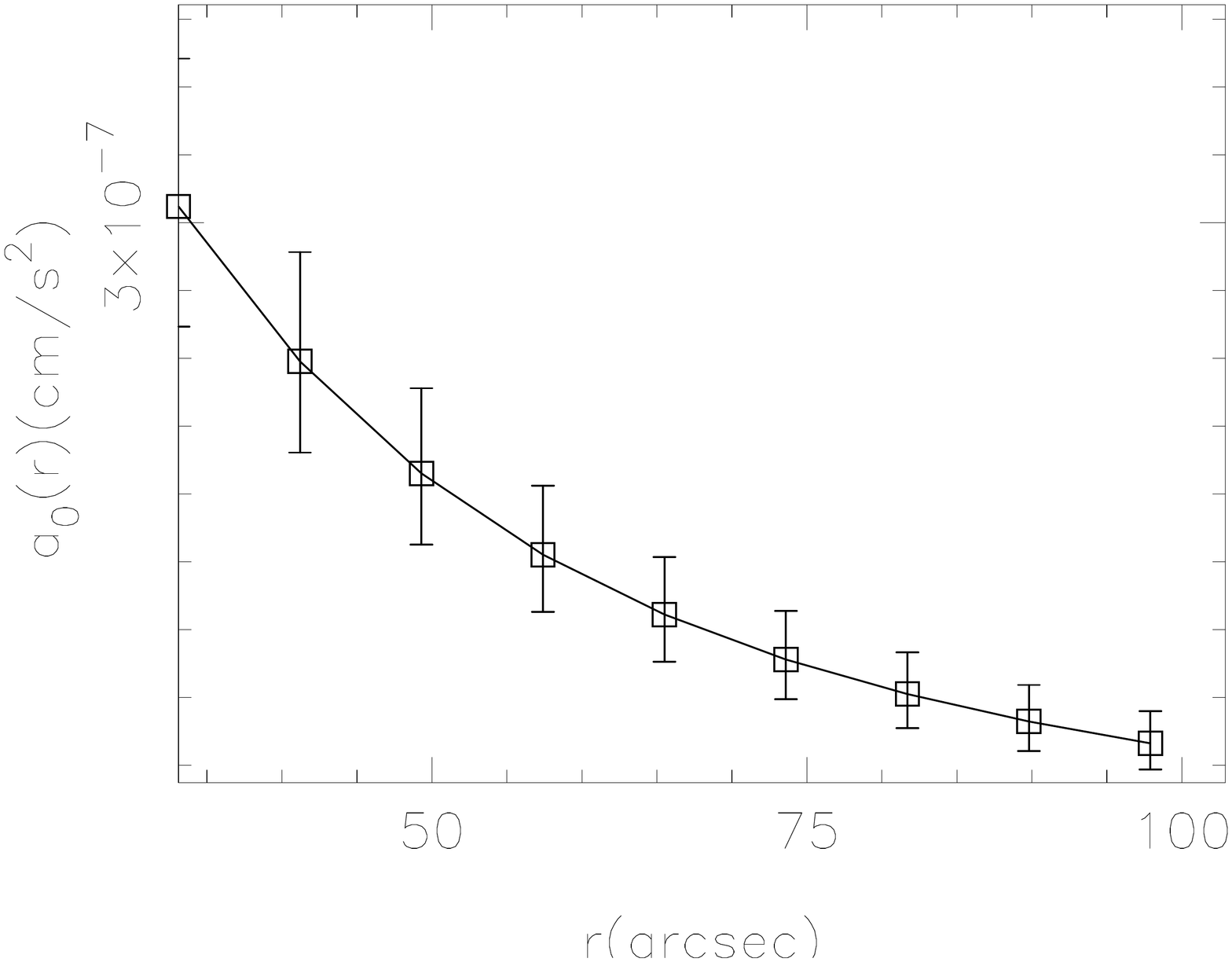}}
\subfloat[MS0451.6-0305]{\label{MS0451.6-0305a0}\includegraphics[width=3.0cm,angle=270]{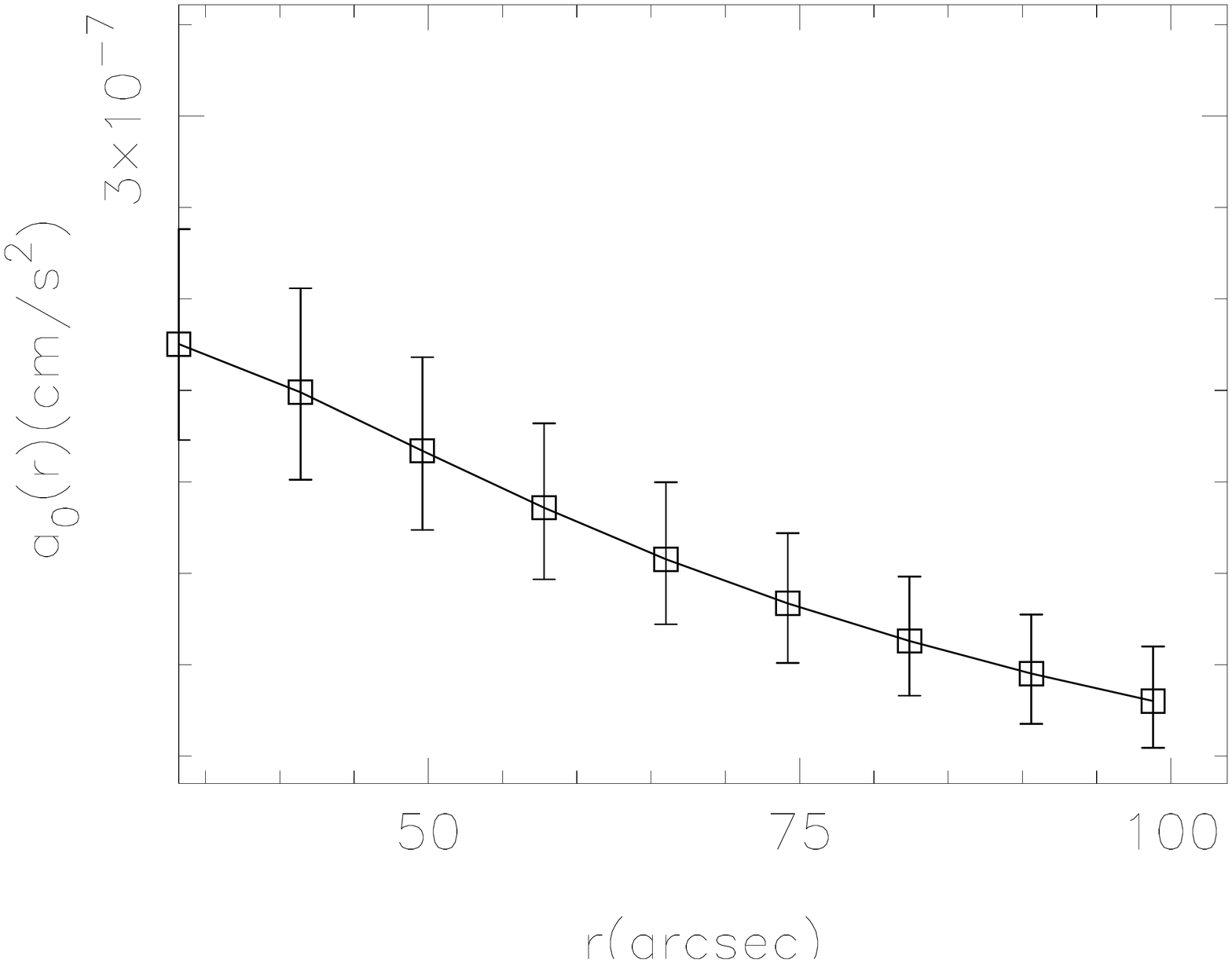}}

\subfloat[MS1054.5-0321]{\label{MS1054.5-0321a0}\includegraphics[width=3.0cm,angle=270]{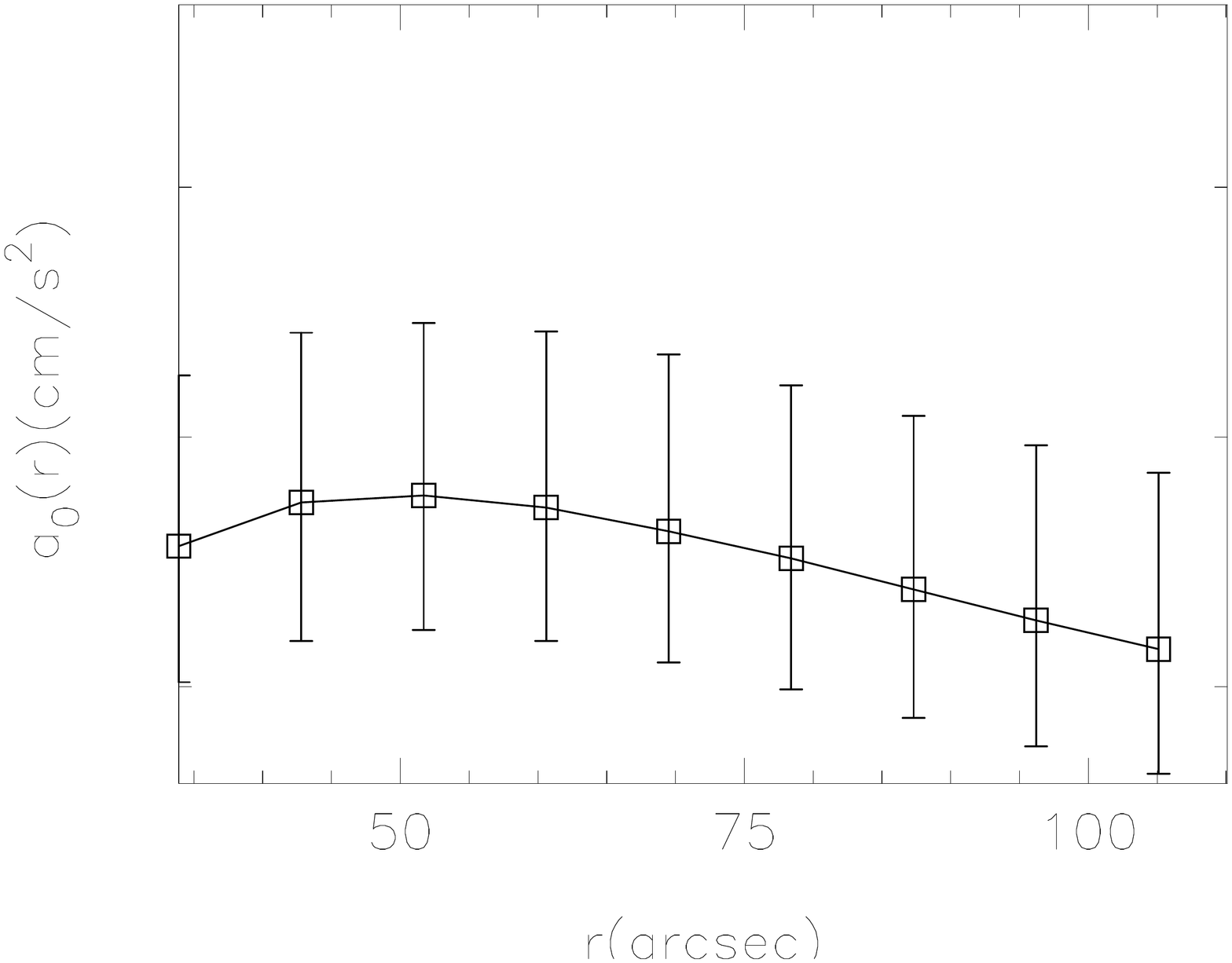}}
\subfloat[MS1137.5+6625]{\label{MS1137.5+6625a0}\includegraphics[width=3.0cm,angle=270]{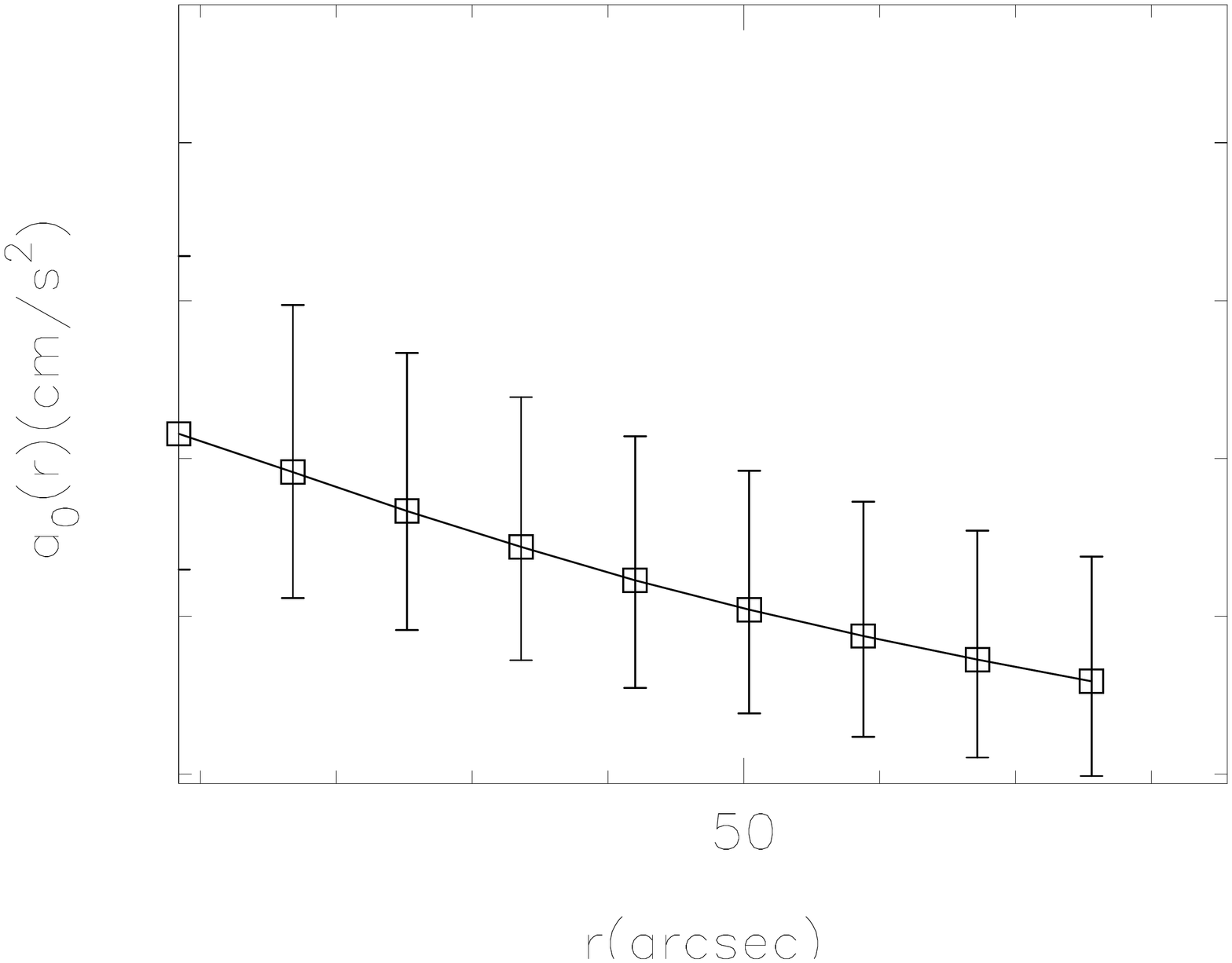}}
\subfloat[MS1358.4+6245]{\label{MS1358.4+6245a0}\includegraphics[width=3.0cm,angle=270]{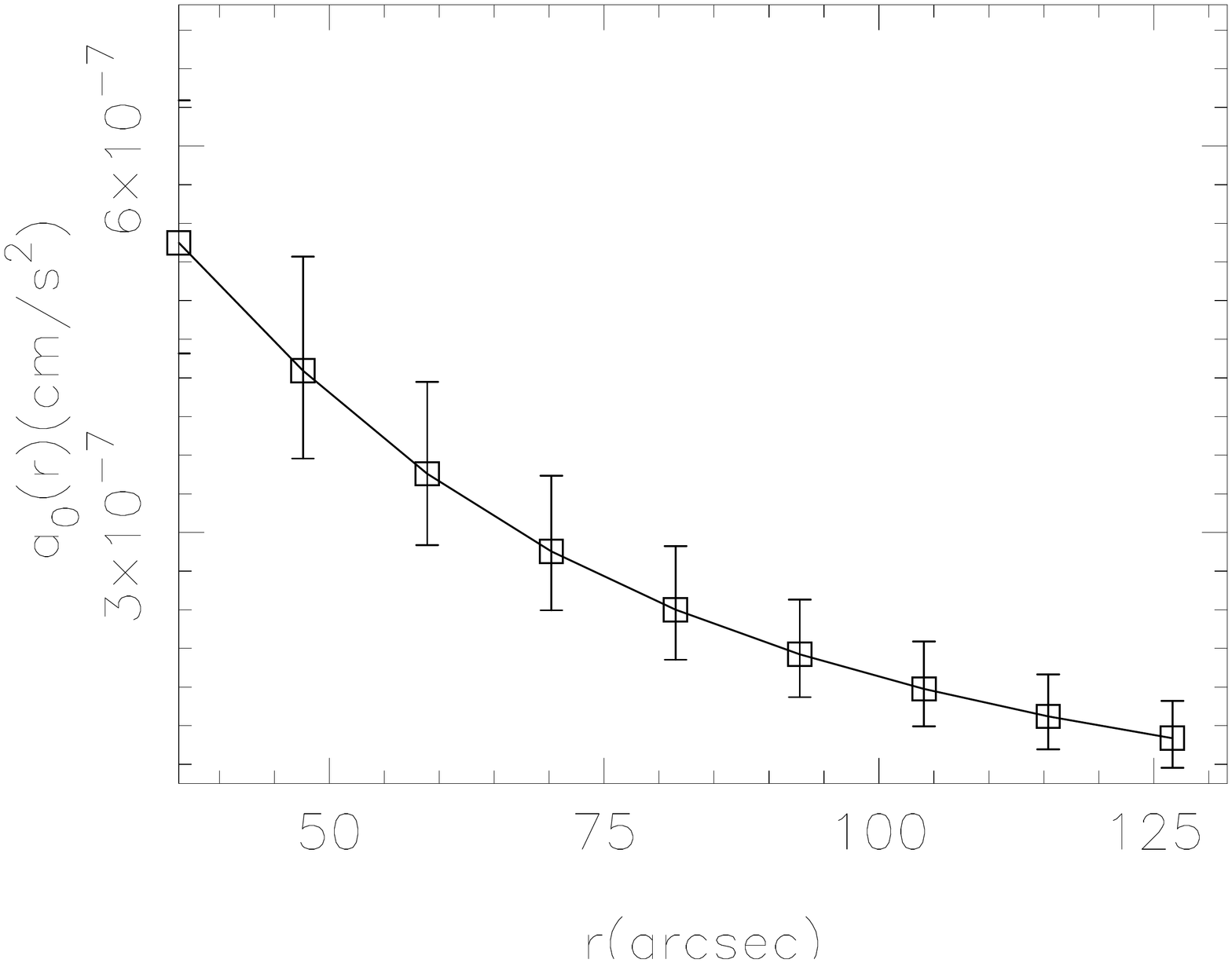}}

\subfloat[MS2053.7-0449]{\label{MS2053.7-0449a0}\includegraphics[width=3.0cm,angle=270]{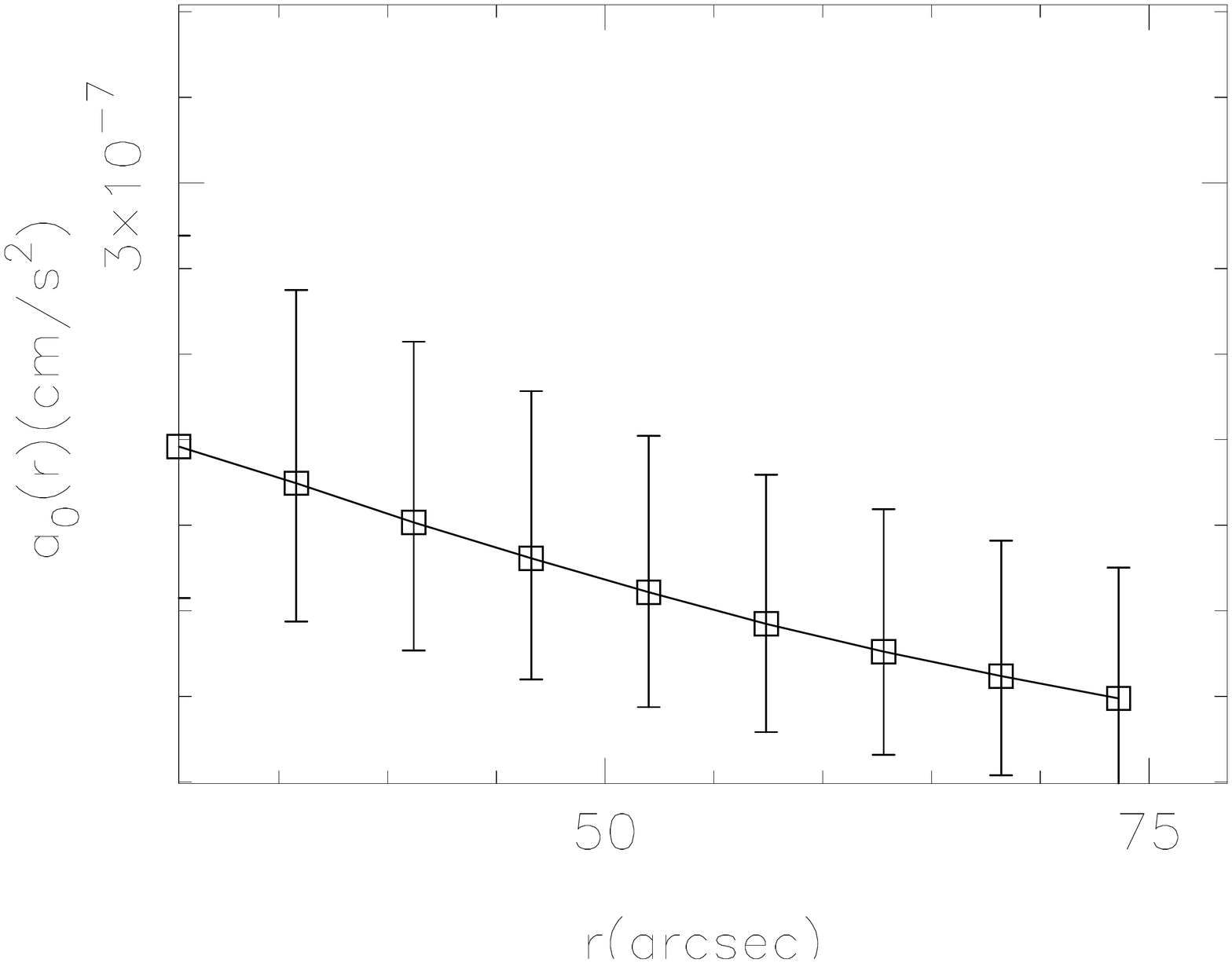}}
\subfloat[RXJ1347.5-1145]{\label{RXJ1347.5-1145a0}\includegraphics[width=3.0cm,angle=270]{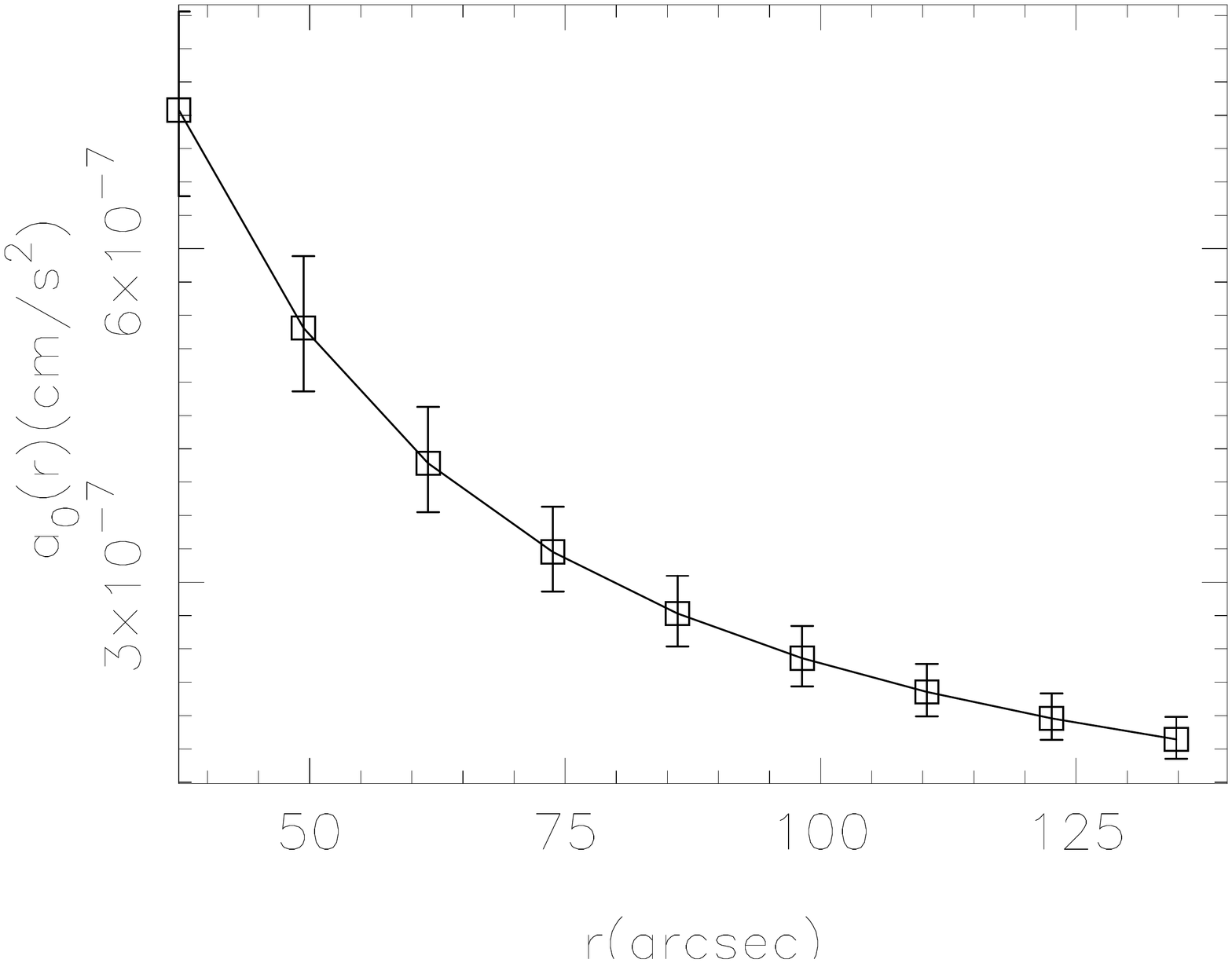}}
\subfloat[RXJ1716.4+6708]{\label{RXJ1716.4+6708a0}\includegraphics[width=3.0cm,angle=270]{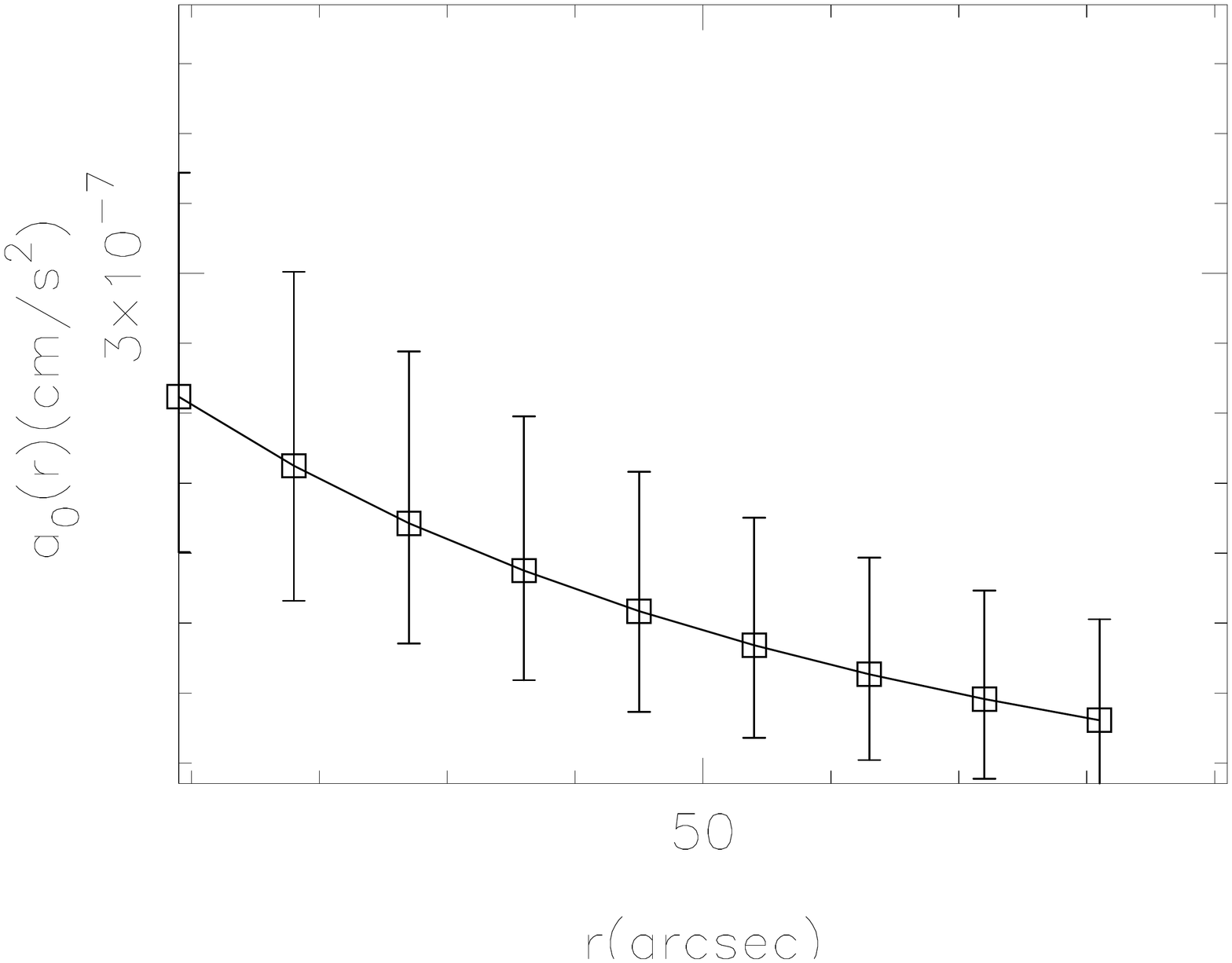}}

\subfloat[RXJ2129.7+0005]{\label{RXJ2129.7+0005a0}\includegraphics[width=3.0cm,angle=270]{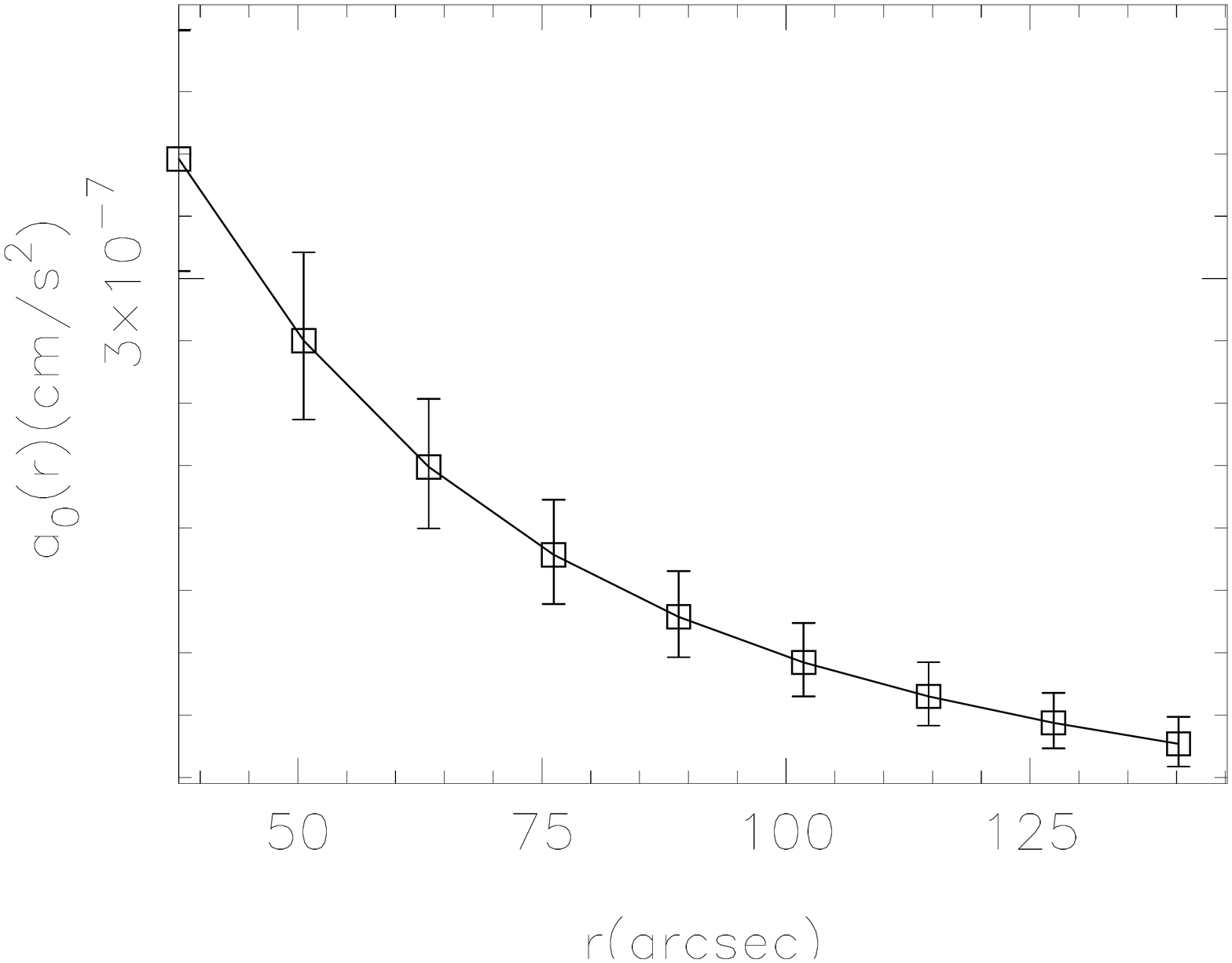}}
\subfloat[ZW3146]{\label{ZW3146a0}\includegraphics[width=3.0cm,angle=270]{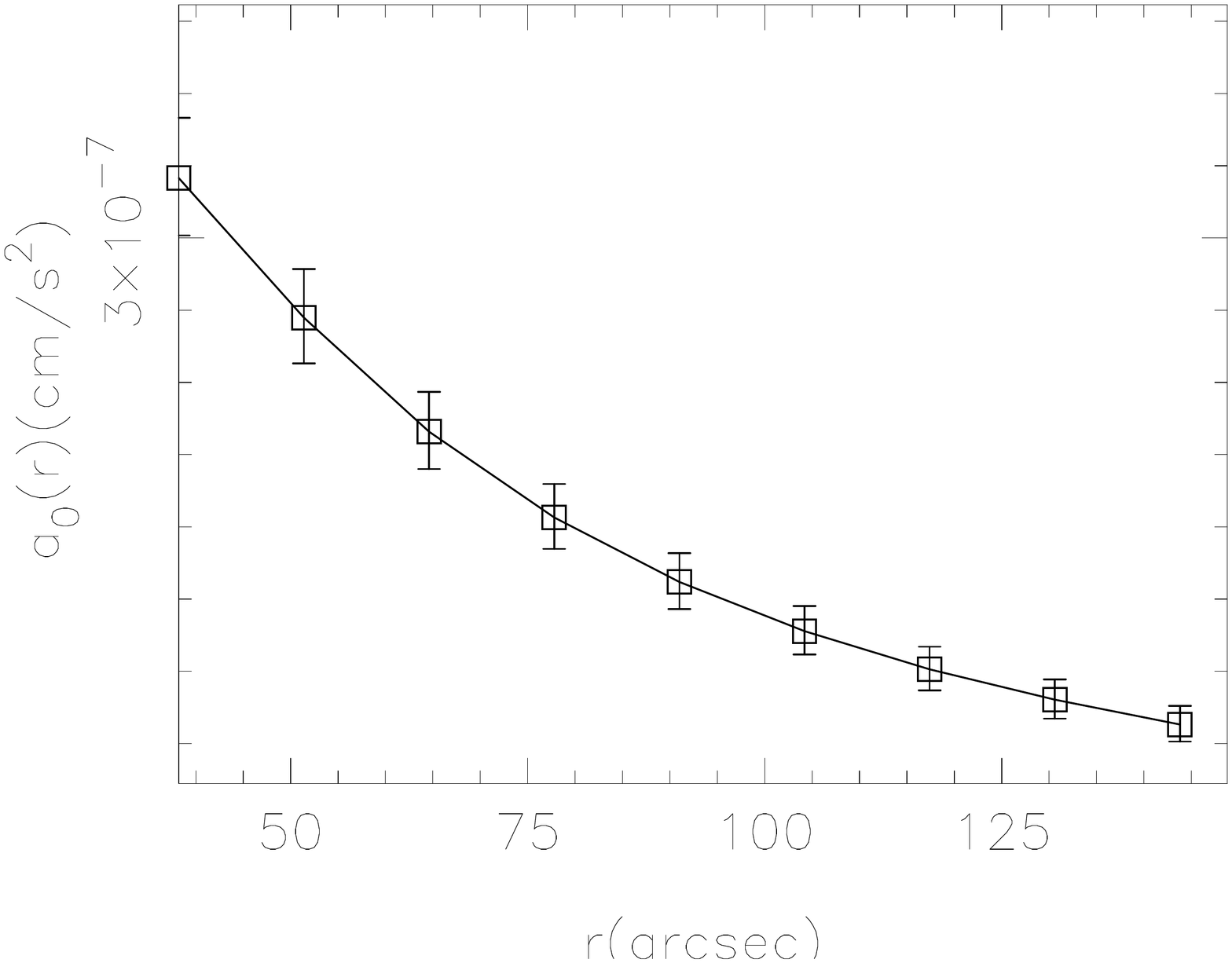}}
\end{figure}

\end{document}